\documentclass[fleqn,usenatbib]{mnras}

\usepackage{newtxtext,newtxmath}

\usepackage[T1]{fontenc}
\usepackage{ae,aecompl}

\usepackage{tabularx}
\usepackage{pdflscape}
\usepackage{graphicx}	
\usepackage{amsmath}	
\usepackage{relsize}    
\usepackage{pifont}

\newcommand{\cmark}{\ding{51}}%
\newcommand{\xmark}{\ding{55}}%

\title[NS pre-merger radio emission]{Pulsar revival in neutron star mergers: multi-messenger prospects for the discovery of pre-merger coherent radio emission}

\author[A. J. Cooper et al.]{
A. J. Cooper,$^{1,2}$\thanks{E-mail: a.j.cooper@uva.nl}
O. Gupta,$^{3,10}$
Z. Wadiasingh,$^{5,6,7}$ 
R. A. M. J. Wijers,$^{1}$
O. M. Boersma,$^{1,2}$ \newauthor
I. Andreoni,$^{4,5,6}$\thanks{Neil Gehrels Fellow.} 
A. Rowlinson,$^{1,2}$
K. Gourdji,$^{8,9}$
\\
$^{1}$ API, Anton Pannekoek Institute for Astronomy, University of Amsterdam, Science Park 904, 1098 XH Amsterdam, the Netherlands \\
$^{2}$ ASTRON, Netherlands Institute for Radio Astronomy, Oude Hoogeveensedijk 4, 7991 PD, Dwingeloo, the Netherlands\\
$^{3}$ Department of Physical Sciences, Indian Institute of Science Education and Research (IISER) Kolkata, Mohanpur 741246, West Bengal, India\\
$^{4}$ Joint Space-Science Institute, University of Maryland, College Park, Maryland 20742, USA \\
$^{5}$ Department of Astronomy, University of Maryland, College Park, Maryland 20742, USA \\
$^{6}$ Astrophysics Science Division, NASA Goddard Space Flight Center, Mail Code 661, Greenbelt, Maryland 20771, USA \\
$^{7}$ Center for Research and Exploration in Space Science and Technology, NASA/GSFC, Greenbelt, Maryland 20771, USA \\
$^{8}$ Centre for Astrophysics and Supercomputing, Swinburne University of Technology, Hawthorn VIC 3122, Australia \\
$^{9}$ OzGrav: ARC Centre of Excellence for Gravitational Wave Discovery, Hawthorn VIC 3122, Australia \\
$^{10}$ Department of Astronomy, University of Texas at Austin, Austin, Texas 78712, USA \\
}

\date{Accepted XXX. Received YYY; in original form ZZZ}

\pubyear{2022}
\usepackage{soul}
\usepackage[dvipsnames]{xcolor}

\begin{document}
\label{firstpage}
\pagerange{\pageref{firstpage}--\pageref{lastpage}}
\maketitle

\begin{abstract}
We investigate pre-merger coherent radio emission from neutron star mergers arising due to the magnetospheric interaction between compact objects. We consider two plausible radiation mechanisms, and show that if one neutron star has a surface magnetic field $B_{\rm s} \ge 10^{12}$G, coherent millisecond radio bursts with characteristic temporal morphology and inclination angle dependence are observable to Gpc distances with next-generation radio facilities. We explore multi-messenger and multi-wavelength methods of identification of a NS merger origin of radio bursts, such as in fast radio burst surveys, triggered observations of gamma-ray bursts and gravitational wave events, and optical/radio follow-up of fast radio bursts in search of kilonova and radio afterglow emission. We present our findings for current and future observing facilities, and make recommendations for verifying or constraining the model. 

 
\end{abstract}

\begin{keywords}
neutron star mergers -- stars: neutron -- fast radio bursts --  gamma-ray bursts -- acceleration of particles -- gravitational waves
\end{keywords}



\section{Introduction}
Compact object mergers involving black holes (BH) and neutron stars (NS) have long been hypothesized to power a range of high-energy, multi-messenger  astrophysical phenomena \citep{LattimerSchramm1974,ClarkEardley1977,Blinnikov1984,Eichlre1989,LiPaczynski1998}. Many of these predictions were confirmed by the discovery of an electromagnetic counterpart to gravitational wave event GW170817, which provided direct evidence of a common origin of gravitational waves from a NS-NS merger \citep{LIGO_GW_2017}, short gamma-ray bursts (sGRBs) powered by bipolar jets \citep{LIGO_GRB_GW_2017,Goldstein2017}, kilonovae \citep{LIGO_MM2017,Tanvir2017,Kasen2017,Pian2017} and late-time afterglow emission (e.g. \citealt{Alexander2017,Hallinan2017,LIGO_MM2017,Lazzati2018,Mooley2018,Dobie2018,Ghirlanda2019}). Prompt radio observations of the localisation region began 30 minutes post-merger \citep{Callister2017}, therefore radio emission mechanisms related to the inspiral or merger itself were not probed.
\par
Two-body electromagnetic interactions are thought to be the sources of coherent radio emission since the seminal work by \cite{GoldreichLydenBell1969} investigating the Jupiter-Io system. The inspiral phase that precedes coalescence during NS-NS mergers has been hypothesised to give rise to pre-merger electromagnetic emission due to the strong surface magnetic fields present in known pulsars and magnetars. A range of mechanisms have been considered as candidates to power precursor electromagnetic emission during NS mergers including: the unipolar inductor model \citep{HansenLyutikov2001,Lai2012,Piro2012}, resonant NS crust shattering \citep{Tsang2012,Suvorov2020}, magneto-hydrodynamic plasma excited by gravitational waves \citep{Moortgat2006} nuclear decay of tidal tails \citep{Roberts2011}, the formation of a optically thick fireball \citep{Vietri1996,MetzgerBerger2012,Beloborodov2021}, wind-driven shocks \citep{Medvedev2013,Sridhar2021} and particle acceleration through the revival of pulsar-like emission during inspiral \citep{Lipunov1996,Lyutikov2019}. Numerical studies of force-free electromagnetic interaction between magnetized NS-NS binaries support the conclusion that flares may be observed before the merger \citep{Palenzuela2013,Most2020,Most2022}. 
\par
High-energy precursor emission has previously been observed in sGRBs, and may be a powerful tool to either infer properties of the merging compact objects (see Section. \ref{sect:temporal}), or enable targeted observations of the merger event by automated or rapid slew. \cite{Troja2010} found that 8-10\% of \textit{Swift} detected sGRBs display precursor emission in the \textit{Swift} bandwidth, compared to around 20\% of long GRBs (lGRBs) detected by Burst and Transient Source Experiment (\textit{BATSE}; \citealt{Lazzati2005}). Furthermore, \cite{Coppin2020} find that lGRBs observed by \textit{Fermi} are 10 times more likely to display precursor gamma-ray emission before the main burst, as compared to sGRBs. However, we note that apparent precursor emission may be merely a manifestation of variable prompt gamma-ray burst emission: \cite{Charisi2015} show that precursor and post-cursor emission are statistically similar, and may share a common origin. The recent claim of a quasi-periodic precursor emission to coincident with a kilonovae-associated lGRB further implies that precursor emission could be used to infer the nature of the progenitor (e.g. probing crust shattering events at small separation; \citealt{Suvorov2022}), as well as the resultant post-merger compact object \citep{Xiao2022}.
\par
Predictions of pre-merger emission are particularly relevant in light of new observational techniques with which theoretical predictions can be tested. For example, automatic triggered observations in response to high-energy alerts, particularly by software interferometers such as the LOw Frequency ARray (LOFAR \citealt{2013A&A...556A...2V}) \& the Murchison Widefield Array (MWA; \citealt{2013PASA...30....7T}) where physical repointing is not necessary, enable rapid observations on source within $\approx$ 4 minutes and $\approx$ 10 seconds, respectively. Observations triggered by GRB or gravitational wave (GW) alerts (e.g. \citealt{Anderson2021,Rowlinson2021,Tian2022}) also benefit from dispersion delay of low-frequency emission during propagation, which at LOFAR frequencies of 144 MHz corresponds to approximately 3 minutes of delay between gamma-ray and radio signals for a redshift of $z\sim$1 (corresponding to a dispersion measure of $\sim$800 pc cm$^{-3}$). Furthermore, these techniques are enhanced if raw data is saved using transient buffers enabling negative latency triggers (e.g. \citealt{terVeen2019}), or making use of early time alerts from gravitational wave detectors \citep{James2019}. Rapid radio observations of sGRBs can test theories of coherent radiation after the merger \citep{Rowlinson2019} predicted in shocks \citep{UsovKatz2000,SagivWaxman2002}, or from a short-lived magnetar remnant \citep{RowlinsonMetzger2013}. The constraints set by past observations of kind are discussed with respect to the model presented here in Section. \ref{sect:ShortGRBs}. 
\par
Theoretically, interest in the electromagnetic interaction between coalescing compact objects has recently been revived to explain a subset of fast radio bursts (FRBs) \citep{Lorimer2007,Thornton2013,Spitler2014,Petroff2016,Petroff2021} which do not appear to repeat \citep{Totani2013,Zhang2014,Wang2016, Gourdji2020}. This was further bolstered by the tentative association of FRB 150418 with a fading radio source, which was originally thought to be consistent with a short gamma-ray burst afterglow \citep{Keane2016}; however the association was later ruled out as unrelated transient activity from an active galactic nucleus (AGN) \citep{Williams2016}. Recent estimates of the rates of NS-NS mergers (see \citealt{Andreoni2021,Mandel2021} and references therein) suggest a local volumetric NS-NS merger rate not more than $\approx 10^{3} \; {\rm Gpc^{-3} \, yr^{-1} }$, which appears to fall short of the volumetric FRB rate \citep{Luo2020}. Comparing the volumetric rates of FRBs and plausible cataclysmic progenitors, \cite{Ravi2019} demonstrates that most apparently one-off FRBs likely repeat albeit with a low repetition rate, and thus NS mergers may power a fraction of truly non-repeating FRBs. It is an open question as to whether sub-populations of FRBs evolve differently as a function of redshift; \cite{Hashimoto2020b} suggest that an older stellar population underlies one-off FRBs, which may support e.g. merger progenitors, although this is disputed by \cite{James2022}. Interactions between magnetized neutron stars and other objects such as black holes \citep{Zhang2016} or asteroids \citep{Dai_2016,Voisin2021} could also induce magnetospheric interactions resulting in coherent emission observable to cosmological distances. This is discussed further in Section \ref{sect:Discussion}, and many of the results of Sect. \ref{sect:particle_acceleration} are relevant for NS interactions with other conducting bodies.
\par
In this work, we take a detailed look at the plausible emission due to the magnetospheric interaction between merging neutron stars, adapting and extending the work of \cite{Lyutikov2019}. We implement plausible radiation mechanisms, and make robust predictions of the temporal evolution and observer inclination angle dependence of pre-merger coherent radiation emission. The paper is organised as follows. In Section \ref{sect:model}, we present the electrodynamic model of the NS-NS merger originally suggested by \cite{Lyutikov2019}, and present adaptations to the model with a full derivation available in Appendix \ref{sect:Om_derivation}. In Section \ref{sect:particle_acceleration} we discuss plausible particle acceleration and radiation mechanisms that result in coherent pre-merger emission, and estimate the radio luminosity. In Section \ref{sect:MM_MW_prospects} we discuss prospects of detecting such emission in FRB surveys, through triggered observations of gravitational wave and gamma-ray burst detected mergers and confirming a NS-NS merger origin after the fact through observations of kilonovae or radio afterglow emission. Each subsection is devoted to a separate observing strategy, in which we discuss prospects for current and future instrumentation.  We conclude with a discussion in Section \ref{sect:Discussion} and present our primary findings in Section. \ref{sect:conclusion}.



\section{Model: Conductor in NS magnetosphere}
\label{sect:model}
In \cite{Lyutikov2019}, the authors consider electromagnetic interaction between two neutron stars in a binary system. In the first case (denoted in \cite{Lyutikov2019} as the 1M-DNS scenario) which we focus on, it is assumed that one NS is highly magnetized (henceforth the primary NS) such that the other neutron star (henceforth the secondary NS, or the conductor) acts as a perfect, spherical conductor due to negligible magnetization. This is a natural scenario expected from evolutionary considerations, as the first-to-form, older secondary NS may be partially recycled, burying its field, while the younger primary NS will retain higher magnetization. It is also in concordance with the double pulsar system \citep{Lyne2004}, where one pulsar is partially recycled. During orbital motion, the secondary NS moves through the magnetosphere of the primary, expunging magnetic field field lines and inducing an electric field outside the surface of the secondary NS, with a significant component parallel to the magnetic field lines, $E_{\parallel}$. We also note that the results in this work will also be applicable to pre-merger emission from double white dwarf (WD) binaries that Laser Interferometer Space Antenna (LISA; \citealt{2017arXiv170200786A}) will observe in our Milky Way. Although they will emit at a much lower luminosity, their proximity may mean that incoherent higher frequency nonthermal emission is within the horizon of optical or X-ray instruments, similar to AR Scorpii \citep{Marsh2016}.
\par
\begin{figure}
  \centering
{\includegraphics[width=.4\textwidth]{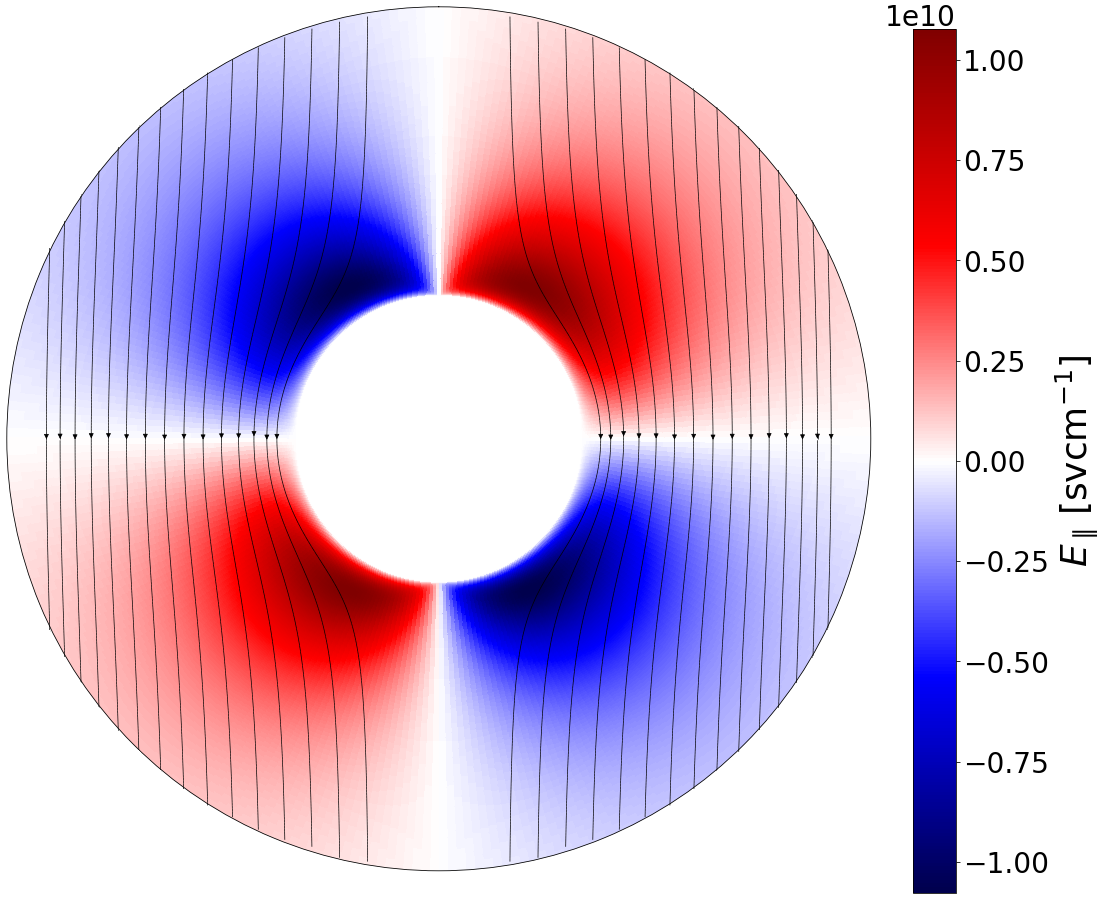}}
\caption{A map of the electric field parallel to the magnetic field lines $E_{\parallel}$ surrounding the conductor in the $x-z$ plane for representative values of $B=10^{11} \rm{G}$ and $\beta=0.5$. The induced $E_{\parallel}$ field has a quadrupole structure, meaning coherent radiation is preferentially emitted in solid angles perturbed from the primary NS' magnetic axis.}
\label{fig:conductor_field_lines}
\end{figure}

To estimate particle acceleration and the pre-merger emission signature, we derive the parallel electric field component due to the motion of the secondary neutron star. We use a spherical coordinate system $(r,\theta, \phi)$ centred on the conductor unless otherwise stated, where $r$ is the radial distance, $\theta$ is the polar angle and $\phi$ the azimuthal angle. $B = B(r,\theta,\phi)$ is the magnetic field strength at the location of the secondary due to the primary, and $\beta = v/c$ where $v$ is the relative velocity of the conductor through the magnetosphere. For the purpose of this simple electromagnetic model, the direction of the magnetic field of the primary NS dipole is assumed to be uniform at the position of the secondary NS (as in \citealt{Lyutikov2019}), as would be the case for large binary separations.
The parallel magnetic field approximation holds reasonably well despite the fact that the conducting secondary and the primary are close together, as the regions with large values of $E_{\parallel}$ are close to the conducting NS surface where expunged magnetic field lines are tangential (see Fig. \ref{fig:conductor_field_lines}). This means that even for small binary separations the emission region and direction are dominated by the motion of the conductor and not by the primary's field topology. 
We use a separation-dependent magnetic field strength in Section \ref{sect:numerical}. 
\par
\par
In Appendix \ref{sect:Om_derivation}, we present a full derivation of the electric field that develops. We find a small correction to the $E_{\parallel}(r, \theta, \phi)$ derived in \cite{Lyutikov2019} (given by Eqs. \eqref{eq:lyutikov19} \& \eqref{eq:om}). The correction is made by first finding the electric field vector in a frame co-moving with the secondary NS. It is in this frame that the electromagnetic interface conditions are solved for, following which a Lorentz transformation to the primary NS frame gives the electric field for an observer stationary with respect to the primary's magnetosphere. The electromagnetic response of the secondary establishes a surface current and a surface charge, which are responsible for eliminating the magnetic field and the electric field respectively from within the secondary unmagnetized NS. While the magnetic field is tangential at the surface in both frames when ignoring second order terms of velocity originating from relativistic correction, the stipulation that the electric field is perpendicular to the conductor's surface is valid only in the co-moving frame, but not in the primary's frame. The parallel component of the electric field is calculated by taking the dot product with the unit vector in the direction of the magnetic field. It becomes evident from Eq. \eqref{eq:om} that the region of maximum parallel electric field is not at the surface of the conductor, but approximately $0.23R_{\rm NS}$ away from the surface (see Figure \ref{fig:conductor_field_lines}).

\begin{equation}
    E_{\parallel, \rm L19} = -\dfrac{3}{2 \sqrt{2}} \dfrac{\sin(\theta) \cos(\theta) \cos(\phi) \big(6 - \frac{R_{\rm NS}^3}{r^3}\big) }{\sqrt{8 \big(1 - \frac{R_{\rm NS}^3}{r^3}\big)^2 + 6 \big(4 - \frac{R_{\rm NS}^3}{r^3}\big) \frac{R_{\rm NS}^3}{r^3} \sin^2(\theta)}}   B \beta \frac{R_{\rm NS}^3}{r^3} 
    \label{eq:lyutikov19}
\end{equation}

\begin{equation}
      E_{\parallel, \rm this \: work} = \frac{3\sin(\theta)\cos(\theta)\cos(\phi) \big(1-\frac{R_{\rm NS}^3}{r^3}\big)}{\sqrt{4\cos^2\theta \big(1-\frac{R_{\rm NS}^3}{r^3}\big)^2 + \sin^2(\theta)\big(2+\frac{R_{\rm NS}^3}{r^3}\big)^2}}B \beta \frac{R_{\rm NS}^3}{r^3}
      \label{eq:om}
\end{equation}

\subsection{Inspiral phase}
Using the post-Newtonian approximation by \cite{Peters1964}, we can write down the binary separation $a$, of two identical spherical objects of mass $M$, as a function of time $t$, due to gravitational radiation:
\begin{equation}
    \frac{1}{a} \frac{da}{dt} = -\frac{128}{5} \frac{G^3 M^3}{c^5 a^4}
    \label{eq:sep}
\end{equation}
As in \cite{Metzger2016}, the merger time until $a=0$ is:
\begin{equation}
\begin{split}
t_{\rm m} = \dfrac{5}{512} \dfrac{c^5 a^4}{G^3 M^3}
\end{split}
\end{equation}
This equation is valid only until the disruption of the conductor at $a_{\rm min}$. We can put a lower limit on $a_{\rm min}$ by considering how the Roche lobe of the conductor evolves as the binary separation decreases \citep{Eggleton1983}. Assuming two identical $1.4 M_{\odot}$ NSs ($q=1$), we find that $a_{\rm min} = 26.4$km, however depending on the equation of state of the NS, tidal disruption may occur sooner. We note that the secondary NS will always move through the magnetosphere of the primary neutron star as the stars will not be tidally locked, but that neglecting tidal forces may effect dynamics and therefore lightcurve morphology (Sect. \ref{sect:temporal}) during the final few orbital periods \citep{Bildsten1992}. 
\par
Solving Eq. \eqref{eq:sep} we find:
\begin{equation}
    \begin{split}
        a(t) = \bigg(a_0^4 \big(1 - \frac{t}{t_{\rm m,0}} \big) \bigg)^{1/4}
    \end{split}
\label{eq:a(t)}
\end{equation}
where $a_0$ is the initial separation, and $t_{\rm m,0}$ is the time to merger at the initial separation. We assume both the primary magnetized NS and secondary conducting NS have radii $R_{\rm NS} = 12$km \citep{LIGO2018_NS_Radii,Lattimer2019} and masses $M = 1.4 M_{\odot}$. Substituting equations for $B(r,\theta, \phi)$ and $\beta$ into Eq. \eqref{eq:om}, and using Eq. \eqref{eq:a(t)} we can write $E_{\parallel}(r,\theta,\phi,t)$ as:
\begin{equation}
    \begin{split}
        E_{\parallel} &= f_1(r,\theta, \phi) B(r,\theta, \phi) \beta \\
        &= f_2(r,\theta, \phi) \dfrac{B_{\rm s} R_{\rm NS}^3 \sqrt{G M}}{c a_0^{7/2}} \big(1 - \frac{t}{t_{m,0}}\big)^{-7/8}
    \end{split}
    \label{eq:E(t)}
\end{equation}
\par
\begin{figure}
  \centering
{\includegraphics[width=.39\textwidth]{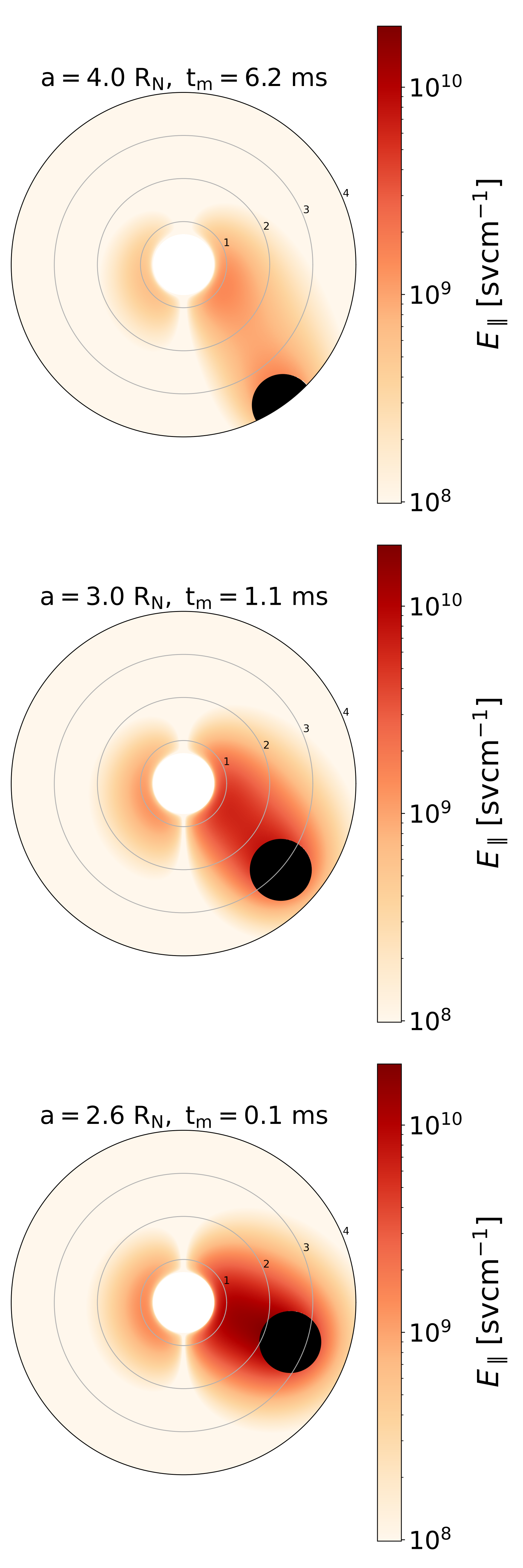}}
\caption{The parallel electric field surrounding the conducting NS (centre, white) and the magnetized primary NS (black) for three different binary separations $a$ during the inspiral. Each panel is centered on the conductor as viewed from above the orbital plane. As the inspiral progresses and orbital separation decreases, regions of high parallel electric field surround the conductor, particularly in the region between the conducting NS and the primary NS due to the large local magnetic field. The view shows the $\theta = \frac{\pi}{4}$ plane as Eq. \eqref{eq:om} tells us that $E_{\parallel} = 0$ if $\theta = \frac{\pi}{2}$, i.e., in the orbital plane. The electric field units are stated in statvolts per centimetre.}
\label{fig:Lyutikov_map}
\end{figure}

Eq. \eqref{eq:E(t)} dictates the induced electric field around the surface of the conductor, where $f(r,\theta, \phi)$ varies at different spatial points around the conductor. It encapsulates the prefactor in Eq. \eqref{eq:om}, but also the depends on the magnetic field strength at each point. We also take into account the orbital motion of the magnetized object in the frame of the conductor, leading to asymmetric $E_{\parallel}$ field as $a \rightarrow a_{\rm min}$ as seen in Fig. \ref{fig:Lyutikov_map}. 
As expected, we find that $E_{\parallel}$ increases as $t \rightarrow t_{\rm m,0}$, and thus particle acceleration and attainable radiation luminosity around the conductor increases as the inspiral progresses.


\subsection{Numerical method: emission directed along field lines}
\label{sect:numerical}
To investigate the time-dependent and viewing angle dependent emission expected from these systems, we calculate the electromagnetic fields during the inspiral in 3 dimensions and map the parallel electric field component given by Eq. \eqref{eq:lyutikov19}. To compensate for our assumption of uniform magnetic field strength around the conductor while building the electromagnetic model, we compute the local magnetic field value $B$ at $(t, r, \theta, \phi)$ surrounding the conductor by finding the distance to the centre of the primary magnetized neutron star, $a(r,\theta,\phi)$, and assuming the field decreases as $B \approx B_{\rm s} \bigg( \frac{R_{\rm NS}}{a(r,\theta,\phi)} \bigg)^{3}$. Assuming $\mathbf{B} = B \mathbf{\hat{z}}$ frees the uniform field condition even though the background field is still assumed to be parallel. Furthermore, the magnetic field lines expunged by the conductor are defined in \cite{Lyutikov2019} as:
\begin{equation}
    \mathbf{B} = -B \cos(\theta) \bigg( 1 - \frac{R^3}{r^3} \bigg) \mathbf{\hat{r}} + B\sin(\theta) \bigg( 1+\frac{R^3}{2r^3} \bigg) \boldsymbol{\hat{\theta}}
    \label{eq:expunged_B}
\end{equation}
This equation is defined for uniform and parallel $B$, however, we perform computations with the assumption of separation-dependent magnetic field strength.

As we will show in Section \ref{sect:particle_acceleration}, we expect particle acceleration and therefore any coherent radiation to be directed along the local magnetic field lines regardless of the specific radiation mechanism. The angle subtended by total field line $\mathbf{B}$ and the radial direction $\mathbf{\hat{r}}$ is given by: $\theta_{\mathbf{\hat{r}},\mathbf{B}} = \arctan{\frac{B_\theta}{B_r}}$. Therefore in the frame of the conductor the direction of a local field line with respect to the $\mathbf{\hat{z}}$ direction at $(r, \theta, \phi)$ is given by: 
\begin{equation}
        \theta_{\rm B} = \theta + \arctan{\frac{B_\theta}{B_r}}
        \label{eq:Local_B_theta}
\end{equation}
In the calculation, each cell in ($r,\theta,\phi$) is assigned a value of $E_{\parallel}$ via Eq. \eqref{eq:om}, a magnetic field vector $\mathbf{B}$ according to Eq. \eqref{eq:Local_B_theta} and a volume element $\delta V$. To estimate the radio flux measured by an observer at $(D, \theta, \phi)$ in the frame of the conductor, we find set of cells with magnetic field vectors whose solid angle subtended by a beaming angle $\theta_{\rm beam, coh} \approx 0.1$ radian \footnote{This choice, although motivated by observations of the pulsar duty cycle (neglecting period dependence e.g. \citealt{Rankin1993}), is somewhat arbitrary in that it depends on the details of plasma EM mode propagation and decoupling within the magnetosphere.} encompasses the observer. We then sum the luminosity of all cells aligned with the observer according to Eqs. \eqref{eq:expunged_B} \& \eqref{eq:Local_B_theta}, to produce lightcurves in Fig. \ref{fig:lightcurves}. We find that emission is primarily observed at angles of $5-45 \deg$ from the background magnetic field, and this viewing angle dependent emission is discussed in Sect. \ref{sect:viewingangle}. We omit general relativistic (GR) effects on the radiation such as gravitational redshift, gravitational lensing, Lense-Thirring precession, frame dragging and relativistic abberation ($\theta_{\rm abb} \approx 0.1 \,$rad) due to the orbital motion, which effect the magnetic field topology and therefore where emission is directed (e.g. \citealt{Wasserman1983,Gonthier1994}). These effects are generally small i.e. on the order of $\beta/2 \approx 20$\% in the centre-of-mass frame, and are therefore neglected in our calculation and in the predictions of Section \ref{sect:temporal}. For edge-on observers, conditions may be met for strong lensing of emission regions of the second by the primary, dependent on the magnetic field geometry. Considering how GR may modify the overall luminosity and the temporal morphology of the signal could be explored in a future work.


\section{Particle acceleration and radiation}
\label{sect:particle_acceleration}
As orbital motion progresses and the motion of the conductor through the magnetic field of the primary induces a large parallel electric field $E_{\parallel}$, charged particles will be pulled from the secondary NS's surface and accelerated along field lines to high energies (e.g. \citealt{Dai_2016}). In any case, \cite{Timokhin2010,Timokhin2013} have shown that a vacuum-like gap will be formed regardless of whether the surface is free to emit charges, as pair creation discharges are non-stationary and pairs are advected out of the acceleration zone. In \cite{Lyutikov2019} it was noted that coherent radiation emitted in the radio band is the most feasible method by which to observe such precursor emission, given the relatively low total power as compared to e.g. gravitational waves or GRBs. This particle acceleration bears resemblances to two theories of coherent radiation: pulsar-like emission that invokes on pair production fronts across gaps developed to explain radio pulsars \citep{Sturrok1971,RudermanSutherland1975,Timokhin2010,Timokhin2013} and coherent curvature radiation which has been recently used to explain the origin of FRBs from magnetars \citep{Katz2014,Kumar2017,LuKumar2019,CooperWijers2021}. In the following section we discuss the former, and in Appendix \ref{app:cohcurv} we discuss the latter, in the context of the model.
\par





\subsection{Pulsar-like emission}
As mentioned in previous works \citep{Lipunov1996,Totani2013,Lyutikov2019}, a NS-NS merger may revive pulsar-like emission. Although pulsar emission is poorly understood (see e.g. \citealt{Melrose2017}), the merger of a $10^{12}$ G neutron star invokes similar electromagnetic conditions to those expected to power coherent radio emission from radio pulsars (see discussion above in Section \ref{sect:model}). The radio luminosity during the inspiral may be much larger than typically observed from pulsars of similar magnetic field strength for two main reasons. Firstly, the spatial extent of the $E_{\parallel}$ acceleration region due to the motion of the secondary is large ($l_{E_{\parallel}} \approx R_{\rm NS} \approx 10^{6} \; {\rm cm}$), in contrast to pulsar cap models where $l_{\rm pc} \approx 10^{4} \, P_{0}^{-1/2} \; {\rm cm}$. Secondly, the required charge density of the magnetosphere may be much higher than in the isolated pulsar. This is because the $B \times v$ motion of the magnetosphere is dominated by the binary orbital period $P_{\rm orb} \approx \frac{2 \pi a}{v_{\rm orb}} \approx 10^{-3} \, a_{30 km}^{3/2} \; {\rm seconds}$ and not by the spin period of the aging magnetized $10^{12}$G NS which is usually $P_{\rm spin} > 1$s. In the following, we explore the basics of pulsar-like emission in gap models, and calculate analytically the expected radio luminosity from NS merger systems. 
\par
\subsection{Acceleration gap}
In the polar cap models of pulsar emission \citep{RudermanSutherland1975,DaughertyHarding1982}, rotation-induced electric fields close to the surface of the NS accelerate particles along open magnetic field lines. The acceleration of these particles along curved magnetic field lines perpendicular to their velocity produces gamma-ray curvature radiation. These high-energy photons interact with magnetic fields through magnetic pair production to produce cascades of secondary pairs, where the ratio of primary to secondary pairs is known as the pair multiplicity: $\kappa = N_{\rm sec}/N_{\rm pri}$ and is $1 < \kappa \lesssim 10^{3}$ (\citealt{Timokhin2019}; although see also \citealt{Harding2011} who find the multiplicity could be as high as $10^{6}$ for multipolar field topologies). The secondary pairs inherit momenta and energy from the primaries, leading to nonstationary discharges that launch superluminal waves with an efficiency $\eta < 1$ \citep{Philippov2020}. We assume that the secondaries (and all subsequent orders) are energetically subdominant as compared to the primary particles that initiate the burst-like cascades. To understand the expected coherent luminosity in a viewing angle dependent manner, we estimate the radio luminosity through a pulsar-like mechanism in Section \ref{sect:radio_estimate} using our numerical set-up described in Section \ref{sect:numerical}. In the following, we discuss likely scenarios of the formation of a one-dimensional and stationary acceleration gap in the limiting cases of this model. In reality, the gaps are non-stationary on sampling a variety of temporal and spatial scales, yet on average ought to be not disparate from the physical scales yielded by the stationary calculation.
\par
Crucial to calculating the gap height is to understand the accelerating electric field in the region and the subsequent particle acceleration. In Eq. \eqref{eq:om} we calculate the unscreened $E_{\rm parallel}$ component due to the bare magnetospheric interaction. To understand the true value of the electric field across the acceleration gap, $E_{\rm gap} = 4 \pi q n l_{\rm acc}$, where $n$ is the number density of charges to sustain required current, we consider two limiting approaches. Firstly, in the maximal case we take this to be equal to $E_{\parallel}$, such that $n \approx 10^{15} \; {\rm cm^{-3}} E_{\parallel,10} \, l_{\rm acc, 3}^{-1}$. The second minimal case is comparing $n$ to the expected Goldreich-Julian density scale required by the co-rotation of the magnetosphere by the orbit: $n = n_{GJ} \sim (2 B)/(q c P_{\rm orb})$, where $B$ is the local magnetic field. We find that for regions of the strongest emission (i.e. corresponding to a similar value of $E_{\parallel} = 10^{10}\, {\rm sv \, cm^{-1}}$), the Goldreich-Julian density is: $n_{GJ} \approx 10^{14} \; {\rm cm^{-3}} \; B_{12} \, P_{\rm orb,-3}^{-1}$. This difference is within one order of magnitude (corresponding to a factor $10^{2}$ in the overall luminosity, see Eq. \ref{eq:coherent_lum_appendix}) could be plausibly attributed to the fact that in the NS merger case the electric field is induced by binary interaction instead of a pulsar's spin. Furthermore, in contrast to the pulsar emission case, current requirements vary significantly on timescale associated with the orbital period.
\par
To understand the plausible luminosity ranges of the NS merger system, we consider the upper limit case of $E_{\rm gap} = E_{\parallel}$ in this Section. We stress that this approximation is an upper bound adopted in this analytic calculation. We also include the analytic gap height and luminosity calculation for what can be considered a conservative, plausible lower bound $n \sim n_{\rm GJ}$ case in Appendix \ref{app:goldreichapprox}, although note that it is possible the radiation reaction limit invoked here may not persist in this case. This is because for $E_{\rm gap} < E_{\parallel}$, particles are accelerated more slowly and therefore may not reach $\gamma_{\rm max}$ \citep[see Appendix B of][]{TimokhinHarding2015} . These two cases bound the range of possibilities, and while we are encouraged by our findings below, future pair cascade simulations will be required to fully understand the details of particle acceleration, pair creation and multiplicity, and radiation reaction in the NS merger case.
\par
\subsubsection{Curvature radiation reaction limited acceleration}
The maximum energy of primary pairs accelerated by the electric field along the B-field is limited by either curvature radiation or resonant inverse-Compton scattering from soft X-rays \citep[e.g.,][]{Baring2011,Wadiasingh2018}. In the absence of other particle acceleration mechanisms, we do not expect a significant X-ray radiation field due to the relatively low magnetic field (compared to magnetars) and slow spin of the primary magnetized neutron stars. However, if there is tidal friction transmitted to the crust (rather than heating the interior and core) or crust shattering before merger this may not be the case. Assuming curvature losses dominate over scattering, the equation of motion of the primary accelerated pairs is:
\begin{equation}
\begin{split}
    \frac{d \epsilon_{\rm e}}{d t} = q E_{\parallel} c - \dot{\epsilon}_{\rm curv}\\ 
    \end{split}
\end{equation}
Note that we used $\epsilon$ when referring to particle energies to make a clear distinction between energy and the electric field. It has been shown for both magnetars \citep{Wadiasingh2020} and high-field, slowly rotating pulsars \citep{TimokhinHarding2015} that the gap terminates before the radiation reaction regime occurs, where the maximum energy of accelerated primary pairs is limited by the curvature radiation. It is not clear in the NS merger case whether radiation reaction will always be obtained, due to the small $P_{\rm orb}$ close to merger. To understand this, we can find the characteristic length and timescales along which free particle acceleration (i.e. no significant curvature losses) by finding the Lorentz factor at which curvature losses become important by equating the acceleration power and loss power $P_{\rm acc} = P_{\rm curv}$ in the upper limit $E_{\rm gap} = E_{\parallel}$ case:
\begin{equation}
    \begin{split}
        q E_{\parallel} c &= \frac{2 q^2 c \gamma^4}{3 \rho_{\rm c}^2} \\
        \gamma_{\rm max} &= \bigg(\frac{3 E_{\parallel} \rho_{\rm c}^2}{2 q}\bigg)^{1/4} = 8 \times 10^{7} \: E_{\parallel, 10}^{1/4} \, \rho_{\rm cm,6}^{1/2}
    \end{split}
\end{equation}
Where $\rho_{\rm c}$ is the curvature radius. We can find the characteristic length scale of free acceleration by comparing $\gamma_{\rm max}$ to $\gamma(l) = \dfrac{q E_{\rm gap} l_{\rm acc}}{2 m_e c^2}$. In the $E_{\rm gap} = E_{\parallel}$ limit (see below):
\begin{equation}
    l_{\rm free, acc} = 25 \: \rho_{\rm c,6}^{1/2} \, E_{\parallel,10}^{-3/4} \; {\rm cm}
\end{equation}
We can compare this to the derived gap height in Eq. \eqref{eq:gap_heights_x} and due to the similar scalings of $E_{\parallel}$ in both $h_{\rm gap}$ and $l_{\rm free, acc}$, the free acceleration length scale is always smaller than the gap height for $B \lesssim 5 \times 10^{12}\,$G. The validity of this calculation relies on the fact that if curvature losses are self-consistently included in the gap height calculation, the energy of emitted curvature photons will decrease and therefore the gap height will always increase. This means that the radiation reaction limited particle acceleration is always reached, and we can neglect curvature losses during the gap height calculation. We note that our approximate luminosity calculations below (Eqs. \ref{eq:analytic_pulsar_flux} \& \ref{eq:analytic_pulsar_flux_2}) do not explicitly depend on the gap height $h_{\rm gap}$. This means that even if the radiation reaction limit is not reached (e.g. in high B limit), the main results are not affected as long as a gap forms. This is likely, given that $\gamma_{\rm max}$ is much higher than the threshold value required to produce curvature photons capable of pair production.

\subsubsection{Gap height}
\begin{figure}
  \centering
{\includegraphics[width=.5\textwidth]{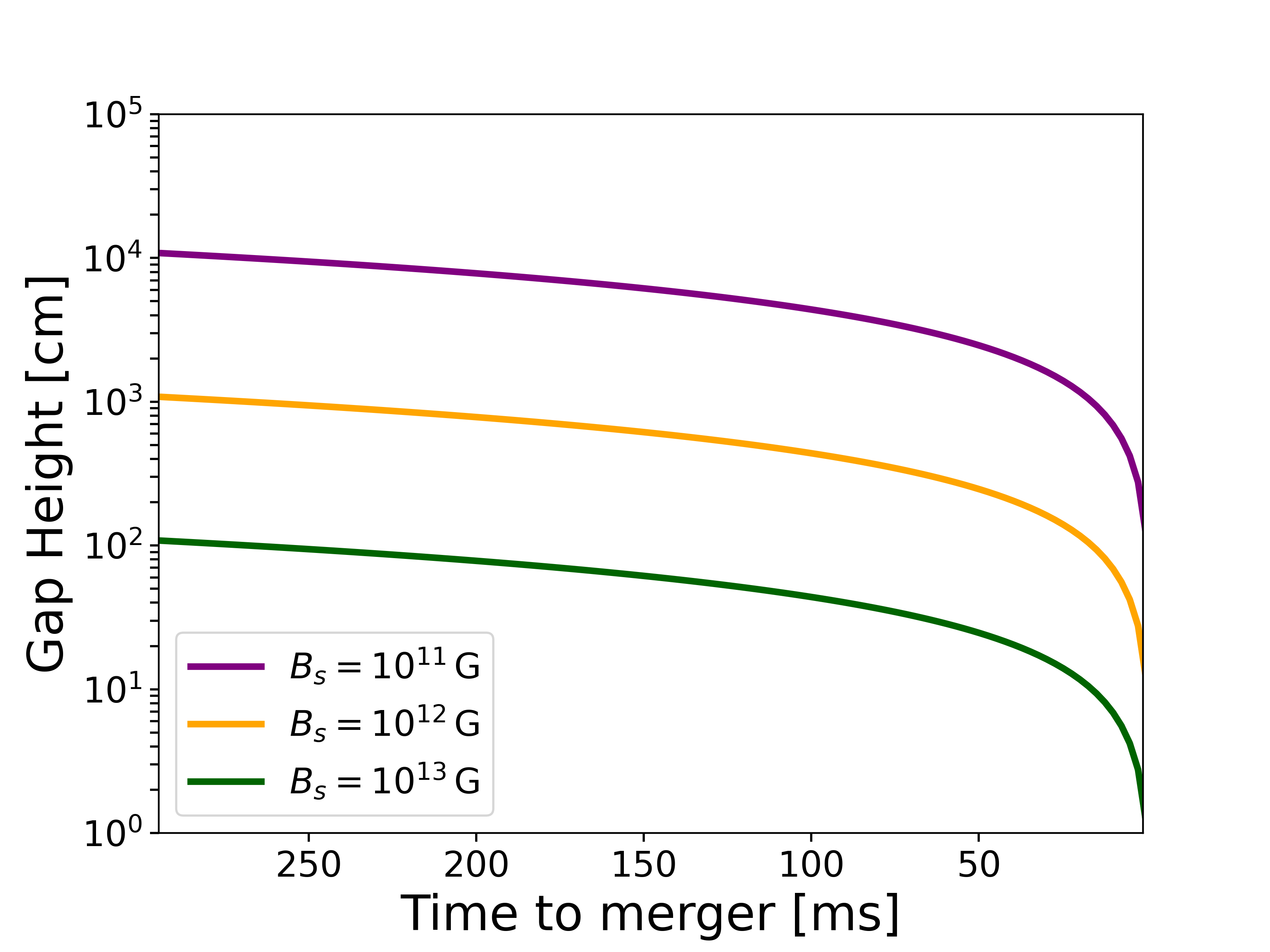}}
\caption{Gap height for a single point close to the secondary neutron star's surface as a function of time.}
\label{fig:gap_times}
\end{figure}

\begin{figure}
  \centering
{\includegraphics[width=.5\textwidth]{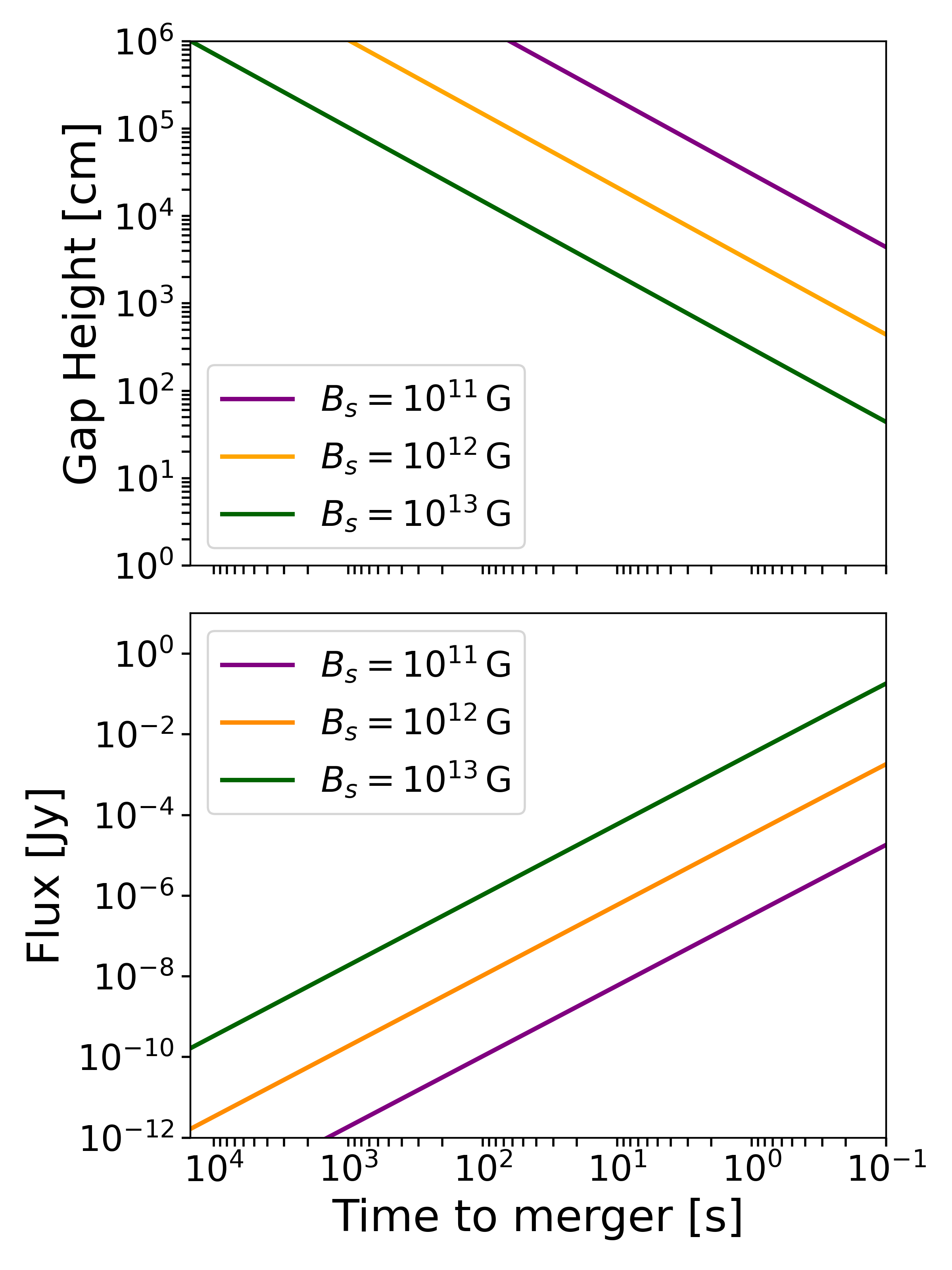}}
\caption{The approximate gap height and flux at $D = 100 \,$Mpc for fiducial model parameters. A gap height $h_{\rm gap} < R_{\rm NS}, \rho_{\rm c} \approx 10^{6} \,$cm (corresponding to a typical separation of $10^{8}$cm) is where radio emission could be begin, albeit at a very low luminosities. }
\label{fig:flux_and_gap_long}
\end{figure}

In the following, we calculate the gap height for the radiation reaction limited case which we consider reasonable in the $E_{\rm gap} = E_{\parallel}$ upper limit. The height of the acceleration gap $h_{\rm gap}$ is the distance from the initial acceleration point $h_0$ to a pair production front, where cascades occur efficiently enough to completely screen the electric field. This gap height is the sum of two length scales: $l_{\rm acc, gap}$ is the distance traversed by accelerated primaries before they attain enough energy such that emitted curvature photons are capable of pair production; and $l_{\gamma,\rm gap}$ is the distance traversed by curvature photons before pair production occurs. For larger values of $l_{\rm acc, gap}$, higher energy curvature photons are produced and therefore smaller values of $l_{\gamma,\rm gap}$ are attained. Therefore to find the $h_{\rm gap}$ = $l_{\rm acc, gap} + l_{\gamma,\rm gap}$, we minimize $h_{\rm gap}$ to find the distance at which the pair creation cascade begins.
\begin{equation}
\begin{split}
        \epsilon_{\rm ph} &= \hbar \omega = \frac{3 \hbar c \gamma^3}{\rho_{\rm c}} = 8 \times 10^{-5} \; {\rm erg} 
        = 500 \; {\rm GeV} \: \gamma_{5}^3 \, \rho_{\rm c, 6}^{-1} \\
\end{split}
\label{eq:crit_freq}
\end{equation}
At $\gamma_{\rm max}$ curvature photons are produced many orders of magnitude above the energetic threshold required for magnetic pair production, $\epsilon_{\rm threshold} \approx 2 m_e c^2$, therefore even in this regime the gap will form. However, under the assumption that radiation reaction does not occur we neglect curvature losses during free acceleration such that the primaries' Lorentz factor as a function of path length $l$ is:
\begin{equation}
\begin{split}
    \gamma(l) &\approx \frac{q E_{\rm gap} l}{m_e c^2}
\end{split}
\label{eq:gamma_l}
\end{equation}
In the $E_{\parallel} = E_{\rm gap}$ limit, the Lorentz factor of the primaries is:
\begin{equation}
    \gamma(l) = \frac{q E_{\parallel} l_{\rm acc}}{m_e c^2}
\end{equation}
Substituting into Eq. \eqref{eq:crit_freq} we find:
\begin{equation}
    \epsilon_{\rm ph} = \frac{3 \hbar c}{\rho_{\rm c}} \bigg(\frac{q E_{\parallel} l_{\rm acc}}{m_e c^2} \bigg)^3
    \label{eq:energy_photons}
\end{equation}
For above-threshold pair production in fields $B \ll B_{\rm c}$ we require that:
\begin{equation}
    \chi \equiv \frac{\epsilon_{\rm ph}}{2 m_e c^2} \frac{B}{B_{\rm c}} \sin(\theta_{k,B}) \gtrsim {\rm max}[0.2, B/B_{\rm c}]
    \label{eq:pair_threshold}
\end{equation}
Where $\theta_{k,B}$ is the angle between the magnetic field $B$ and the photon momentum, and can be approximated as $\sin(\theta_{k,B}) \approx \frac{l}{\rho_{\rm c}}$ due to small angles and curved magnetic field lines diverging linearly from the photons' paths. Therefore the distance photons must travel before pair production is:
\begin{equation}
    l_{\gamma, \rm gap} \approx 0.2 \rho_{\rm c} \frac{2 m_e c^2}{\epsilon_{\rm ph}} \frac{B_{\rm c}}{B}
\end{equation}
By substitution of Eq. \eqref{eq:energy_photons}, we find:
\begin{equation}
\begin{split}
    l_{\gamma, \rm gap} = \frac{2 \rho_{\rm c}^2 m_e^4 c^7 B_c}{15 \hbar B q^3 E_{\parallel}^3 l_{\rm acc}^3}
\end{split}
\end{equation}
Let $k = \frac{2 \rho_{\rm c}^2 m_e^4 c^7 B_c}{15 \hbar B q^3 E_{\parallel}^3}$.
By expressing $l_{\gamma, \rm gap}$ in terms of $l_{\rm acc}$, we can minimize $h_{\rm gap} = l_{\gamma, \rm gap} + l_{\rm acc}$ with respect to variations in $l_{\rm acc}$  to find values for both length scales that satisfy $\frac{\delta h_{\rm gap}}{\delta l_{\rm acc}} = 0$:
\begin{equation}
\begin{split}
    l_{\rm acc} &= (3 k)^{1/4} \\
    l_{\gamma, \rm gap} &= \frac{k}{l_{\rm acc}^3} = \frac{k^{1/4}}{3^{3/4}}
\end{split}
\end{equation} 
Therefore the $h_{\rm gap}$ is:
\begin{equation}
    \begin{split}
    h_{\rm gap} &= \frac{8 k^{1/4}}{3^{3/4}} = \frac{8}{3^{3/4}} \bigg(\frac{2 \rho_{\rm c}^2 m_e^4 c^7 B_c}{15 \hbar B q^3 E_{\parallel}^3} \bigg)^{1/4} \\
    &= 40 \; {\rm cm} \: \rho_{\rm c, 6}^{1/2} \, B_{11}^{-1/4} \, E_{\parallel,10}^{-3/4}
    \end{split}
    \label{eq:gap_heights_x}
\end{equation}
Where we have included a factor of 2 to account for relative motion of pairs as in \cite{Wadiasingh2020}. The approximate gap height as a function of time until merger is shown in Fig. \ref{fig:gap_times} for short timescales, and Fig \ref{fig:flux_and_gap_long} on a longer timescale. The analytic gap height derivation assuming $E_{\rm gap} = 4 \pi q n l_{\rm acc}$ where $n = n_{GJ} = (2 B)/(q c P_{\rm orb})$ referred to here as the lower limit, is described in Appendix \ref{app:goldreichapprox}. 

\par
We show in Figs. \ref{fig:gap_times} \& \ref{fig:flux_and_gap_long} that for our timescales of interest this length scale is smaller than characteristic length scales for variations in both $B$ and $E_{\parallel}$ as required for efficiency pair cascades \citep{TimokhinHarding2015}, which are both on the order of the neutron star radius $R_{\rm NS} \approx 10^{6} \, {\rm cm}$. Furthermore, the fact that the gap length scale is smaller than the $B$ variation length scale means that radio emission generated from pair cascades will be directed along local B-field lines. The superluminal O-modes will couple to plasma downstream of the cascades, be advected along B by adiabaticity and decouple at higher altitudes, transforming into vacuum EM modes. 

Given this gap height, we can compute the potential difference across the gap:
\begin{equation}
    \Phi = E_{\parallel} h_{\rm gap} \approx 10^{12} \; {\rm statvolt} \: E_{\parallel,10} \, h_{\rm gap, 2}  \\
    \label{eq:voltage}
\end{equation}
We note that this is higher than the minimum voltage of $\Phi = 10^{10} \; {\rm statvolt}$ that is thought to be required for pulsar emission, which has been used to explain the pulsar `death line' \citep{Timokhin2013}.

\subsection{A Radio Luminosity Proxy}
\label{sect:radio_estimate}
As mentioned above, the radio emission is presumed to result from single-photon pair cascades from a significant $E_{\parallel}$ field component during the inspiral. In the Timokhin-Arons mechanism \citep{Timokhin2013}, the pair creation is a necessary and sufficient condition for generation of superluminal electromagnetic modes, while other mechanisms of the Ruderman-Sutherland type \citep{RudermanSutherland1975} require additional possibly unrealistic constraints and caveats. An upper limit for the radio luminosity in any scenario is the power furnished to free-accelerating primaries in the gap. The pulsar mechanism is broadband; in the Timokhin-Arons mechanism, this is due to the sum of a self-similar spectrum of non-stationary discharges in a scale invariant range of wavenumbers. For this simple estimate, we assume the entire radio luminosity is emitted across a bandwidth of $\delta \nu = 10 \, {\rm GHz}$, although in a future work considering typical pulsar spectral index would provide better estimates for frequency dependent luminosities. In Appendix \ref{app:cohcurv} we discuss a Ruderman-Suderland type coherent curvature radiation as an alternative radiation mechanism.




\par
Below we show two similar methods of obtaining the luminosity of primaries (i.e. two renditions of the involved wave-particle processes), which in turn may be used as a proxy for the radio luminosity with the inclusion of an efficiency factor, $\eta$. This proxy is expected to capture the gross parameter scaling of coherent radio emission's luminosity involved in neutron stars, i.e. $L_{\rm R} \approx \eta L_{e^+e^-}$. The efficiency $\eta<1$ moderates this estimate and depends on local conditions such as shape and extent of current regions with space-like or time-like regions (e.g. $J/(\rho_{\rm q} c)$ value and sign), the angle or shape of the pair formation front, and varying field curvature radii. In both derivations below, within factors of unity associated with geometric factors, the primary luminosity is $ L_{e^+e^-} \sim 4 \pi \rho_{\rm q}^2 h_{\rm gap}^2 A c$ where A is a characteristic cross sectional area of flux tubes associated with the accelerating region, $\rho_{\rm q}$ is the required charge density to satisfy the transient conditions, and $h_{\rm gap}$ is the characteristic gap height appropriate to physical conditions for the cascades. For canonical rotation-powered radio pulsars, this calculation implies $\eta \sim 10^{-2}$ and is compatible with the voltage-like scaling of pulsar luminosity inferred by population studies with beaming models \citep[e.g.][]{2002ApJ...568..289A}. Likewise, for seismically oscillating magnetars \citep{2019ApJ...879....4W,2019MNRAS.488.5887S,Wadiasingh2020,2020ApJ...903L..38W} the pair luminosity estimate yields the correct energy scale $L_{e^+e^-}  \sim 10^{39}-10^{43}$ erg~s$^{-1}$ observed in cosmological FRBs (as well as the low-luminosity Galactic FRB observed from SGR 1935+2135 in April 2020). Correspondingly, as shown below, NS-NS inspirals where on NS has a large magnetic field $B_{\rm s} > 10^{13}$G also yield energy scales commensurate with observed FRBs albeit with possibly wider range in allowed luminosities for varying parameters. These varied luminosities, in addition to multi-messenger signals and chirps in FRB quasi-periodicity, may be a distinguishing characteristic of NS-NS mergers from magnetar progenitors in a sub-population of one-off FRBs.


\subsubsection{Estimate due to energy in primaries}
First, we compute the power of the primary particles accelerated across the gap: $P_{\rm particles} = q \Phi_{\rm gap} \dot{N}$ where $\dot{N}$ is the rate of primaries and $\Phi_{\rm gap} = E_{\parallel} h_{\rm gap}$ is the voltage drop across the gap. $\dot{N}$ scales linearly with the local plasma density $n$, which can be estimated using: $\dot{N} = n A c$ where $A$ is the cross-sectional area of the acceleration region. For the analytical estimate, we use the characteristic size of the particle acceleration region of $A \approx 4 \pi R_{\rm NS}^2$, and use the fact that $n = \frac{E_{\rm gap}}{4 \pi q l_{\rm acc}} \approx \frac{E_{\parallel}}{4 \pi q h_{\rm gap}}$. Therefore the total luminosity, inclusive of the efficiency factor $\eta$ is:
\begin{equation}
    \begin{split}
        L_r &= \eta q \Phi_{\rm gap} \dot{N} = \eta q E_{\rm gap} h_{\rm gap} n A c \\
        &= \eta E_{\parallel}^2 A c \approx 10^{40} \; {\rm erg s^{-1}} \: \eta_{-2} \, E_{\parallel,10}^2 \, R_{\rm NS, 6}^2
    \end{split}
    \label{eq:analytic_pulsar_flux}
\end{equation}
We note that the radio luminosity estimate in the $E_{\rm gap} = E_{\parallel}$ limit has no explicit dependence on the gap height. However at the timescales of interest the gap height is smaller than the characteristic size of the spatial extend of $E_{\parallel}$ and field line curvature radius $\rho_{\rm c}$ (Fig. \ref{fig:gap_times}).



\subsubsection{Estimate due to energy in field}
An alternate method of estimating the luminosity is by consideration that a fraction $\eta$ of the total energy in the parallel electric field is converted to coherent radio luminosity each time the gap discharges. The total energy density in the gap electric field is $\frac{\epsilon_E}{V_{\rm gap}} = \frac{E_{\rm gap}^2}{8 \pi} \approx \frac{E_{\parallel}^2}{8 \pi}$. The gap volume is the cross-sectional area times the gap height: $V_{\rm gap} = h_{\rm gap} A$. 
\par
In the pulsar gap model, curvature photons emitted by the primary accelerated particle population will produce secondary pairs, with the efficiency of the cascade and therefore multiplicity $\kappa$ dependent on the specific gap physics \citep{TimokhinHarding2015,Wadiasingh2020}. The gap discharge is likely non-stationary, occurring on timescales longer than $h_{\rm gap}/c \sim 10^{-7}$s \citep{Timokhin2010}, but still shorter than the orbital timescale, thus we expect quasi-continuous emission whenever a large $E_{\parallel}$ field is present. Given this, we take $h_{\rm gap}/c$ as an estimate for the gap discharge timescale in the lower number density limit. Therefore the radio luminosity is estimated as:
\begin{equation}
    \begin{split}
    L_r &= \eta \frac{\epsilon_E}{V} h_{\rm gap} A \frac{c}{h_{\rm gap}} \\
    &=  \frac{\eta}{8 \pi} E_{\parallel}^2 A c = \frac{\eta}{2} E_{\parallel}^2 R_{\rm NS} c \\
    \end{split}
    \label{eq:analytic_pulsar_flux_2}
\end{equation}
We see the two methods of approximating the coherent radio luminosity agree to within a factor of 2. The above calculation assumes implicitly that a single gap is sufficient to supply enough charge to satisfy current requirements along field lines with $E_{\parallel}$. If this is not the case, multiple gaps could develop in the longitudinal direction along field lines, meaning for a length scale $L \approx R_{\rm NS}$ the number of gaps and thus radio luminosity would scale linearly with a filling factor $f = \frac{L}{h_{\rm gap}}$. We assume that current provided by the gap discharge is sufficient and thus take $f=1$, but an upper limit for this dimensionless factor is: $f = 10^{4} \; L_{6} \, h_{\rm gap,2}$, and thus would represent a luminosity increase by a factor $10^4$. 



\begin{figure}
  \centering
{\includegraphics[width=.5\textwidth]{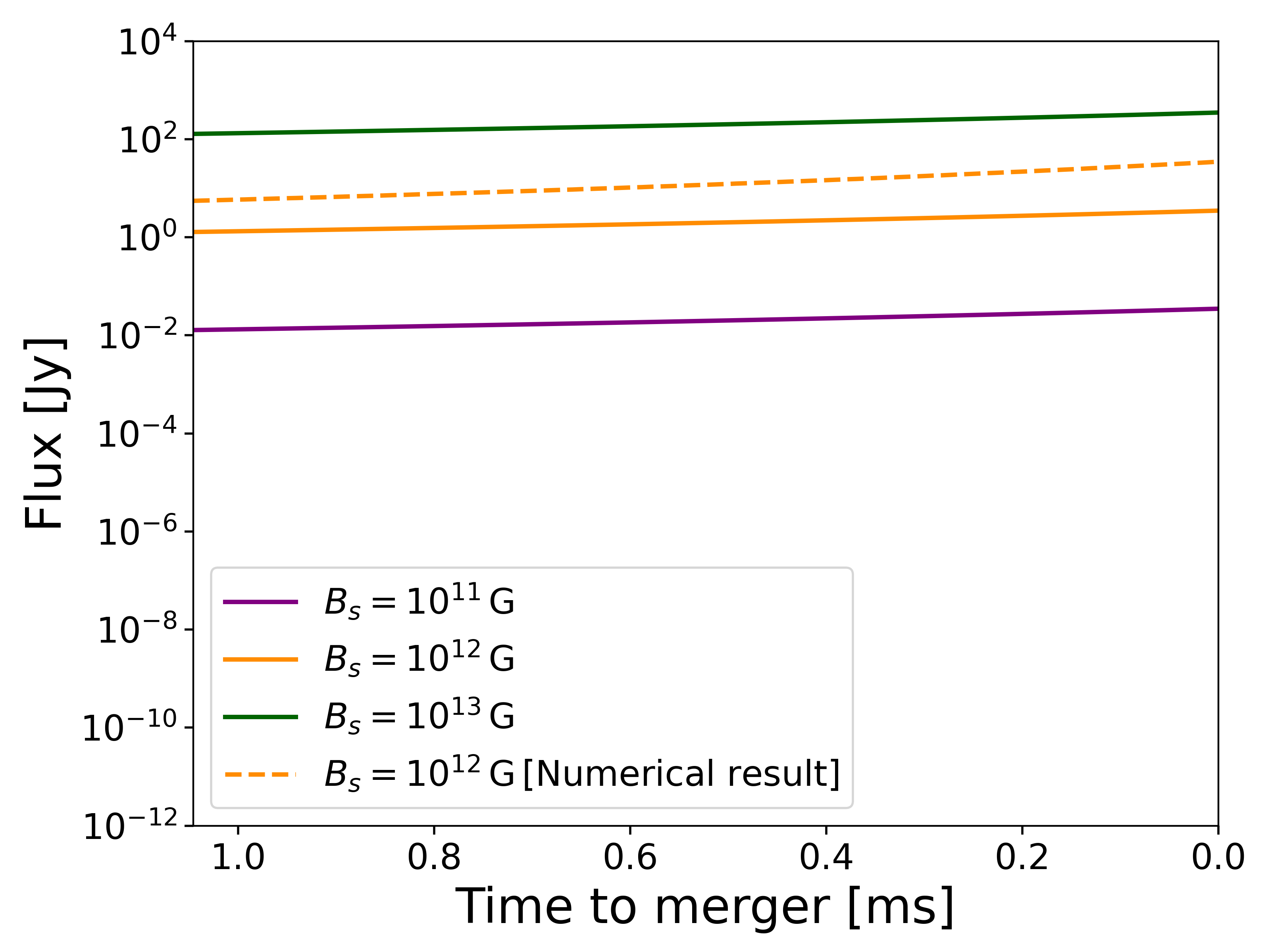}}
\caption{An analytic estimate of the total flux as a function of time to merger given by Eq. \eqref{eq:analytic_pulsar_flux}, assuming an emission bandwidth of $\delta \nu = 10^{10} \,$Hz, $D = 100 \,$Mpc and $\eta = 10^{-2}$. For $B_{\rm s} = 10^{12}$G, we include both the analytic (solid) \& numerical (dashed) fluxes as described in the text.}
\label{fig:analytic_pulsar_flux}
\end{figure}

\begin{figure}
  \centering
\includegraphics[width=.5\textwidth]{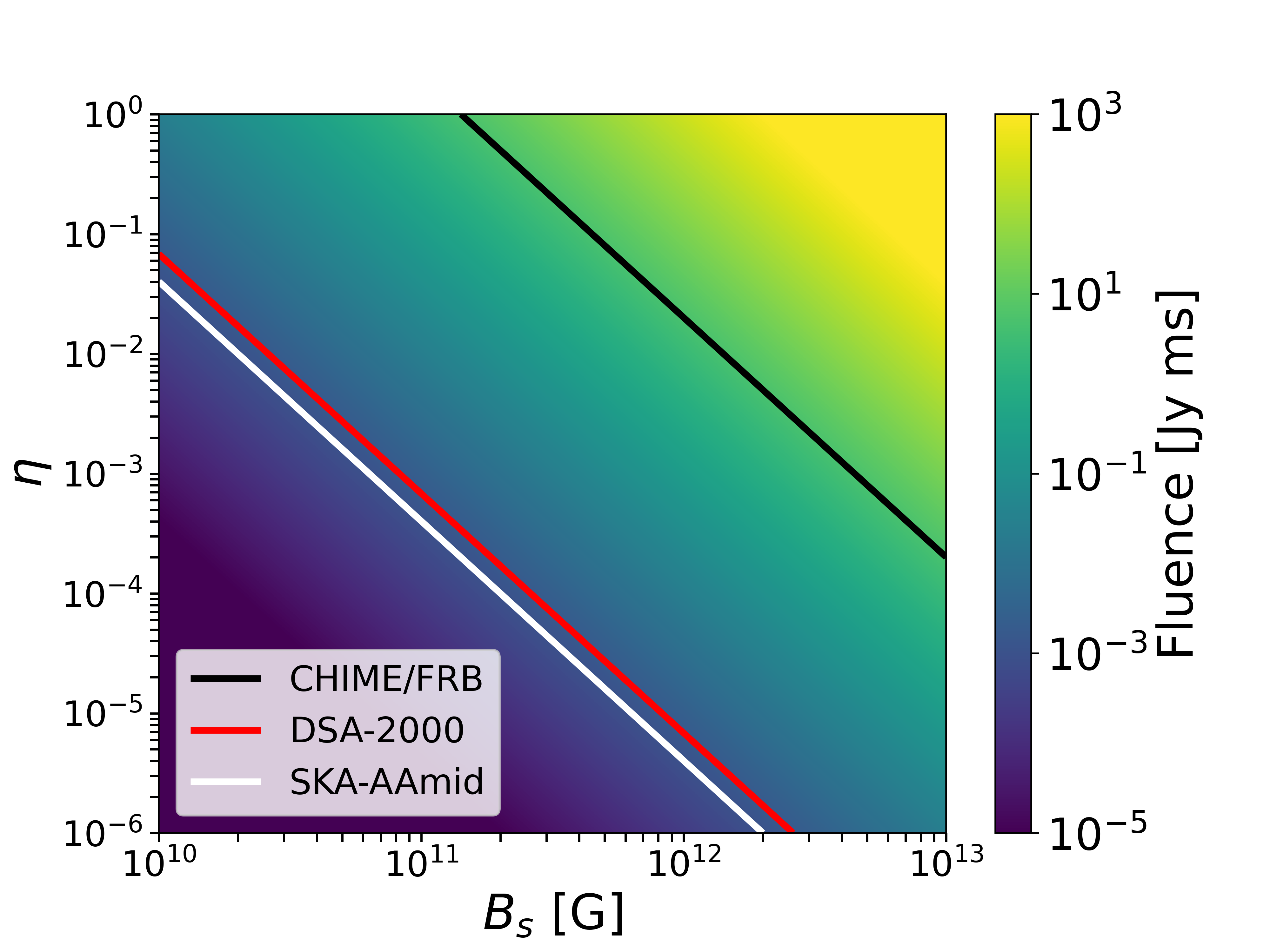}
\caption[1]{Total fluence of the last 3ms of the inspiral, corresponding to 2 orbital periods and thus 2 peaks of emission, for various values of efficiency $\eta$ and surface magnetic field $B_{\rm s}$. We assume an emission bandwidth of $\delta \nu = 10^{10} \,$Hz, and distance to source of $D = 100 \,$Mpc. The assumed fluence limits are: CHIME/FRB 5 Jy ms  \citep{Josephy2021}, DSA-2000 1.6 mJy ms\footnotemark \& SKA-mid 1 mJy ms \citep{Torchinsky2016}. The fluence predictions in this plot refer to the case where $E_{\rm gap} = E_{\parallel}$, and assume an optimal viewing angle.}
\label{fig:fluence_pulsar}
\end{figure}
\footnotetext{\url{https://www.deepsynoptic.org/instrument}}
\par
It is useful to compare the radio luminosity proxy to the Poynting luminosity of the binary. For an equal mass system where the magnetic moment of the primary $\mu_{\rm pri} \gg \mu_{\rm sec}$, the Poynting luminosity $L_{\rm P}$ is:
\begin{equation}
    L_{\rm P} = \frac{2 \mu_{\rm pri}^2}{3 c^3} \bigg(\frac{\Omega}{\Omega_{\rm g}}\bigg)^{2/3} \Omega^4 \propto \Omega^{14/3} \propto a^{-7}
\end{equation}
Where $\mu_{\rm pri} = B_{\rm s, pri} R_{\rm NS}^3$, $\Omega$ is the angular frequency of the binary defined in Eq. \ref{eq:binary_freq} and $\Omega_{\rm g}$ is the angular frequency defined at a point $R_{\rm g} \approx 3 R_{\rm NS}$ \citep{Medvedev2013}. In the $E_{\rm gap} = E_{\parallel}$ limit $L_{\rm R} \propto E_{\parallel}^2 \propto a^{-7}$ via Eq. \ref{eq:E(t)}, such that the radio luminosity scales similarly to the Poynting luminosity but it is always the case that $L_{\rm R} < L_{\rm P}$. 
\par
In the lower limit $E_{\rm gap} = 4 \pi q n_{\rm GJ} h_{\rm gap}$ case, $L_{\rm R} \propto n_{\rm GJ}^2 h_{\rm gap}^2 \propto a^{-30/7}$  (Eqs. \ref{eq:gap_appendix} \& \ref{eq:coherent_lum_appendix} in Appendix \ref{app:goldreichapprox}). In this case, we can define the time at which radiation mechanism ought to commence by finding the separation and time at which $L_{\rm R} = L_{\rm P}$, while $\eta = 1$:
\begin{equation}
    \begin{split}
        a_{R = P} = 5 \times 10^{8} \: {\rm cm} \: B_{\rm s, 12}^{8/19} \, \eta_{-2}^{-7/19} \, \rho_{\rm c,6}^{-4/19} \, A_{12}^{-7/19} \\
        t_{R = P} = 2 \times 10^{5} \: {\rm seconds} \: B_{\rm s, 12}^{32/19} \, \eta_{0}^{-28/19} \, \rho_{\rm c,6}^{-16/19} \, A_{12}^{-28/19}
    \end{split}
\end{equation}
The above equation suggests that in the $E_{\rm gap} = 4 \pi q n_{\rm GJ} h_{\rm gap}$ limit, radio emission should turn on approximately one day before merger, albeit at a low luminosity. Despite this, the necessary condition that $h_{\rm gap} < R_{\rm NS}$ (e.g. Fig. \ref{fig:flux_and_gap_long}) is in general more constraining and therefore it is unlikely the radiation mechanism turns on before $10^{3} \, B_{\rm s, 12}$ seconds before the merger for NS-NS binaries.

\subsection{Numerical Implementation}
As mentioned in Section \ref{sect:numerical}, we use a numerical set-up to estimate viewing angle dependence of the radio luminosity from the system. We calculate $E_{\parallel}$ surrounding the secondary conducting neutron star out to distance $R = 5 R_{\rm NS}$, and compute the radio luminosity using Eq. \eqref{eq:analytic_pulsar_flux}. We calculate the gap height for each cell for each timestep, to ensure that pair production occurs within a fraction of the neutron star radius, as is required by our assumption of emission along field lines and to ensure the voltage is sufficient for pulsar-like emission using Eq. \eqref{eq:voltage}. We replace the cross-sectional area $A$ with the cross sectional area of each cell, estimated as $(\delta V)^{2/3}$. As aforementioned we include an efficiency factor $\eta < 1$ to capture both the uncertainty related to the number of gaps that form and contribute to emission, but also the conversion  of primary particle power to coherent radio radiation. We further assume an observing frequency of $\nu_{\rm obs} = 10^{9} \, {\rm Hz}$ and a total spectral bandwidth of emission of $\delta \nu_{\rm obs} = 10^{10} \, {\rm Hz}$. In Fig. \ref{fig:analytic_pulsar_flux}, we also show the total radio luminosity integrated over all viewing angles by way of comparison to the analytic calculation. The luminosity found using the numerical approach is slightly higher, attributed to the fact that we calculate emission from a larger volume than assumed in the analytic calculation, and thus the total cross-sectional area $A$ in the analytic calculation under-estimates the total area from which pulsar emission is expected.
\par
We also estimate the maximum luminosity of the system assuming coherent curvature radiation, discussed in Appendix \ref{app:cohcurv}. We discuss only upper limits to the bunch luminosity based on electromagnetic considerations, and therefore instead of summing all emission in the direction of an observer as in the pulsar-like case, we simply find the maximum value associated with the set of field lines aligned with each observer.

\subsection{Viewing angle dependence}
\label{sect:viewingangle}
The approach we take to the calculation of $E_{\parallel}$ in Section \ref{sect:model} assumes a uniform background magnetic field $\mathbf{\hat{B}} = - B \mathbf{\hat{z}}$ stemming from the primary NS's dipole magnetic field. This approximation limits a full understanding of the viewing angle dependence of emission, for two reasons. Firstly, the calculation of the expunged magnetic field will be different in a realistic dipole magnetic field, thus yielding a $E_{\parallel}$ field map having a different spatial morphology. Secondly, the perturbed magnetic field lines along which particle acceleration and radiation is directed may be offset to the directions described here. We expect the emission to be emitted in a slightly wider range of observing angles due to the dipole nature of the magnetic field, particularly at small values of orbital separation $a$ where the dipole's deviation from a uniform field is greatest. However, the strongest $E_{\parallel}$ fields occur close to the NS surface (maximal value at $R = 1.23 R_{\rm NS}$) and thus the direction of magnetic field lines is more strongly influenced by the perturbation of the field lines caused by the moving secondary NS, and not the background field orientation.
\par
The perturbations of the magnetic field lines from their background orientation also result in variations of the radio luminosity at different observing angles. The maximum magnetic field line deflection occurs is quadrapole in nature (corresponding to maximal values of the absolute value of $\sin(\theta)$; see Fig. \ref{fig:conductor_field_lines} \& Eq. \ref{eq:om}), which means emission is suppressed at larger angles to the background field. For field lines that are unperturbed (i.e. at $\theta = \pi/2$) there is no strong $E_{\parallel}$ field component, meaning radiation is suppressed for observers on-axis to the background field, as seen in Section \ref{sect:MM_MW_prospects}. Corresponding to this, we see that the radio luminosity drops off substantially for observers at angles to the background field smaller than 10 degrees. As such, we find that almost all of the emission is emitted within 5-45 degrees of the magnetic axis of the field of the primary magnetized neutron star, with a peak of emission occurring at an angle offset from the background magnetic field $\approx$ 10 degrees.
\par
It is crucial to remember that we do not necessarily expect the magnetic axis of the primary magnetized neutron star to be perpendicular to the orbital plane, as has been assumed throughout this work. However, our viewing angle dependent results need only be rotationally transformed to represent cases where the angle between the orbital plane and the magnetic axis of the primary NS is not $90$ degrees, as shown in Fig. \ref{fig:rotated_alpha}. This rotational symmetry for the uniform magnetic field at the secondary's position comes from the fact that the motion of the conductor is orthogonal to the magnetic field lines. In the Figs \ref{fig:GW}-\ref{fig:kilonovae_g} in Section \ref{sect:MM_MW_prospects}, we assume the magnetic obliquity $\alpha_{\rm B, orb} = 90 \deg$. Observing the coherent pre-merger emission could aid in constraining the magnetic obliquity of NS merger sources, and thus provide insights into the binary evolution of merging neutron stars. 
\par
As we consider that the primary NS has a dipole magnetic field, the value of the magnetic field strength B at a point ($r, \theta, \phi$) will change depending on the orientation of the axis to the point in question.  For values of $\alpha_{\rm B, orb} < 90 \deg$, the strength of this magnetic field at the secondary's position, for a separation $a$, will range between:
\begin{equation}
   B_{\rm s} \frac{R_{\rm NS}^3}{a^{3}} \leq B \leq B_{\rm s} \frac{R_{\rm NS}^3}{a^{3}} \sqrt{1 + 3 \cos^2(\alpha_{\rm B,orb})}
\end{equation}
where the maximal case occurs when the magnetic axis is tilted exactly towards the secondary $\alpha_{\rm B, orb} = 0 \deg$. In this case, the magnetic field can increase by up to a factor of 2, resulting in a coherent luminosity increase by a factor of 4. These considerations have not been numerically implemented in any section, given their relatively small increase to the overall luminosity.

\subsection{Temporal Morphology}
\label{sect:temporal}
In Section \ref{sect:MM_MW_prospects} we discuss how one may confirm a NS-merger origin of coherent precursor bursts discussed in this paper. One way in which these precursor bursts can be distinguished from FRBs from other sources is through analysis of the temporal morphology of the burst, which we discuss here (see also \citealt{Gourdji2020}). 
\par
\begin{figure}
  \centering
{\includegraphics[width=.5\textwidth]{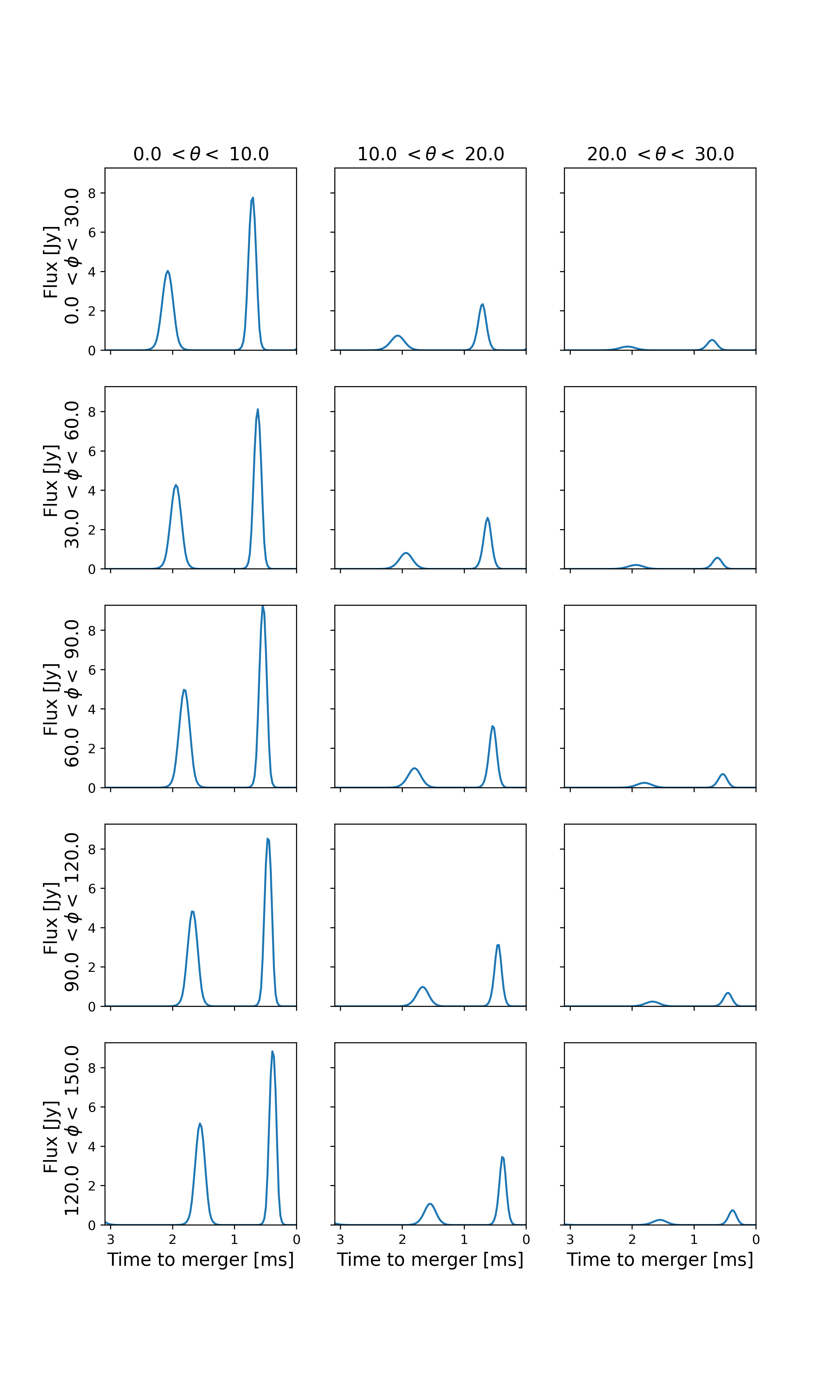}}
\caption{Example light curves for the final 3ms of the inspiral, covering the final 2 orbital periods. We show pulsar-like emission from NS-merger $\eta = 10^{-2}$ \& $B_{\rm s} = 10^{12}$G at $D = 100 \,$Mpc, for various observers positioned at different azimuthal and polar angles.}
\label{fig:lightcurves}
\end{figure}

In the model presented here, radio precursors of NS-NS mergers are modulated by the orbital period of the binary. In the Newtonian approximation, the angular frequency of the binary is:
\begin{equation}
    \Omega = \frac{(G (M_1 + M_2))^{1/2}}{a^{3/2}} \approx 10^4 \: M_{\rm NS}^{1/2} \, a_{6}^{-3/2} \; {\rm Hz}
    \label{eq:binary_freq}
\end{equation}
 where $M_1$ \& $M_2$ are the masses of the neutron stars and $a$ is the separation. This is seen in Fig. \ref{fig:lightcurves}, where observers at different azimuthal and polar angles to the orbital plane observe emission from different phases and magnetic field lines respectively. If coherent radio emission is bright enough to be observed at more than one orbital period, progressively brighter sub-millisecond bursts are observed with a decreasing separation between bursts as dictated the decreasing orbital period $P \approx \frac{2 \pi a}{\beta c} \propto \sqrt{a} \propto \big(1 - \frac{t}{t_{\rm m,0}} \big)^{1/8} $ using Eq. \eqref{eq:a(t)}. Sub-millisecond periodicity has been claimed in a four FRB sources \citep{CHIME_ms_2021,Pastor-Marazuela2022}, however explaining the periodicity as modulated by a compact object binary inspiral is generally disfavoured, as the period between sub-bursts does not appear to decrease as expected. We note that one could plausibly invoke eccentric orbits or unequal mass ratios in order to change the expected sub-burst morphology. 
\par
The characteristic increase of flux can also be used to distinguish such bursts, and the increase depends on the exact nature of the emission mechanism. In the upper limit assumption that $n \propto \frac{E_{\parallel}}{4 \pi q h_{\rm gap}}$, the radio luminosity scales as $L_r \propto E_{\parallel}^2$. This is also the case for the coherent curvature radiation mechanism (unless the coherent emission is limited by the magnetic field constraint of \cite{Kumar2017}, see Appendix \ref{app:cohcurv}). By inspection of Eq. \eqref{eq:om} \& \eqref{eq:E(t)}, we find that $E_{\parallel}^2 \propto B^2 \beta^2 \propto a^{-7} \propto \big(1 - \frac{t}{t_{\rm m,0}} \big)^{-7/4}$. In the lower limit estimate where $n=n_{GJ}$ (Appendix \ref{app:goldreichapprox}), the characteristic flux increase is instead $L_r \propto h_{\rm gap}^2 B^2 P_{\rm orb}^{-2} \propto a^{-22/7} \propto \big(1 - \frac{t}{t_{\rm m,0}} \big)^{-11/14}$. However, there may be spectro-temporal variations (with prejudice towards increasingly higher frequencies as the plasma density increases during inspiral) which may require broadband observations to ascertain this anticipated temporal dependence. Where sub-millisecond temporal resolution is available, we suggest that matched template technique could be used to identify NS-merger origin FRBs, and possibly to distinguish between the different emission mechanism limits discussed in this Section. Furthermore, identification of coherent radio emission modulated by the orbital period and phase will provide a measurement of the (combined) neutron star masses similarly to gravitational wave emission \citep{Cutler1994} which may inform the neutron star equation of state. Close-by NS-mergers where one source has a high magnetic field are rare, but may be detectable over many orbital periods and may allow for detailed estimates of the neutron star masses. 
\par

\subsection{Absorption}
As discussed in \cite{Lyutikov2019}, it is possible that the radio signal predicted in this section may not escape the source. In particular, predictions of high-energy precursors to NS mergers invoke a dense shroud of particles surrounding the merging neutron stars \citep{Metzger2016}. The dense pair production front may prevent the propagation of radio emission, unless generated EM waves are superluminal relative to the plasma (e.g. \citealt{Timokhin2013}). The absorption frequency of electromagnetic waves in a plasma is modified due to the magnetic environment \citep{AronsBarnard1986}, and emission may propagate if $\omega > \frac{\omega_p^2}{\omega_B}$. 
\begin{equation}
    \begin{split}
        \omega &> \dfrac{4 \pi q n_{\rm sec} c}{B} = \dfrac{E_{\parallel} \kappa c}{B h_{\rm gap}} \\
        &= 300 \: E_{\parallel,10} \, \kappa_1 B_{11}^{-1} \, h_{\rm gap,2} \; {\rm Mhz} 
    \end{split}
\end{equation}
Where we have used the fact that the density of the secondaries from which coherent emission is radiated is $n_{\rm sec} = \kappa n_{\rm primaries} \approx \frac{E_{\parallel} \kappa}{4 \pi q h_{\rm gap}}$. Thus it is plausible that GHz emission escapes in regions of highest electric and magnetic field for typical multiplicity and gap height values, with a caveat that \cite{AronsBarnard1986} assume a homogeneous and stationary plasma which is not the case. These regions are also where we expect the strongest particle acceleration and therefore emission. \cite{Kumar2017} also argue that free-free absorption of GHz radiation will be negligible as long as the number density does not exceed $\sim 10^{19} \rm{cm^{-3}}$ for a source size corresponding to a gap height $\approx 10^{3} \, {\rm cm}$. This implies that pair multiplicity in the secondary cascade region should be $\kappa \lesssim 1000$, matching the range predicted by \cite{TimokhinHarding2015}. Furthermore, as the photons propagate through the magnetosphere of the primary, the magnetic field strength $B$ and density $n$ will both decrease linearly assuming a Goldreich-Julian charge density, meaning emission that escapes the immediate vicinity is expected to propagate to the observer. In \cite{Wang2016}, the authors find that coherent radio emission of approximately $\nu \approx 1$ GHz may freely escape the magnetosphere during a NS-NS inspiral. 


\subsection{High-energy emission}
Regardless of the specific mechanism of coherent emission, high-energy radiation will also be emitted. In the polar cap model of pulsar emission, this is explained by curvature \& synchrotron photons emitted by accelerated pairs that do not meet energy requirements to interact with the magnetic field to produce pairs, and thus contribute to the gamma-ray flux \citep{DaughertyHarding1982,DaughterHarding1996}. Gamma-ray emission modulated by the spin period has been observed for hundreds of isolated neutron stars \citep{Abdo2013a,Caraveo14}. In the NS-NS merger case, the most likely production mechanism would resemble those in observed gamma-ray pulsars, where radiation emerges from charged current sheets outside the light cylinder in the equatorial plane (relative to rotation) where large electric fields are likely realized for curvature radiation. The pulsar gamma-ray luminosity functions of \cite{2019ApJ...883L...4K,2022arXiv220313276K} provide a gross baseline estimate, with the identification of spin period to orbital period, corresponding to $L_{\gamma} \approx 10^{37} \, {\rm erg s^{-1}}$, assuming $\epsilon_{\rm cut-off} = 1$ GeV, $\dot{P} = 10^{-12}$ and $P = P_{\rm orb} \approx 1$ ms. In the coherent curvature radiation model, high-energy emission may be emitted by the coherently radiating particles themselves due to the twisting of magnetic field lines by coherent bunches \citep{CooperWijers2021}, or by a trapped fireball associated with a crustal trigger event \citep{Yang2021}. In both cases, high-energy radiation is far too weak to be probed to extra-Galactic distances with current facilities. 
\par
Finally, \cite{Metzger2016} consider an unspecified mechanism which converts a large fraction of available electromagnetic energy during the inspiral to gamma-ray radiation. In all cases, the lower sensitivity of gamma-ray detectors means that precursor emission is difficult to observe. \cite{Metzger2016} found that even for very efficient conversion of electromagnetic energy to high-energy radiation, precursors are only observable to a distance $D \approx 10 \, {\rm Mpc} \; (B_{\rm s}/10^{14} {\rm G})^{3/4}$ with current instruments.

\section{Multi-wavelength \& multi-messenger detection prospects}
\label{sect:MM_MW_prospects}
In this Section we discuss the feasibility of (co-)detection of the coherent pre-merger emission discussed in this work in blind searches, triggered observations of multi-wavelength and multi-messenger signatures of neutron star mergers, and follow-up observations. We do not discuss all-sky radio telescopes such as the Survey for Transient Astronomical Radio Emission 2 (STARE-2; \citealt{Bochenek2020}), the planned Galactic Radio Explorer (GReX; \citealt{Connor2021}), and the Amsterdam-ASTRON Radio Transients Facility And Analysis Center (AARTFAAC; \citealt{Prasad2016}) in detail as they are not in general sensitive enough to detect extra-Galactic coherent radio emission from the model presented here.  
\begin{table}
\centering 
\begin{center}
\begin{tabular}{c|c|c|c|c|c} 

 & $B_{\rm s}$ & GRB & GW & Radio & Kilonova \\
&[G] &trigger & trigger & afterglow & \\
\hline
Current & $10^{12}$ & \xmark & \xmark & \xmark &  \xmark \\ 
generation& $10^{14}$ & ? & \cmark & ? & ? \\
\hline
Next & $10^{12}$ & \cmark & \cmark  & \cmark & ? \\ 
generation & $10^{14}$ & \cmark &\cmark  & \cmark & \cmark \\

\end{tabular}
\end{center}
\caption{Simplified summary of this section. For current/next-gen detectors and two surface magnetic field strengths, we show whether the coherent emission model is likely to be probed from NS-NS mergers (GRB \& GW triggers) or given an observed radio burst, can a NS-NS merger origin be verified (afterglow \& kilonova). The afterglow \& kilonova possibilities rely on the current \& future generation FRB facilities as detailed in Section. \ref{sect:frb_survey}. We have assumed an optimal viewing angle and magnetic obliquity $\alpha_{\rm B, orb}$ for each observing method, and question marks represent uncertainties in the merger rate and/or model parameters. Much more detail is given in the text of each section, including current observational status, methodology \& expected next generation capabilities.}
\label{tab:frb_MWMM_prospects}
\end{table}

\par
\begin{figure}
  \centering
{\includegraphics[width=.5\textwidth]{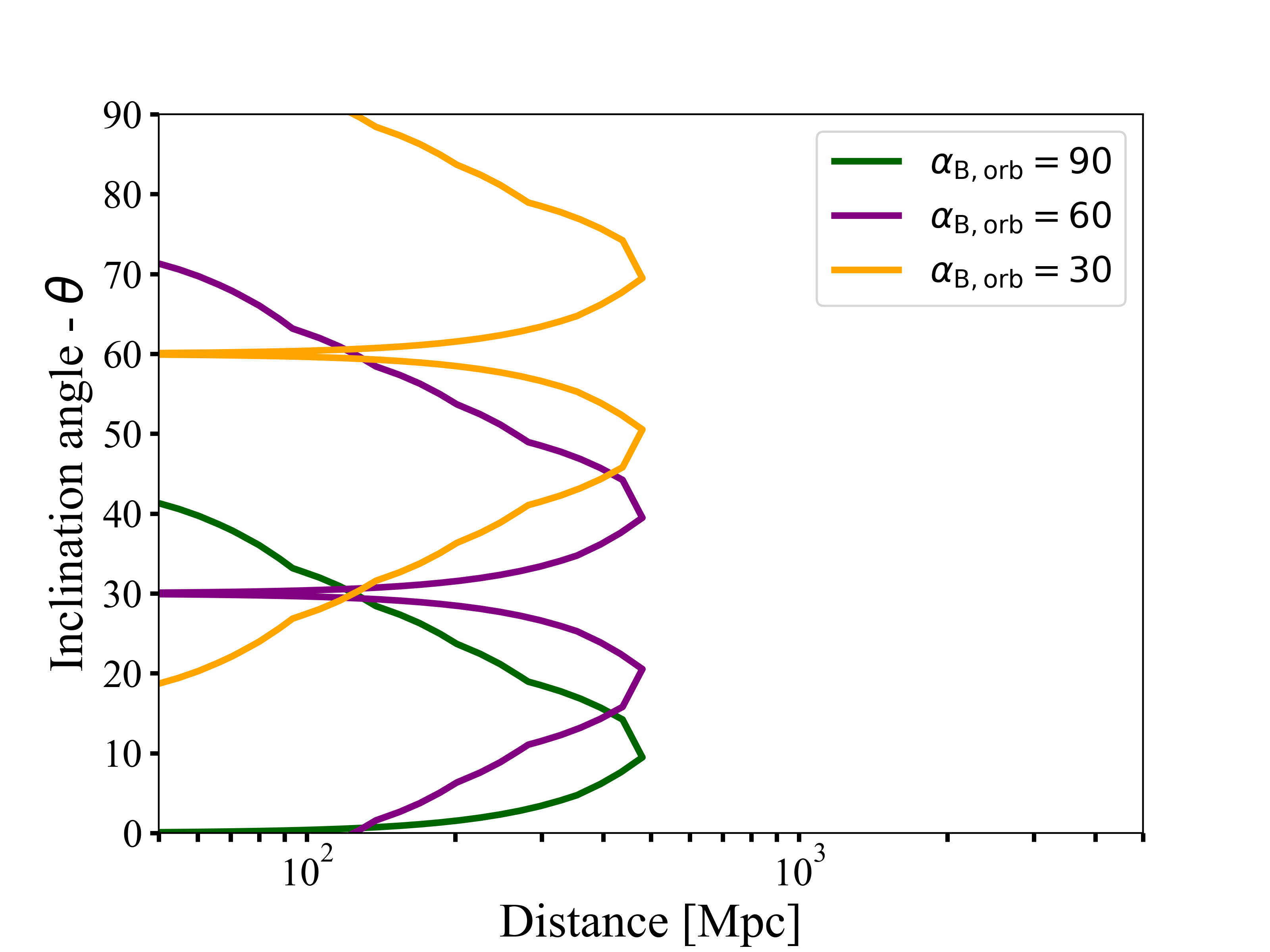}}
\caption{Inclination angle dependent observing horizon for SKA-mid for NS merger radio bursts from mergers with different magnetic obliquities $\alpha_{\rm B, orb}$; the angle between the magnetic axis of the primary magnetized neutron star and the orbital plane. The symmetry is due to the $E_{\parallel}$ azimuthal symmetry, i.e. the two peaks refer to particle acceleration in the red and blue regions of Fig. \ref{fig:conductor_field_lines}. We assume pulsar-like emission with parameters, $\eta = 10^{-4}$, $B_{\rm s} = 10^{12}$ G as explained in Section. \ref{sect:particle_acceleration}. For the rest of the plots in this section, we show only the $\alpha_{\rm B, orb} = 90$ horizon for readability.}
\label{fig:rotated_alpha}
\end{figure}

In Section \ref{sect:frb_survey}, we discuss the prospects for detecting coherent pre-merger emission from NS mergers through blind FRB surveys. In Section \ref{sect:ShortGRBs} we discuss rapid and triggered observations of sGRBs, putting past rapid observations of sGRBs with MWA and LOFAR in the context of this work, and make predictions for future observations with SKA. In Section \ref{sect:gw} we discuss prospects for rapid observations of gravitational-wave detected mergers and detection of pre-merger emission. Finally in Sections \ref{sect:afterglow} \& \ref{sect:kilonovae} we discuss how follow-up observation of one-off FRBs without a GRB or GW counterpart could be confirmed to be of merger-origin using radio and optical facilities. A full description of the instruments discussed in the text is available in Table \ref{tab:telescopes}. In all cases, we show the co-detection space as a function of binary inclination angle (assuming the magnetic axis and orbital plane are perpendicular: $\alpha_{\rm B, orb} = 90 \deg$) and distance, assuming fluence sensitivity for coherent bursts with next-generation radio telescopes. In Fig. \ref{fig:rotated_alpha}, we show how the magnetic obliquity $\alpha_{\rm B, orb}$ changes the inclination angle dependent detectability horizon. In Table \ref{tab:frb_MWMM_prospects} we provide a highly simplified summary of this Section, and in Table \ref{tab:telescopes} we provide a description of the properties of the current and future instrumentation considered in this study.
\par
\par

\par

\begin{table*}
    \centering
    \caption[1]{Here we list the properties of the telescopes discussed in the text and figures of this work including CHIME/FRB \citep{CHIME2018,CHIME_FRB_CAT2021}, CHORD \citep{Vanderlinde2019}, DSA-2000 \citep{Hallinan2019}, LOFAR \citep{2013A&A...556A...2V}, MWA \citep{2013PASA...30....7T}, ZTF \citep{Bellm2019}, LSST \citep{Tyson2002}, MeerKAT \citep{Jonas2016}, SKA1\footnotemark \citep{Dewdney2009,Braun2019} and SKA-AAmid \citep{Torchinsky2016}. In each case we list specifications and assumed values given the use of the instrument within the context of this work, as a blind-FRB survey instrument, a GW/GRB rapid response instrument, and optical and radio afterglow follow-up. The assumed sensitivity, field-of-view and localization refer to those values at typical observing frequencies in the centre of bandwidths. For SKA-AAmid, we assume coherent bursts are are searched in real time over the entire $200 \deg^2$ FoV as discussed in \citep{Torchinsky2016,Hashimoto2020}. Typical LOFAR and MWA fluence sensitivities are used based on past successful triggers of GRBs, and the FoVs are taken from \cite{Chu2016} but vary depending on observing setup. Transient buffer board capabilities of MWA's voltage capture system are detailed in \cite{Tremblay2015}. Fluence detection thresholds for future instruments are estimated by scaling continuum sensitivities to millisecond integration. Finally we note that in 2023, the CHIME/FRB localization is expected to reach 50 milliarcseconds for a majority of sources thanks to the outrigger project \citep{Leung2021} and the beginning of CHORD. For CHORD sensitivity, we assume a 1ms burst that emits over the entire observing bandwidth.} 
\label{tab:telescopes}
\begin{tabular}{ccccccccc}
\hline
\hline 
\\
Type & Telescope & Frequency & Sensitivity  & FOV & Localization/Resolution & Trigger time & TBB   \\
   &        &    [MHz]    & &       [$\deg^2$]        &     [arcsec]                 &          &            &   \\ \\
\hline 
\\
FRB Survey & CHIME & 400-800 &  5 Jy ms  & 250  & 1 $\deg$ & n/a & n/a \\
    & DSA-2000  & 700–2000 & 1.8 mJy ms  & 10.6  & 3.5 & n/a & n/a \\ 
    & CHORD & 300-1500 &  60 mJy ms  & 65  & 0.05 & n/a & n/a \\
    & SKA-AAmid  &  450-1450 & 1 mJy ms & 200  & 0.22 & n/a & n/a&  \\ \\

GW/GRB Trigger & LOFAR (imaging) & 120-240 & 3000 Jy ms & 48 & 6 & 3-4 min & 5 s &\\      
        & LOFAR (beamformed)  & 120-240 & 25-1000 Jy ms & 0.05-16 & n/a & 3-4 min & 5 s &\\
        & MWA & 80–300 & 1000 Jy ms & 610  & 100 & 20-30 s & $\leq 100$ s  \\
        & SKA1-low & 50-350 & 4 mJy [1ms int] & 27  & 11 & <20 s & 30 s \\ \\

Follow-up & ZTF & 464-806 [nm] &  m19.9 [i] - m20.8 [g] & 47.7  & 2 & n/a & n/a& \\
         & LSST  & 0.3-1 [$\mu$m] & m24.0 [i] - m25.0 [g] & 9.6  & 0.7 & n/a & n/a & \\
         & MeerKAT & 580–2500& 700 $\mu$Jy [2hr int]& 0.85 & 10 & n/a & n/a \\ 
         & SKA1-mid  &  350-1500 & 2 $\mu$Jy [1hr int] & 0.48 & 0.22 & 1-10 min & >9 min \\ \\
\hline
\\
\hline 
\end{tabular}
\end{table*}
\footnotetext{see also: \url{https://www.skatelescope.org/wp-content/uploads/2012/07/SKA-TEL-SKO-DD-001-1_BaselineDesign1.pdf}}

\subsection{Fast radio burst surveys}
\label{sect:frb_survey}
The simplest method of detecting coherent radio bursts from NS merger events is through blind FRB surveys. FRBs are extra-Galactic, (sub)-millisecond duration radio bursts, and many hundreds of bursts have now been seen since their discovery \citep{Lorimer2007,CHIME_FRB_CAT2021}. FRBs are classified as either repeating or non-repeating sources, and the burst properties of these two classes appear quite different, particularly the spectral bandwidth and burst duration \citep{Pleunis2021a}, which may be suggestive of different progenitors. The all-sky FRB rate is large, with the latest estimate from FAST \citep{2006ScChG..49..129N} putting the rate at $1.24^{+1.94}_{-0.99} \times 10^5 \, {\rm sky^{-1} \, day^{-1}}$ above 0.0146 Jy ms (95\% confidence interval; \citealt{Niu_FRB2021}). Instruments with large fields of view and sufficient sensitivity are best suited to finding them, including CHIME/FRB \citep{CHIME_FRB_CAT2021}, ASKAP \citep{Macquart2010} \& DSA \citep{Hallinan2019} to name three of the most prolific. Of current FRB instruments, FRBs are found most frequently by the Canadian Hydrogen Intensity Mapping Experiment (CHIME)/FRB team \citep{CHIME2018}, reporting over 500 FRBs in the first catalog \citep{CHIME_FRB_CAT2021}, and the Australian Square Kilometre Array Pathfinder (ASKAP) \citep{Macquart2010} has had success in localizing one-off bursts to their host galaxies \citep{Bhandari2020,Heintz2020,Day2021,Bhandari2022}. Despite this, poor localizations of CHIME FRBs and high redshift sources present difficulties in finding persistent or variable counterparts to non-repeating FRBs \citep{Gourdji2020}. Many authors have suggested that coherent bursts originating from NS mergers may be detected as one-off FRBs (e.g. \citealt{Totani2013}) without an observed multi-messenger or multi-wavelength counterpart, but the volumetric rate of NS-NS mergers appears to be a factor of 10-100 too low to explain all FRBs \citep{Ravi2019,Lu2019,Luo2020,Mandel2021}. In this subsection, we consider whether pre-merger coherent emission could be probed by the CHIME/FRB radio telescope, but also future FRB survey instruments, namely the upcoming Deep Synoptic Array (DSA-2000; \citealt{Hallinan2019}), the CHIME/FRB successor the Canadian Hydrogen Observatory and Radio-transient Detector (CHORD; \citealt{Vanderlinde2019} and the Square Kilometre Array (SKA; \citealt{Dewdney2009,Torchinsky2016}) observatories. 

\subsubsection{Current generation}
In the model presented here, the number of pre-merger coherent radio bursts that would be detectable as one-off FRBs depends sensitively on the surface magnetic field of the primary neutron star $B_{\rm s}$ and the radio efficiency $\eta$. We show the fluence of radio bursts for a range of parameters $B_{\rm s}$ and $\eta$ in Fig. \ref{fig:fluence_pulsar}, assuming a distance to the source of $D = 100\,$Mpc. In order to estimate the rate of FRBs, we assume that all NS mergers contain a $10^{12} \, {\rm G}$ neutron star, and pulsar-like emission occurs with an efficiency of $\eta = 10^{-2}$ (henceforth the fiducial parameters), but include parameter scalings for the horizon and rate. In this case, the CHIME/FRB fluence horizon of the pulsar-like emission at optimal viewing angles is: $D_{\rm horizon, CHIME} \approx 70 \; \eta_{-2}^{1/2} \, B_{s, 12} \;$Mpc. Assuming a universal volumetric NS-NS merger rate of $\mathcal{R} = 10^3 \, {\rm Gpc^{-3} yr^{-1}}$ (see e.g. \citealt{Mandel2021}) and CHIME/FRB FoV of 250 $\deg$, this corresponds to just $N_{\rm CHIME} = 0.002 \; \eta_{-2}^{3/2} \, B_{s, 12}^3 \, \mathcal{R}_3$ events per year and thus cannot explain observed CHIME events, without invoking magnetar strength magnetic fields. We note that if just 20\% of all mergers involved a magnetar with $B_{\rm s} = 10^{14}$G, the observed CHIME rate could be explained, however this contradicts two observed facts. Firstly, the volumetric NS-NS rate is too low \citep{Ravi2019} and as most of the mergers capable of producing FRBs would have dispersion measures too high to be compatible with the observed population. Secondly, one would be required to explain why the characteristic temporal morphology is not observed in the brightest FRBs where multiple orbital periods would be bright enough to be observed if allowed by temporal resolution\footnote{It is plausible this is explained by a radiation mechanism only producing the cosmological radio bursts at threshold electromagnetic conditions, which we aim to explore in a future extension to this work.}. We therefore suggest it is unlikely that a significant population of the current observed CHIME/FRBs are powered by this mechanism, but searches for the temporal morphology suggested in Sect. \ref{sect:temporal} could yield a sub-population of mergers containing NS with $B_{\rm s} \approx 10^{14}$G.



\par

\begin{table}
\centering 
\begin{center}
\begin{tabular}{ |c|c|c| } 
\hline
 Telescope & Horizon [Mpc] & Event rate [yr$^{-1}$] \\
 &  $\big(\eta_{-2}^{3/2} \, B_{s, 12}^3\big)$ &  $\big(\eta_{-2}^{3/2} \, B_{s, 12}^3 \, \mathcal{R}_3 \big)$  \\
\hline
CHIME & 70 & 0.002\\ 
\hline
CHORD & 650 &  0.4\\
\hline
DSA-2000 & 3700 & 15 \\ 
\hline
SKA-AAmid & 5000 & 600\\ 
\hline
\end{tabular}
\end{center}
\caption{Observing horizon and the 100\% duty cycle detection rate for current leading and future FRB facilities. We assume fiducial model parameters for the efficiency $\eta$ and surface magnetic field $B_{\rm s}$, as well as a volumetric NS-NS merger rate $\mathcal{R} = 10^{3} \; {\rm Gpc^{-3} \, yr^{-1}}$}
\label{tab:frb_survey_detections}
\end{table}
\par
\subsubsection{Next generation}
In Table \ref{tab:frb_survey_detections}, we show the detection horizons and expected event rate of the fiducial coherent pre-merger emission of current and next generation FRB facilities. For DSA-2000, the smaller FoV is greatly offset by the expected increase in sensitivity, and thus the observed event rate is much larger than for either CHIME/FRB or CHORD. For SKA-AAmid, an unconfirmed extension to SKA-mid, the large FoV coupled with sensitivity produces many hundreds of detectable events per year. We note that these values are larger than expected by a factor of a few due to viewing angle dependencies discussed in Sect. \ref{sect:viewingangle}. However, the sensitive dependence on the magnetic field strength means that if just a few merging neutron stars have magnetic fields $B_{\rm s} > 10^{12}$G, the event rate will increase dramatically. 
\subsubsection{Other considerations}
The temporal resolution of CHIME/FRB intensity data (i.e. without triggering raw baseband data recording; see \citealt{CHIME_period2020,Michilli2021}) is approximately $1$ms \citep{CHIME2018}, although simulations have shown that sub-burst timescales down to $0.1$ms can be probed in a few cases \citep{CHIME_FRB_CAT2021}. The temporal morphology of coherent pre-merger bursts predicted in this paper is sub-millisecond peaks separated by the orbital period and increasing in intensity (as $\propto (1 - t/t_{\rm m})^{-7/4}$; see Sect. \ref{sect:temporal}). Such morphology may be detectable by CHIME/FRB depending on the signal-to-noise, scattering due to multi-path propagation \citep{Chawla2022} and bandwidth of bursts. SKA-mid not only has a higher sensitivity such that many peaks could be observed from orbital phases before merger, but is also expected to have temporal resolution on the order of 1-100 nanoseconds. If these coherent burst from NS-NS mergers are observed with SKA, they will likely be identifiable by their temporal morphology. 

\subsection{Short gamma-ray bursts}
\label{sect:ShortGRBs}
There have been many successful rapid radio observations of GRBs dating back decades (e.g. \citealt{Green1995}), but to detect pre-merger emission instruments must be on source of sGRBs extremely quickly. Some radio telescopes employ rapid-response modes, such that repointing can occur automatically in response to transient alerts issued by other facilities on platforms such as the VOEvent network \citep{Williams2006}, which allow machine-readable astronomical event distribution. In particular software telescopes that do not require physical repointing can respond to alerts and be on source within minutes, and sometimes seconds \citep{Hancock2019}. Rapid radio observations of sGRBs observed by \textit{Fermi-GBM} and \textit{Swift-BAT} are possible for a few reasons: a high GRB event rate owing to a large field of view (1.4 steradians and 2 $\pi$ steradians respectively; \citealt{Meegan2009,Gehrels2004}); a large horizon of detection resulting in large dispersion delays (although sGRB tend to have redshift z < 2 \citealt{Fong2013}); rapid notification of detections through the GCN \citep{1998Barthelmy} and VOEvent systems; and the precise localization of sources to within 1-4 arcmin for \textit{Swift-BAT} and 1-10 deg for \textit{Fermi-GBM}. Furthermore, upgrades to the \textit{Swift-BAT} pipeline are expected to increase the number of localized nearby sGRBs by a factor 3-4 in the near future (\citealt{DeLaunay2021}, see also \citealt{Tohuvavohu2020}). Thus far, rapid response observations of NS-NS mergers have been triggered in response to sGRBs as reported by \textit{Swift-BAT} by the Low Frequency Array (LOFAR; \citealt{Rowlinson2021}), the Murchison Widefield Array (MWA; \citealt{Anderson2021,Tian2022}), Arcminute Microkelvin Imager (AMI; \citealt{Anderson2018}, the Australian Compact Telescope Array (ATCA; \citealt{Anderson2021a}), and a 12m dish at the Parkes radio observatory \citep{Bannister2012}.
\par
The small opening angles of collimated GRB jets mean that triggered radio observations will probe NS merger systems with small viewing angles with respect to relativistic jets, which we assume to be perpendicular to the orbital plane. The opening angles of long GRBs are often determined through jet breaks in the afterglow emission \citep{Sari1999} and range between approximately $\theta_{\rm core} \approx 3-10 \deg$ \citep{Berger2014}. Afterglow observations of sGRBs are sparse, but jet breaks are thought to have been observed in a few sources, corresponding to estimated opening angles of $\theta_{\rm core} \approx 4-8 \deg$  \citep{Soderberg2006,Fong2012}. \cite{Aksulu2022} perform multi-wavelength afterglow modelling of 4 sGRBs and 3 out of the 4 sources have opening angles $\theta_{\rm core} \lesssim 6 \deg$, and one source is found to have a much larger opening angle of $\theta_{\rm core} \approx 34 \deg$. If the orbital plane and primary magnetic axis are perpendicular ($\alpha_{\rm B, orb} = 90 \deg$) as discussed in Sect. \ref{sect:viewingangle}, it is likely that the set of mergers from which prompt emission is observable does not substantially overlap with the set of mergers from which coherent radio bursts are luminous enough to be observed. Rapid radio observations of sGRBs will probe NS mergers with specific magnetic obliquities which direct coherent radiation along the jet axis; i.e. when $\alpha_{\rm B, orb}$ is 10-30 degrees misaligned with the jet axis (see. Fig. \ref{fig:rotated_alpha}). In any case, the radio emission predicted in this paper is radiated into a much larger solid angle than the prompt GRB emission. 

\subsubsection{Current generation}
LOFAR has performed successful triggered observations of GRB 180706A \citep{Rowlinson2019} \& GRB 181123B \citep{Rowlinson2021}. The former was a long GRB but the latter was a short GRB, and its afterglow has been associated with a galaxy at z=1.8 \citep{Paterson2020} with a chance alignment of 0.44\%. Assuming a DM-redshift relation ({\rm DM = 1200 z} pc cm$^{-3}$; \citealt{Ioka2003}, and the NE2001 Galaxy model; \citealt{CordesLazio2002}), it is very likely that the dispersion delay from the source to the telescope was large enough ($\tau_{\rm delay} = \frac{DM}{241 \nu_{\rm GHz}^2} > 400$ seconds) such that LOFAR probed pre-merger radio emission. The attainable FRB fluence limits depend sensitively on the dispersion measure (i.e. Fig. 3 of \citealt{Rowlinson2021}), primarily due to the extent to which the dispersed burst fills each snapshot image. Assuming the galaxy association suggested by \cite{Paterson2020}, fluence limits of $2 \times 10^{4}$ Jy ms can be placed for millisecond FRB emission and can be used to constrain our model. Assuming standard cosmological parameters
($H_0 = 70$, $\Lambda_{\rm M} = 0.286$, flat universe; \citealt{Wright2006}), a redshift of $z=1.8$ corresponds to a luminosity distance of $\sim14$ Gpc. This means that the LOFAR observations of GRB 181123B can constrain the pre-merger radio emission in the present model to a primary NS magnetic field strength of $B_s < 10^{16}$ G, assuming $\eta = 10^{-2}$ and optimal magnetic obliquity.
\par
The Murchison Widefield Array (MWA) successfully triggered on sGRB 180805A \citep{Anderson2021}, and was on source 84 seconds after the burst. For most typical sGRB redshifts \& DMs, coherent emission during the inspiral would have been probed. The resulting fluence limits ranged from $570$ Jy ms to $1750$ Jy ms depending on the assumed dispersion measure, but a reliable constraint for our model cannot be placed without a distance measurement. In \cite{Tian2022}, the authors present a catalog of rapid radio limits with MWA for a total of 9 sGRBs. \cite{Tian2022} make use of image-plane de-dispersion techniques to report triggered observations with deep fluence limits of $80-12000$ Jy ms, with most limits clustered around $1000$ Jy ms. GRB 190627A was the only event in this sample with an associated redshift ($z = 1.942$; \citealt{2019GCN.24916....1J}), corresponding to an approximate luminosity distance of $15$ Gpc assuming the same cosmological parameters as before. The authors were able use their derived fluence limit of $\approx 80$ Jy ms to constrain efficiency parameters for various models of prompt radio emission during GRB 190627A to reasonable values for the first time. For optimally aligned pulsar-like emission in the model presented here, the fluence limit presented by \cite{Tian2022} constrains the primary NS magnetic field to $B_{\rm s} \lesssim 10^{15}$G, assuming an efficiency $\eta \approx 10^{-2}$. 
\par
Making use of direction-dependent calibration and source subtraction techniques, \cite{Rowlinson2019} were able to place deep fluence limits, corresponding to $3 \times 10^3$ Jy ms for a typical sGRB of redshift $z=1$\footnote{In the near future, image-plane de-dispersion will be implemented in the LOFAR rapid response pipeline, significantly improving sensitivity.}. Given this, and the results of \cite{Tian2022}, we find that triggered observations by LOFAR and MWA will detect pre-merger coherent emission to a distance of approximately $3-5 \; \eta_{-2}^{1/2} \, B_{s,12} \,$ Mpc, using detection limits in Table \ref{tab:telescopes}. Although this distance may be an under-estimate, as the low DM expected for mergers at this short distance will aid snapshot sensitivity, such a low DM would also mean that LOFAR will not be on source fast enough to observe pre-merger emission. However, MWA's rapid triggering could probe dispersed bursts at $\approx 300$ Mpc, and would be sensitive to pulsar-like emission from NS mergers at this distance if $B_{\rm s} \gtrsim 6 \times 10^{13}$ G. Such a close sGRB would be rare, but \textit{Swift-BAT} pipeline upgrades discussed above may provide triggering opportunities in the near future for off-axis sGRBs.
\par
Other instruments that have performed rapid observations of GRBs, namely Parkes, AMI, \& ATCA, \citep{Bannister2012,Anderson2018,Anderson2021a} nominally begin observations with delay times incompatible with pre-merger observation, instead probing early radio afterglow emission. 

\subsubsection{Next generation}
The upgraded LOFAR 2.0 will allow simultaneous imaging and beam-formed triggered observations, which will allow more sensitive high-time resolution burst searches for well-localized GRBs \citep{Gourdji2022}. A tied-array beam (TAB) set-up using the LOFAR core stations could be utilized to perform time-domain search for dispersed radio bursts across the most likely localization region of \textit{Swift} GRBs. \cite{Pleunis2021} used such a set-up to search for FRBs, achieving a fluence limit of 26 Jy ms for bursts with an assumed 50ms duration. Importantly, the $\sim 3$ arcmin full-width half maximum (FWHM) of the TAB is approximately the same as the \textit{Swift-BAT} localization region and therefore could be used for prompt GRB searches. The large scattering timescale of FRBs at LOFAR frequencies will reduce the signal-to-noise of bursts, but 100 Jy ms is a reasonable fluence target for the coherent pre-merger bursts which may be observable for many milliseconds, as the fluence limit scales as  $\sqrt{\frac{t_{\rm scatt}}{t_{\rm burst}}}$. This would allow LOFAR 2.0 to probe NS-NS merger emission to $15 \; \eta_{-2}^{1/2} \, B_{s,12} \,$ Mpc, notably probing mergers with $B_{\rm s} = 10^{14}$G to Gpc distances.
\par
Comparatively, the fiducial horizon distance for SKA-mid at full sensitivity of 1 mJy ms is $5000 \; \eta_{-2}^{1/2} \, B_{s,12} \,$Mpc, corresponding to a redshift $z \approx 0.8$. The large dispersion measure expected from these sources, coupled with the precise localization particularly by \textit{Swift-BAT} mean that triggered observations may allow deep radio observations of sGRBs. The dispersion delay to $z\approx 0.8$ to 770 MHz and 110 MHz (i.e. lowest nominal observing frequencies of SKA-mid and SKA-low) is 7 seconds and 330 seconds respectively, assuming a DM-redshift relation \citep{Ioka2003}. SKA-mid is expected to respond to alerts within seconds, but assuming a similar slew speed to its precursor MeerKAT ($\approx 2 \deg \, {\rm s^{-1}}$) repointing could take 0.1-10 minutes and thus is unlikely to be able to detect pre-merger emission via triggered observations. SKA-low, assuming a similar triggering performance to MWA, should be on source within 20 seconds and thus should be sensitive to radio emission from sGRBs. To estimate the rate for SKA, we assume that SKA-low telescope triggers rapid observation on half of all \textit{Swift-BAT} and \textit{Fermi-GBM} detections of likely sGRBs: 10 and 45 per year respectively \citep{Burns2016}. \cite{Gompertz2020} find that in a sample of 39 \textit{Swift-BAT} observed (likely) sGRBs with known redshifts, three quarters have redshifts of $z <0.8$. SKA-low's large FoV means the entire \textit{Swift-BAT} localization region and most (if not all) of \textit{Fermi-GBM} can be probed. Given this, and extrapolating to \textit{Fermi-GBM} sGRBs, we could expect SKA to be sensitive to fiducial pulsar-like emission from $\approx 20-30$ sGRB events per year. 

\subsubsection{Other considerations}
We note that \cite{Gourdji2020} search for sGRB counterparts to two well-localized, non-repeating FRBs: FRB 180924 and FRB 190523. Non-detections of sub-threshold \textit{Fermi} counterparts in both cases constrain FRB models in which coherent radiation is emitted along the same axis as a GRB. We note that in model we present here, sGRBs may not be aligned along the same axis as the emitted coherent radio emission, thus non-detection of gamma rays may not preclude a NS-merger origin. Moreover, the authors disfavour FRB models where coherent radio emission is powered by the inspiral, as the predicted flux as a function of time is not compatible with the observed temporal morphology of the FRBs. This severely constrains the present models ability to reproduce FRB lightcurves (see discussion at the end of Sect. \ref{sect:frb_survey}).


\subsection{Gravitational wave events}
\label{sect:gw}
NS merger events are also observable using gravitational wave (GW) detections of compact object mergers, where many of the rapid, multi-wavelength observing techniques discussed in Sect. \ref{sect:ShortGRBs} can be utilized. The fourth observing run (O4) of the GW detector network is expected to start in March 2023\footnote{\url{https://observing.docs.ligo.org/plan/}}. 

\subsubsection{Current generation}
The three detectors of the third observing run (O3) will be operational during O4 at improved sensitivities: the two Advanced Laser Interferometer Gravitational-Wave Observatory (aLIGO; \citealt{Abramovici1992,Abbott2009}) detectors near their design sensitivities with a BNS range $D_\mathrm{BNS} \approx$ 160 - 190 Mpc, and the advanced Virgo (AdV; \citealt{Caron1997}) detector with $D_\mathrm{BNS} \approx $ 80 - 115 Mpc~\citep{2020LRR....23....3A}. A fourth detector the Kamioka Gravitational Wave Detector (KAGRA; \citealt{2019NatAs...3...35K,2022arXiv220307011K}) will be starting operations as well but the anticipated sensitivity is limited with $D_\mathrm{BNS} \approx$ 1 - 10 Mpc. In~\citet{2020LRR....23....3A}, it is estimated that $10^{+52}_{-10}$ BNS detections will occur in O4 with a median 90\% localisation area of 33 $\mathrm{deg}^2$. These predictions assume $D_\mathrm{BNS} =$ 80 Mpc for KAGRA, however this is likely well above the horizon that will be obtained in O4 for that detector. We therefore use a more likely 100 $\mathrm{deg}^2$ for O4. 
\par
\begin{figure}
  \centering
{\includegraphics[width=.5\textwidth]{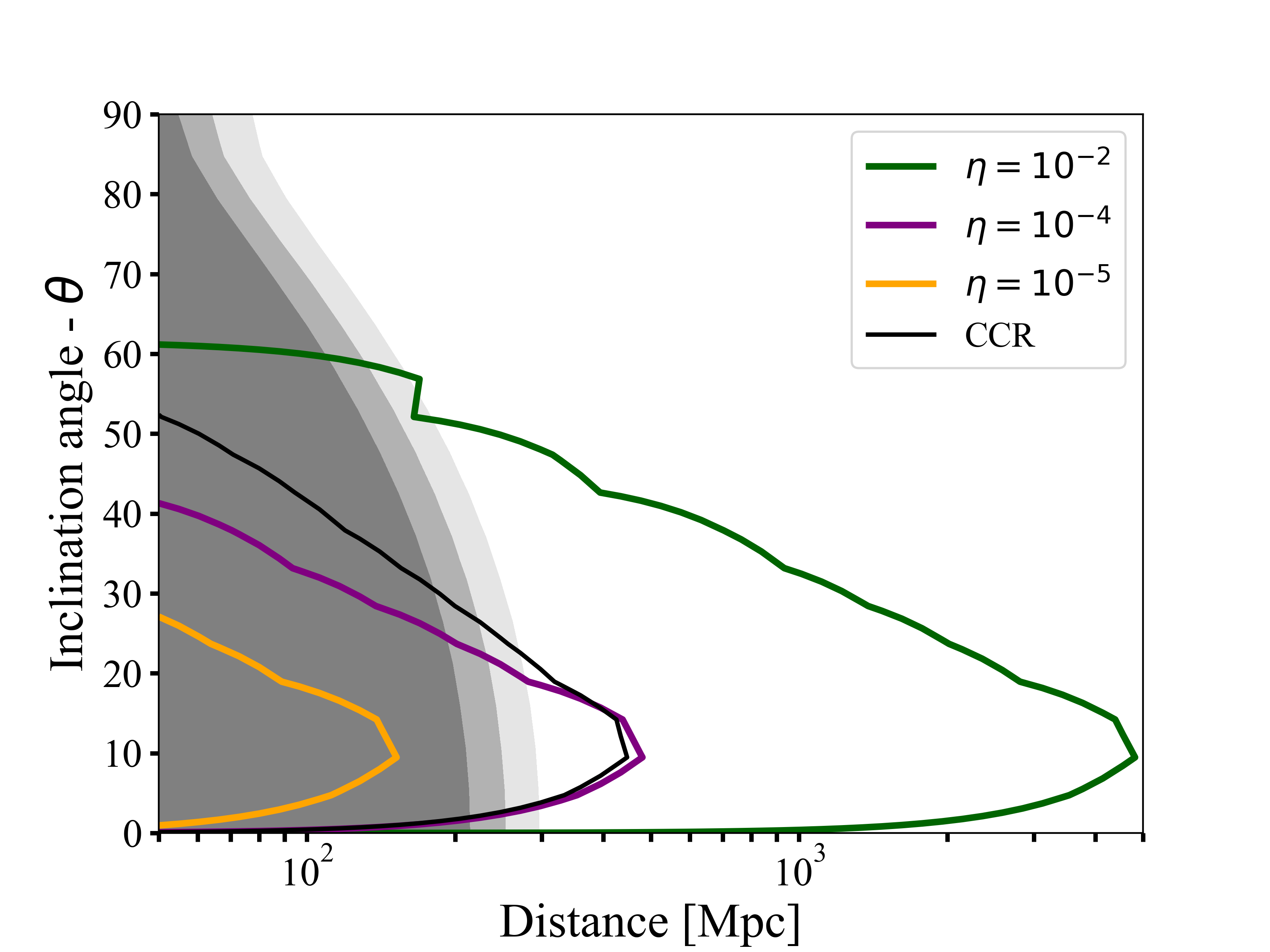}}
\caption{We show the inclination angle dependent observing horizon for SKA-mid for NS merger radio bursts powered by pulsar-like emission (for three efficiency values $\eta$, Sect. \ref{sect:particle_acceleration}) and coherent curvature radiation (Sect. \ref{app:cohcurv}). The apparently increase in horizon for $\eta = 10^{-2}$ at $50 \deg$ is a artifact related to the way in which inclination angle dependent horizons are calculated. In the background grey are detection limits (assuming detections for SNR =8) for the fourth observing run of 2nd generation gravitational wave instruments: LIGO (Hanford + Livingston) (darkest shade), LIGO-VIRGO (medium shade) and LIGO-VIRGO-KAGRA (lightest).}
\label{fig:GW}
\end{figure}

In this Section, we estimate the distance to which BNS GW signals can be detected, averaged over sky position but, not over the inclination angle, due to the strong inclination angle dependence of the pre-merger emission. We compute the inclination angle dependent GW signal-to-noise ratio (SNR), averaged over thirty random sky positions, using various GW detector network setups (see Figs. \ref{fig:GW} and \ref{fig:GW_3rdgen}). We calculate the SNR with the \texttt{pycbc} Python package using the prescription of \cite{Cutler1994}, using the standards PSDs in \texttt{pycbc} for the current generation detectors (aLIGO, AdV, KAGRA) at design sensitivity. For the PSDs of future detectors (LIGO A+, LIGO Voyager, AdV+, KAGRA+, ET, CE), we follow~\citet{2022arXiv220211048B} and take them from \texttt{ce\_curves.zip}\footnote{\url{https://dcc.cosmicexplorer.org/public/0163/T2000007/005/ce_curves.zip}}.  For our template waveform $h$, we use the de facto standard waveform model for BNS mergers "IMRPhenomD\_NRTidalv2"~\citep{2019PhRvD.100d4003D}. We vary the component masses of the BNS merger but set the NS spins to zero.
\par

\subsubsection{Current generation}
To make predictions for the fourth gravitational-wave observing run, we assume $10^{+52}_{-10}$ BNS gravitational wave detections all with localization area of 100 $\deg^2$ as mentioned. The dispersion delay to the maximum BNS range of $190$ Mpc ($z\approx 0.043$) is approximately 10 seconds at 144 MHz and 30 seconds at 80 MHz. This means that although LOFAR would not be on source quick enough for pre-merger detection, we note that if LOFAR 2.0 can be on source within 10-15 seconds, precursor emission from NS mergers in O5 could be probed. However, MWA's quicker triggering time will probe the most distant events in O4, at least at the lowest frequencies. Assuming MWA is well-positioned to respond to half of all alerts (neglecting sensitivity drop off towards a non-optimal declination), this corresponds to 5 BNS events, where the large FoV means the entire localization region can be covered. Assuming a distance of 190 Mpc and $\eta = 10^{-2}$, MWA will be sensitive to emission for a surface magnetic field of $B_{\rm s} \approx 4 \times 10^{13}$G (assuming optimal viewing angle), therefore some limits attainable during the O4 run will be constraining for this model. 

\subsubsection{Next generation}
After O4, all current detectors are planned to undergo significant upgrades towards the fifth observing run (O5). For the aLIGO detectors, this entails the A+ upgrade targeting a maximum $D_\mathrm{BNS} \approx$ 325 Mpc, similarly AdV will be upgraded to the AdV+ configuration with a maximum $D_\mathrm{BNS} \approx$ 260 Mpc~\citep{2020LRR....23....3A}. KAGRA \citep{Kuroda2010,Kagra2019} is expected to reach $D_\mathrm{BNS} \approx$ 130 Mpc and could potentially gain more sensitivity with the KAGRA+ upgrade. Furthermore, a fifth detector, LIGO-India, is planned to join the global GW network starting operations around $\sim$2025\citep{2020LRR....23....3A}. LIGO-India is expected to have identical specifications to the other two LIGO detectors and thus very similar sensitivity. \citet{2022arXiv220211048B} make predictions for the scientific capabilities of such a five detector network at optimal sensitivity. They find that this network will detect around 200 BNS mergers per year with approximately six of those detections having a 90\% localisation area of $\leq \ 1 \ \mathrm{deg}^2$, and the median 90\% localization region is 9–12 $\deg^2$ \citep{Corsi2019}.
\par
As discussed in Section \ref{sect:ShortGRBs}, LOFAR 2.0 will allow for faster triggering, and simultaneous imaging and beam-formed triggers, allowing it to probe a wider variety of the predicted radio emission associated with NS mergers. In terms of detecting millisecond radio bursts, beam-formed observations would allow improve sensitivity, as well as grant greater flexibility de-dispersion techniques. For gravitational wave events where the localization is general much poorer than GRBs, the one possible observing set-up would be to use the LOFAR Tied-Array All-Sky Survey (LOTAAS) survey \citep{Sanidas2019} beam structure in addition to interferometric imaging beams. In this case, coherent tied-array beams cover 12 $\deg^2$ allowing sensitive high time resolution searches, with additional incoherent beams providing lower sensitivity coverage up to 68 $\deg^2$. The latter of which were used for an FRB search in \cite{terVeen2019} to achieve a fluence limit of approximately 1600 Jy ms. The larger field-of-view of the incoherent beams would allow better coverage of the localization region, where roughly half of the 90\% localization region may be covered by incoherent beams, and a smaller portion by the tied-array beams. If LOFAR 2.0 can trigger within 15 seconds, the dispersion measure expected from the O5 observing horizon distance of $\sim 300$ Mpc, would be compatible with pre-merger detection. In this case, incoherent beams would be able to cover the entire localization region and probe emission from a $B_{\rm s} = 10^{14}$G merger, and the tied-array beam could cover a portion of the $1 \deg$ 90\% region, and be sensitive to lower magnetic field.
\par
First light for SKA is predicted to be in 2027, when we expect a 5-detector network to be running with 200 mergers per year with a horizon of $D_\mathrm{BNS} \approx$ 325 Mpc, and a typical $10 \deg^2$ localization. The dispersion delay to 150 MHz at 325 Mpc is approximately 15 seconds, and the delay across the entire 50-350MHz band is $\approx 130$ seconds. These values may be higher depending on the local source and Milky Way contributions. SKA1-low will be able to save $\sim 30$ seconds of raw voltage data per station. Assuming repointing can occur within 10 seconds upon receipt of a gravitational wave alert, pre-merger emission at low frequencies should be observed. Assuming $\eta = 10^{-2}$, SKA-low will be able to test the model presented here for NS mergers with $B_{\rm s} = 10^{11}$ G for $D = 300$ Mpc, and therefore will be able to verify the model. 
\par
\begin{figure}
  \centering
{\includegraphics[width=.5\textwidth]{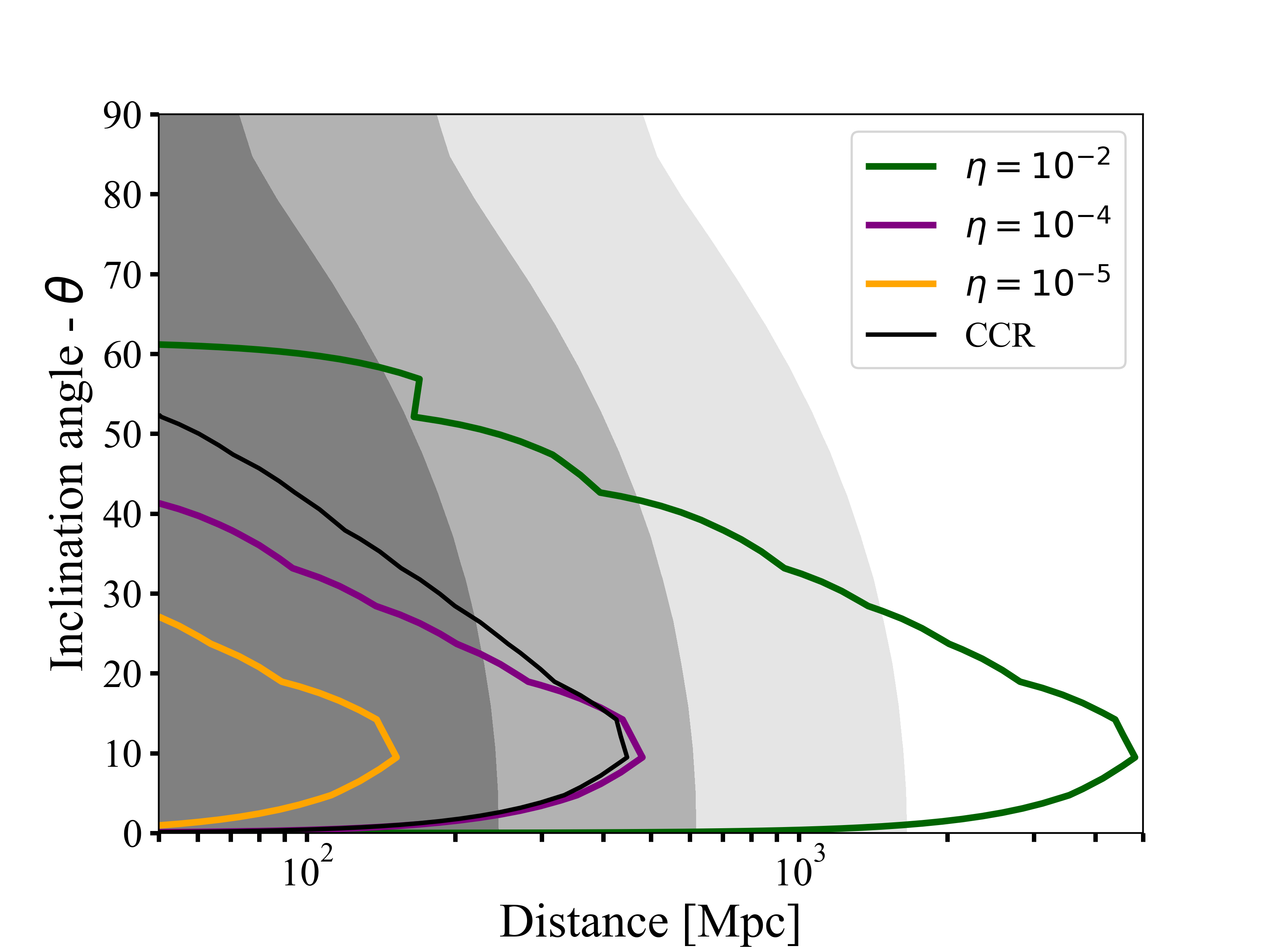}}
\caption{We show the inclination angle dependent observing horizon for SKA for NS merger radio bursts powered by pulsar-like emission (for three efficiency values $\eta$, Sect. \ref{sect:particle_acceleration}) and coherent curvature radiation (black line, Sect. \ref{app:cohcurv}). In the background grey are detection limits (assuming detections for SNR =8) for 2nd and 3rd generation gravitational wave instruments for single detector setups of LIGO+ (darkest shade), LIGO Voyager (medium shade) and the Einstein Telescope (lightest shade).}
\label{fig:GW_3rdgen}
\end{figure}

The capabilities of future post O5 GW detectors are promising, though speculative. LIGO Voyager, expected around 2030, will first upgrade the three LIGO detectors to a range of $D_\mathrm{BNS} \approx$ 1 Gpc~\citep{2020CQGra..37p5003A}. The Einstein Telescope (ET; \citealt{2010CQGra..27s4002P}) and Cosmic Explorer (CE; \citealt{2017CQGra..34d4001A}) will instead be completely new, next generation GW detectors and bring tremendous, ten-fold increases in sensitivity. The definitive range of LIGO Voyager, ET and CE will strongly depend on the final design configurations, which are subject to uncertainty in the technological improvements reached in the next decade. Still, preliminary sensitivity curves exist for these future detectors and can give insight into their potential~\citep{2022arXiv220211048B}. A network of three LIGO Voyager detectors, AdV+ and KAGRA+ will detect on the order of 2000 BNS mergers per year with one percent of the detections having a 90\% localisation area of $\leq 1 \mathrm{deg}^2$. Astonishingly, if three next generation detectors become operational, i.e. ET and two CEs, on the order of 300,000 BNS mergers will be detected per year with again, around one percent of the detections having a 90\% localisation area of $\leq 1 \mathrm{deg}^2$ corresponding to roughly 2000 precisely localised BNS mergers per year. The inclination angle dependent horizons for these next-generation GW detectors are shown in Fig. \ref{fig:GW_3rdgen}. 

\subsubsection{Other considerations}
\begin{figure}
  \centering
{\includegraphics[width=.5\textwidth]{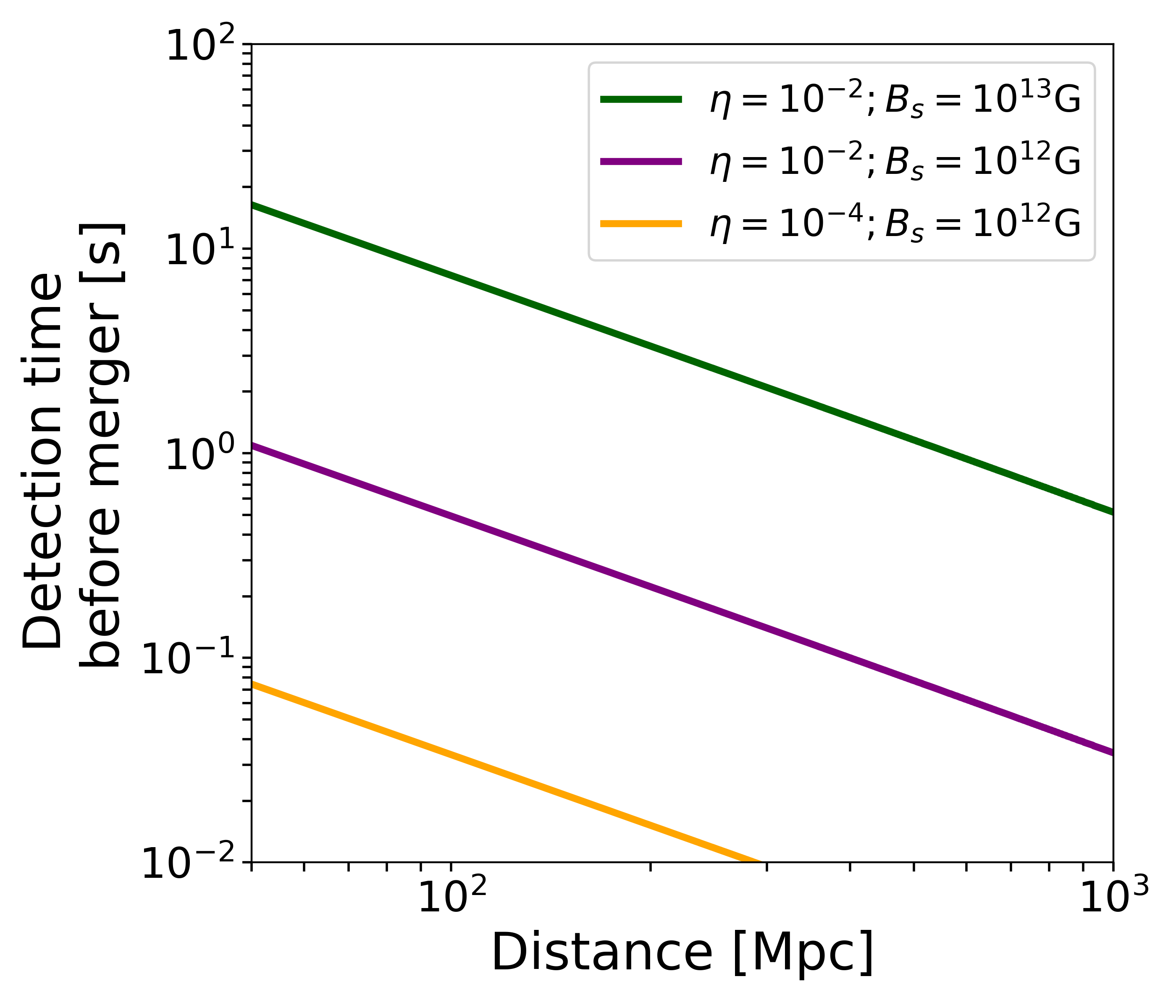}}
\caption{Expected detection time before the merger time in seconds, assuming an SKA fluence sensitivity of 1 mJy ms and averaging over phase dependence.}
\label{fig:pre-merger}
\end{figure}

Pre-merger alerts will be issued by gravitational wave observatories in the coming observing runs \citep{Chu2016,Magee2022}, and radio emission could be observable seconds before the merger event (Fig. \ref{fig:pre-merger}). Assuming 100\% duty cycle and design sensitivities, \cite{Magee2022} find that during a four detector network O4/O5 run there will be 3.5 and 1.5 events per year that can be reported 1 second and 10 seconds before the merger respectively. For a five detector network during O5, these values rise to 30 and 15 events per year respectively. Such pre-merger alerts will be invaluable for radio observatories to begin slew and repointing, even if only approximate localizations can be provided in early reporting \citep{Chu2016}. Finally, we note that mergers that occur at a smaller distances (and therefore with shorter dispersion delays) will be detected with more GW detectors and therefore will have better localizations. 

\subsection{Gamma-ray burst radio afterglows}
\label{sect:afterglow}
The relativistic jetted outflows powering sGRBs plough into the circumburst medium, resulting in shocks and particle acceleration. This particle acceleration leads to broadband synchrotron emission visible first at higher energies with X-ray telescopes and then at lower frequencies at later times \citep{ReesMeszaros1992}. Such afterglow emission usually detected during follow-up observations triggered by gamma-ray burst detectors, or occasionally as so-called `orphan' afterglows without prompt gamma ray emission \citep{Levinson2002,Ghirlanda2015,Law2018} or by follow-up observations of kilonovae \citep{Ho2020ApJ...905...98H,Andreoni2021}. It is expected that future radio facilities such as SKA-mid will detect afterglows both serendipitously during transient surveys, and in gravitational wave event follow-up \citep{Dobie2021} to Gpc distances (Fig. \ref{fig:afterglow_gaussian_MeerKat}). Prospects for detection of radio afterglows from compact object merger events is covered in much more detail, including specific follow-up observing strategies, in the excellent work by \cite{Dobie2021}, to which we refer the interested reader.
\par
Late-time radio follow-up of non-repeating FRBs to look for persistent and long-term variable counterparts has been performed for a few well-localized sources. \cite{Bhandari2018} report on multi-wavelength and multi-messenger follow-up observations of four apparently non-repeating FRBs as discovered by the Parkes telescope. No radio afterglow emission was observed, although the large inferred luminosity distances (ranging from 4.8 Gpc to 17.2 Gpc) mean that afterglow emission would be difficult to detect. Similarly, \cite{Bhandari+2020} report on four FRBs localized by ASKAP and find no significant radio transient sources, and observed emission is attributed to the host galaxy. Lastly, \cite{Bhandari2020b} report on precursor and follow-up of a well-localized FRBs, again with a non-detection that cannot meaningfully constrain radio afterglow models.
\par
\par
To investigate whether the NS-NS merger origin of coherent radio bursts could be confirmed by detection of late-time afterglow emission, we used the python package \textsc{afterglow.py} \citep{Ryan2020}. In Table \ref{tab:afterglow_parameters}, we list the assumed fiducial parameters for the radio afterglows as in \cite{Ryan2020}, where reasonable values are chosen given multi-wavelength fits to GRB afterglows (e.g. \citealt{Aksulu2022}). In particular, the circumburst density $n_0$ is quite uncertain and has a large effect on the detectability horizon of radio afterglows and therefore we take a broad range of values of between $10^{-5}$ and $10^{-1}$ to give an approximate idea of the uncertainties involved. We consider both structured (gaussian) jets and tophat jets. 
\begin{center}
\begin{tabular}{|c|c|c|} 
\hline
Parameter & Fiducial value & Meaning [units] \\
\hline
$E_0$ & $10^{53}$ & Isotropic-equivalent energy [erg] \\ 
$\theta_{\rm core}$ & 0.05 & Half-opening angle [rad] \\ 
$\theta_{\rm wing}$ & 0.2 & Wing angle for gaussian jet [rad] \\ 
$n_0$ & $10^{-5}$-$10^{-1}$ & Circumburst density [${\rm cm}^{-3}$] \\ 
$p$ & 2.2 & Electron energy powerlaw index \\ 
$\epsilon_e$ & 0.1 & Fraction of energy in electrons \\ 
$\epsilon_B$ & 0.01 & Fraction of energy in magnetic field \\ 
$\xi_N$ & 1.0 & Fraction of electrons accelerated\\ 
\hline
\end{tabular}
\label{tab:afterglow_parameters}
\end{center}

\subsubsection{Current generation}
\begin{figure}
  \centering
{\includegraphics[width=.5\textwidth]{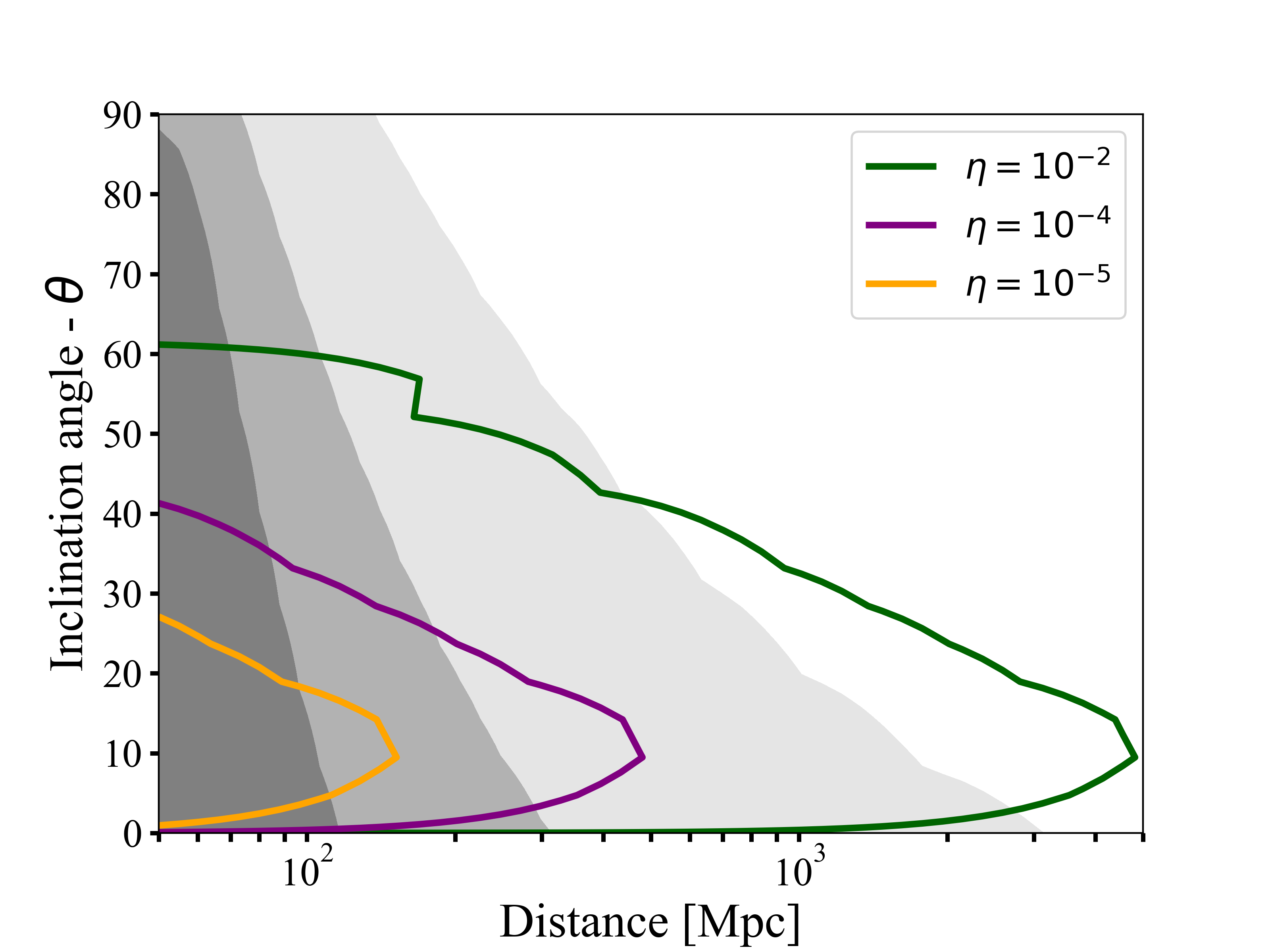}}
\caption{Background colours show MeerKAT detection horizons for peak flux of gaussian jets at $\nu = 1.43 \; {\rm Hz}$ assuming detection threshold of $700 \; {\rm \mu Jy}$. Shades of gray correspond to varying circumburst densities (Darkest $n=10^{-5}$; medium $n=10^{-3}$; lightest $n=10^{-1}$).}
\label{fig:afterglow_gaussian_MeerKat}
\end{figure}

We find that for circumburst densities of $10^{-5}$ and $10^{-1}$ respectively, MeerKAT should be sensitive to many afterglows from NS-mergers from distances of 200 Mpc to 2 Gpc for gaussian jets (Fig. \ref{fig:afterglow_gaussian_MeerKat}) and tophat jets (Fig. \ref{fig:afterglow_tophat_MeerKat}), assuming optimal magnetic obliquity for pre-merger radio bursts. For off-axis jets, the horizon is reduced drastically to 10-200 Mpc, dependent on the assumed afterglow parameters, in particular the circumburst medium density. One of the primary limitations with the current generation of instrumentation is the low number of localized one-off FRBs at distances close enough such that an afterglow can be definitively constrained.

\subsubsection{Next generation}
\begin{figure}
  \centering
{\includegraphics[width=.5\textwidth]{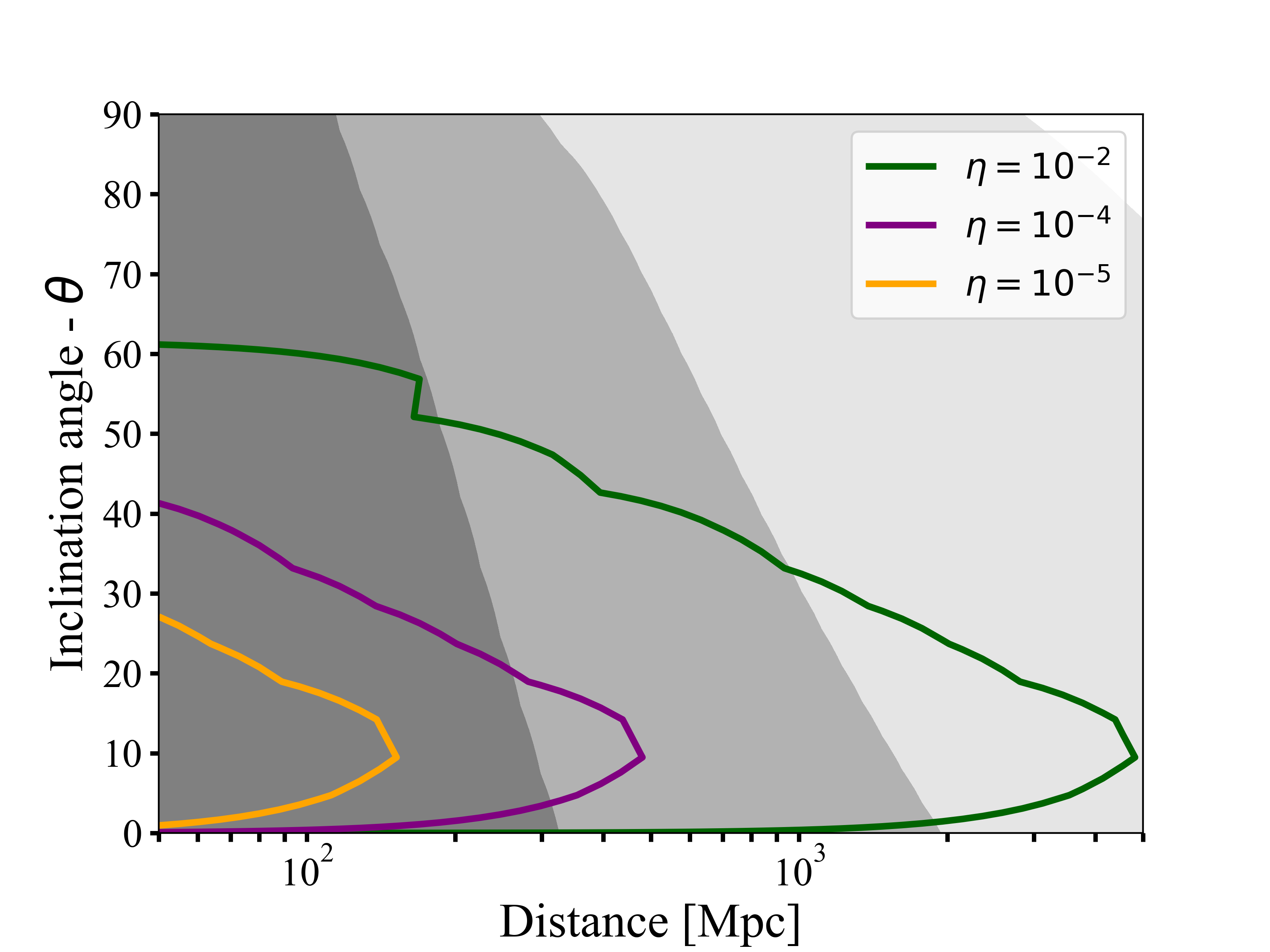}}
\caption{Background colours show SKA detection horizons for peak flux of gaussian jets at $\nu = 1.43 \; {\rm Hz}$ assuming detection threshold of $2 \; {\rm \mu Jy}$. Shades of gray correspond to varying circumburst densities (Darkest $n=10^{-5}$; medium $n=10^{-3}$; lightest $n=10^{-1}$). Coloured contours show viewing angle-dependent SKA detection horizon of pulsar-like coherent pre-merger emission assuming $\eta = 10^{-2}$, $B_{\rm s} = 10^{12}$G.}
\label{fig:afterglow_gaussian_SKA}
\end{figure}

If $n_0 > 10^{-3}$, SKA-mid should be sensitive to almost all afterglows from NS-mergers that can power coherent bursts with fiducial efficiency and surface magnetic field values ($\eta = 10^{-2}$; $B_{\rm s} = 10^{12}$G), as shown in Figs. \ref{fig:afterglow_gaussian_SKA} \& \ref{fig:afterglow_tophat_SKA} for gaussian structured and tophat jets respectively. As we have shown in Sect. \ref{sect:frb_survey}, it is unlikely that all non-repeating FRBs are powered by NS mergers. Nevertheless, follow-up radio campaigns to trigger sensitive observations on localized FRBs (hundreds per year with the CHIME/FRB outrigger and CHORD projects; \citealt{Leung2021,Vanderlinde2019}) could certainly confirm the NS-merger origin of Gpc coherent radio bursts by detecting late-time afterglow emission. We note that in our calculation, detection criteria depends only on the peak flux of the afterglow, and therefore well-timed observations with multiple epochs or frequencies would be required in practice to confirm an afterglow. Finally, if early-time triggered observations of CHIME localized FRBs are also performed, SKA may be sensitive to radio afterglow from maser-shock models of FRBs \citep{2022MNRAS.tmp.2746C}, providing additional discovery space.

\subsection{Kilonovae}
\label{sect:kilonovae}
An alternative method of identifying coherent radio bursts of NS merger origin (if the burst is sufficiently localized), is to perform optical follow-up in search of a kilonova counterpart. Thus far, optical follow-up of fast radio bursts has largely focused on well-localized repeating sources \citep{Andreoni2019,2020ApJ...896L...2A,2021ApJ...907L...3K,2021ATel14666....1A}, and not the one-off bursts discussed in this work. For non-repeating sources, \cite{Petroff2015MNRAS} performed multi-wavelength follow-up, including optical facility DECAM, of FRB 140514 and were able to rule out supernovae and long GRB progenitor, but were not sensitive enough to rule out kilnova models. \cite{Marnoch2020} \& \cite{Nunez2021} performed follow-up observations with the Very Large Telescope (VLT) and Las Cumbres Observatory Global Telescope (LCOGT) respectively on well-localized, one-off FRBs as discovered by ASKAP \citep{Bhandari2018}. \cite{Marnoch2020} find that it is unlikely that the bursts that were observed are coincident with Type Ia or Type IIn supernova explosions, and similarly \cite{Nunez2021} rule out bright supernovae. Neither work is able to rule out coincident kilonovae, due to a lack of sensitivity. \cite{Tominaga2018} report on optical follow-up with the Subaru/Hyper Suprime-Cam on week-long timescales of FRB 151230 and again find no evidence of Type Ia supernovae, but lack the sensitivity to probe kilonovae emission. 
\par
\begin{figure}
  \centering
{\includegraphics[width=.5\textwidth]{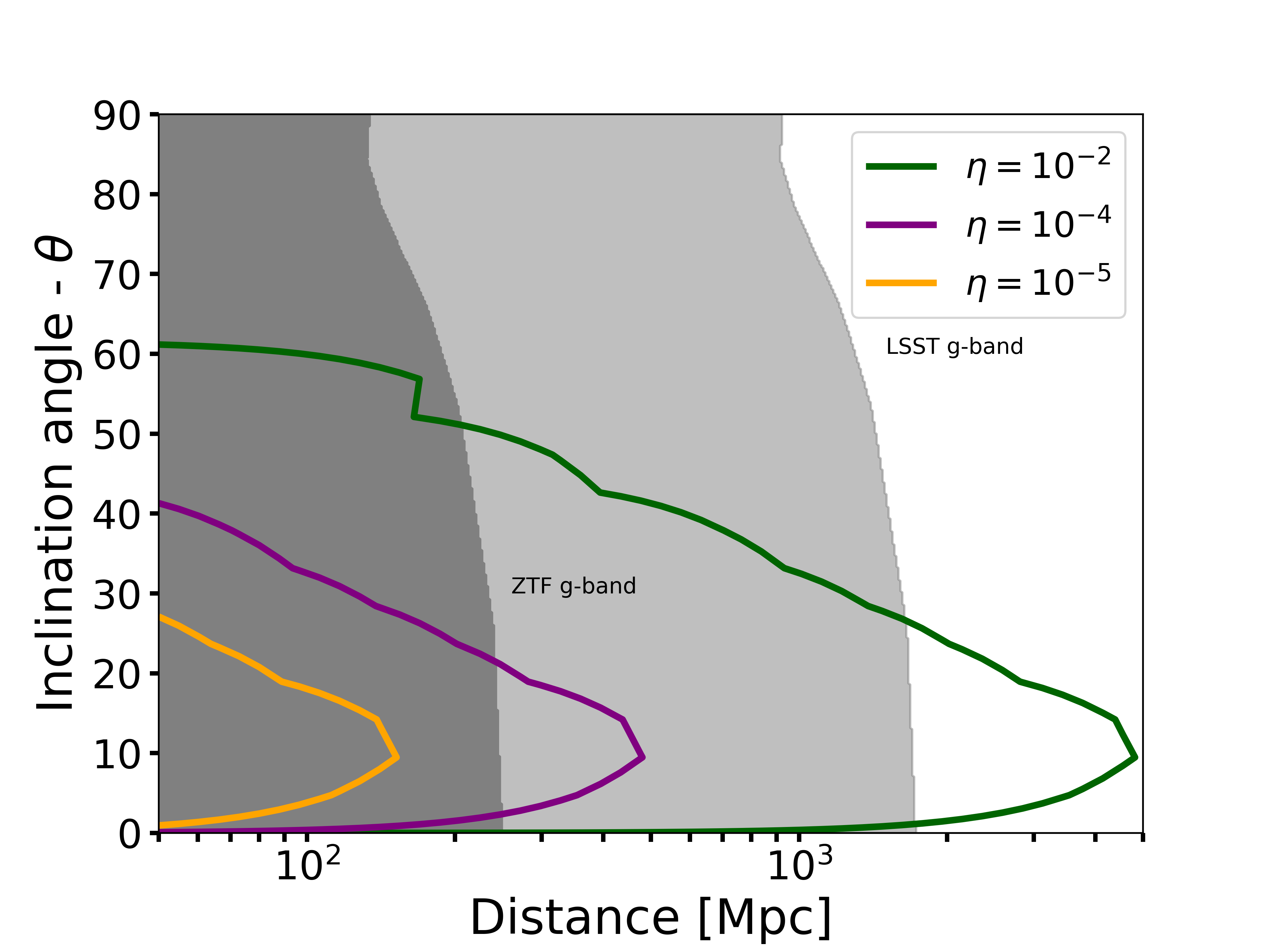}}
\caption[1]{Background colours show ZTF and LSST g-band (darker and lighter shades respectively) detection horizons for peak flux of GW170817-like kilonovae models using models from \cite{Bulla2019}. Coloured contours show viewing angle-dependent SKA detection horizon of pulsar-like coherent pre-merger emission assuming $\eta = 10^{-2}$, $B_{\rm s} = 10^{12}$G.}
\label{fig:kilonovae_g}
\end{figure}

In order to examine this method of confirming a merger origin of one-off FRBs, particularly in expectation of many more localizations through the CHIME/FRB outrigger project \citep{Leung2021,Vanderlinde2019} we make use of the kilonova model grid presented in \cite{Bulla2019}. The grid was generated via radiative transfer simulations sensitive to the viewing angle parameter, allowing viewing angle dependent lightcurve predictions. We select the kilonova model that best fits the GW170817 observations, with dynamic ejecta mass $M_{\rm dyn} = 0.005$ M$_\odot$, disk wind ejecta mass $M_{\rm wind} = 0.05$ M$_\odot$, and half opening angle of the lanthanide-rich dynamical ejecta component $\theta_{\rm core, kn} = 30$ deg. In Figs. \ref{fig:kilonovae_g}, \ref{fig:kilonovae_r} \& \ref{fig:kilonovae_i} show viewing-angle dependent detector horizons for the GW170817-like kilonova model as observed by Zwicky Transient Facility (ZTF; \citealt{Bellm2019}) \& Vera C. Rubin Observatory's Legacy Survey of Space and Time (LSST; \citealt{Ivezic2019}) in g, r and i bands respectively. ZTF can detect kilnovae from NS-NS mergers to $100-200$ Mpc, and LSST to $\approx 1$ Gpc, with a modest drop off for highly inclined viewing angles. As in the previous section, we suggest that sensitive optical follow-up of close-by (low DM) one-off FRBs, or those with quasi-periodic structure (timescale of the orbital period of short separation binaries $\sim$ 1ms) may yield a kilonova detection that could confirm NS-merger origin. For well-localized one-off FRBs, as expected in 2023 with the CHIME outrigger project, pointed target-of-opportunity optical observations may probe kilonova at cosmological distances. In lieu of triggered observations, large FoV optical survey data (e.g. ZTF) could also be searched for kilonova-like emission on the position of reported one-off FRBs. 

\section{Discussion}
\label{sect:Discussion}

\subsection{Binary evolution of highly magnetized NS mergers}
The evolution of a neutron star's magnetic field is complicated, making it difficult to estimate the surface magnetic field strength at the time of merger (see \cite{Igoshev2021} for a recent review). Observable emission through the mechanism described in this work requires one of the two merging neutron stars to have a surface magnetic field $B_{\rm s} \gtrsim 10^{12}$G for detection of precursor emission. Traditional binary evolution channels for NS-NS mergers (e.g. \citealt{Battacharya1991}) suggest that these systems take 100 Myr-Gyr to merge due to gravitational radiation \citep{Tauris2017}. It is expected from evolutionary arguments that one old, recycled $\approx 10^{9}$G millisecond pulsar will merge with a younger, more slowly spinning $10^{12}$G neutron star \citep{Battacharya1991}, as is observed in Galactic double pulsar systems \citep{Burgay2003,Lyne2004}. 
\par
The characteristic lifetime of a pulsar is:
\begin{equation}
    \tau = \frac{P}{\dot{P}} \propto \frac{P^2}{B_{\rm s}^2}
    \label{eq:timescale_pulsar}
\end{equation}
where $P$ is the spin period, $\dot{P}$ is the spin period time derivative, $B_{\rm s}$ is the surface magnetic field. In Eq. \eqref{eq:timescale_pulsar} we have used the fact that the characteristic magnetic field of a neutron star is $B_{\rm s} \propto P^{1/2} \dot{P}^{1/2}$, and the spin down luminosity $L \propto P^{-3} \dot{P} \propto B^2 P^{-4}$. We know of around $25$ magnetar strength neutron stars in the Galaxy, with typical periods and magnetic fields $P = 10$s and $B_{\rm s} = 10^{14}$G, in contrast to approximately 2500 non-recycled radio pulsars with  $P \approx 1$s and $B_{\rm s} = 10^{12}$G. Assuming spin-down occurs on the characteristic timescale of Eq. \eqref{eq:timescale_pulsar}, this implies the birthrate of $10^{12}$G and $10^{14}$G neutron stars may be of the same order of magnitude (see also \citealt{Beniamini2019}). A highly magnetized neutron star's magnetic field are thought to decay primarily through Ohmic decay and Hall drift \citep{Goldreich1992}, where typical timescales are on the order of $100$s Myr. This within a factor of 3 of the merger timescale of the Hulse-Taylor neutron star binary system \citep{Weisberg1981}. We therefore may not necessarily expect the field to substantially decay before the merger, and a portion of NS mergers likely contain a highly magnetized neutron star as required for bright FRB-like emission during the inspiral \ref{sect:frb_survey}. The possibility that high magnetic field neutron stars may retain strong fields for at least 10-100s Myrs is bolstered by recent discoveries of ultra-long period magnetars \citep{Hurley_Walker2022,Caleb2022}, as well as theoretical work on spin period evolution of highly magnetized neutron stars \citep{Beniamini2020,2022arXiv221009323B}.
\par
Moreover, alternative evolutionary channels have been proposed in which one NS may be much younger during merger, and share properties with observed Galactic magnetars. To explain the observed short-period low-mass X-ray binaries, \cite{Kalogera1998} suggested that some initially wide binaries could be brought together by the supernova kick itself, in a formation channel known as direct supernova. Very wide binaries would bypass the common envelope phase during the first supernova, and a small proportion of these systems would undergo a fortuitous kick resulting in a tight orbital separation. \cite{VossTauris2003} noted that if the first supernova kick results in a close X-ray binary, and the second star's demise also results in a NS, the resulting compact object binary could have a relatively small separation and thus could merge relatively quickly. They estimate that up to 5\% of all NS-NS mergers could be produced in this way. Specifically, they predict a small number of NS-NS mergers requiring just $10^{2-4}$ yrs to merge, which may imply a subset of mergers containing young, highly magnetized neutron stars. Furthermore, in \cite{VignaGomez2018}, the authors' fiducial model contains a few NS-NS binaries created with periods of around $10^{-2}$ days, representing systems that would merge in $10^{2-4}$ years. These outliers represent just a few systems in a simulation containing $\sim 10^{3}$ binaries, and the authors note they are the product of fortuitous kicks as discussed in \cite{Kalogera1998}. Another possible evolutionary channel leading to a fast merger time was proposed by \cite{DewiPols2003}. In this scenario, a binary containing a $2.8 - 6.4 M_{\odot}$ helium star and a NS could produce a NS-NS binary with very short periods ($P \sim 0.01$ days), if the NS has sufficient time to spiral in the helium stars' envelope before core collapse. Various models in \cite{DewiPols2003} suggest merger timescales of between $10^4$ and $10^6$ yrs, such that a significant magnetic field may still be present. We note that dynamical channels such as N-body interactions in globular clusters that may significantly contribute to BH-BH merger events, are disfavoured to contribute to the rate of lower mass compact object mergers such as NS-NS and NS-BH events \citep{Ye2020}. 
\par

\subsection{Neutron star - black hole mergers}
The model outlined in this paper required the presence of a magnetized NS merging with an object that acts as a conductor, which we have assumed throughout to be a secondary, less magnetized NS. However, a stellar-mass BH will also act as conductor and result in similar emission. The rates of of NS-BH mergers are generally thought to be lower than NS-NS mergers \citep{Mandel2021}, however typical binary evolution suggests that the NS in a NS-BH merger would be highly magnetized. This is because the more massive star in a binary has a shorter lifetime, will undergo supernovae first and is more likely to result in a BH, such that the younger compact object is the (unrecycled) NS. Therefore such systems may still contribute significantly to the volumetric rate of coherent bursts from compact object mergers, and will also emit multi-wavelength and multi-messenger signals discussed in Section \ref{sect:MM_MW_prospects} that can provide paths for detection \citep{mcwilliams2011,Boersma2022}. Lastly, the gravitational wave detection horizon for NS-BH mergers is larger than for NS-NS mergers by a factor of 2 (for a recent detection see \citealt{Ligo_ns_bh_2021}), which may allow automatic triggering of buffer boards or observations on a much larger number of gravitational wave events than discussed in Section \ref{sect:gw} \footnote{\url{https://dcc.ligo.org/public/0161/P1900218/002/SummaryForObservers.pdf}}. 


\section{Conclusions}
\label{sect:conclusion}
We have presented an model for coherent pre-merger bursts from NS mergers, based on an adapted version of the model presented by \cite{Lyutikov2019} (Section \ref{sect:model}). We primarily consider pulsar-like emission expected as the inspiral revives electromagnetic conditions required for gap particle acceleration. We find that bursts are observable to Mpc - Gpc distances depending on the efficiency, if the mechanism operates maximally ($E_{\rm gap} = E_{\parallel}$) and one NS has a significant magnetic field $B_{\rm s} \approx 10^{12}$G as expected from evolutionary arguments (Section \ref{sect:particle_acceleration}). Radio emission is emitted along magnetic field lines, and the inclination angle between the background dipole magnetic field of the primary NS determines the set of observers for whom the radio burst is visible. (Section \ref{sect:viewingangle}). Coherent precursor bursts can be distinguished from FRBs from other sources by way confirmation of modulation on the orbital period and characteristic flux increase, which may aid in inferring properties of merging neutron stars (Section \ref{sect:temporal}). We have made predictions for detecting these bursts through fast radio burst surveys and triggered observations of short gamma-ray bursts and gravitational wave events (Section \ref{sect:MM_MW_prospects}). We further suggest follow-up of some fast radio bursts in the optical and radio wavelengths to confirm merger origin. Our main observational conclusions are listed below:

\begin{enumerate}
    \item Coherent pre-merger emission is directed along the magnetic field lines of the primary magnetized neutron star, and will be detected by multi-wavelength \& multi-messenger campaigns with next generation of instruments, particularly if the magnetic obliquity of the system is $\alpha_{\rm B, orb} \gtrsim 45$
    \item Coherent emission is expected to turn on at a time $t \approx 10^{3} B_{\rm s, 12}$ seconds before merger
    \item A sub-population of CHIME/FRB detected bursts may involve mergers with $10^{14}$ G fields (Section \ref{sect:frb_survey})
    \item DSA-2000 and SKA FRB surveys will detect tens of bursts per year assuming fiducial parameters $\eta_{-2} \, B_{s, 12}^2$ (Section \ref{sect:frb_survey})
    \item Triggered observations of sGRBs and GW events by MWA \& LOFAR have already probed the most optimistic pre-merger emission parameters, and LOFAR 2.0 \& especially SKA will detect fiducial emission (Sections \ref{sect:ShortGRBs} \& \ref{sect:gw})
    \item Late-time observations in optical and radio of low redshift, localized one-off FRBs with quasi-periodic or increasing temporal structure is highly recommended (Section \ref{sect:afterglow} \& \ref{sect:kilonovae})
\end{enumerate}

\section*{Acknowledgements}
We would like to thank the referee for providing useful comments which improved the clarity of this work. AJC would like to thank Pragya Chawla for explaining the specifics of the CHIME/FRB pipeline and the anticipated specifications of CHORD. AJC also acknowledges useful discussions with Pawan Kumar, and with Jason Hessels about FRB searches with LOFAR (2.0) \& SKA.  
\par
AJC is supported by the Netherlands Research School for Astronomy (NOVA). OG acknowledges ASPIRE\footnote{https://aspire.science.uva.nl/} 2021 at the Anton Pannekoek Institute (API) for supporting part of this research. OMB acknowledges funding through Vici research program 'ARGO' with project number 639.043.815, financed by the Dutch Research Council (NWO). Research visits that contributed to this work were funded by the Leids Kerkhoven-Bosscha Fonds (LKBF). ZW is supported by the Fermi Guest Investigator program and the NASA Theory Program. The material is based upon work supported by NASA under award number 80GSFC21M0002. KG acknowledges Australian Research Council (ARC) grant DP200102243.

\section*{Data Availability}
A reproduction package providing the scripts required to reproduce the figures of this paper will be available upon publication.



\bibliographystyle{mnras}
\bibliography{references} 

%



\appendix

\section{Correction to the parallel electric field calculation}
\label{sect:Om_derivation}

The assumption that the magnetic field of the primary at the position of the secondary is uniform leads to the form of the magnetic flux density given by $\mathbf{B} = - B \mathbf{\hat{z}}$. This further assumes that the magnetic dipole moment of the primary is perpendicular to the orbital plane. In the inertial reference frame co-moving with the non-rotating primary, designated as frame $S$, the electric field $\mathbf{E} = 0$. We initially consider that only the primary is present in frame $S$.

We transform to a frame of reference $S'$ which is moving with velocity $\boldsymbol{\beta'} = \beta \mathbf{\hat{y}}$ with respect to frame S. The magnetic field in this frame $\mathbf{B'} = \mathbf{B} + O(\beta^2)$, where we will ignore second-order terms in $\beta$. Lorentz transformation of the electric field  for non-relativistic velocities follows 
\begin{equation}
\mathbf{E'}_{\rm LT} = \mathbf{E} + \boldsymbol{\beta'} \times \mathbf{B} = (\beta \mathbf{\hat{y}}) \times (-B \mathbf{\hat{z}}) = - \beta B \mathbf{\hat{x}}
\end{equation}

The secondary is considered to be a perfect electrical conductor with negligible magnetization and a vanishing internal magnetic field intensity. When the secondary is placed in the approximately uniform field $\mathbf{B'}$, it will expel the magnetic field from the inside by generating surface currents. The magnetic field outside the secondary in frame $S'$ is then given by, 
\begin{equation}
    \mathbf{B'} = -B \cos \theta \big( 1 - \frac{R^3}{r^3} \big) \mathbf{\hat{r}} + B\sin\theta\big(1+\frac{R^3}{2r^3}\big) \boldsymbol{\hat{\theta}} \label{eq:apptotB}
\end{equation}

Given that a uniform static electric field exists anti-parallel to the x-axis in $S'$, the conductor will generate a dipole field oriented along the x-direction in response to keep the sphere at uniform potential. Taking the dipole moment to be $\mathbf{p}$, we can write,
\begin{align*}
    &V'_{\rm dipole} = \frac{\mathbf{p}.\mathbf{\hat{r}}}{r^2} = \frac{p\sin\theta\cos\theta}{r^2} \nonumber\\
    &V'_{\rm uniform} = \beta Bx = \beta Br\sin\theta\cos\phi \nonumber\\
    &V'_{\rm tot} = V'_{\rm uniform} + V'_{\rm dipole} = \big( \frac{p}{r^2} + \beta Br)\sin\theta\cos\phi
\end{align*}
As the secondary has an equipotential surface we can choose $V'_{\rm tot} = 0$ at $r=R$. This implies that $p = -\beta BR^3$. As such,
\begin{equation}
    V'_{\rm tot} = \big(r - \frac{R^3}{r^2}\big)\beta B\sin\theta\cos\phi
\end{equation}
Therefore, electric field outside the secondary in frame $S'$,
\begin{align}
    \mathbf{E'} &= - \nabla V'_{\rm tot} \nonumber\\ &= - \Big\{ \big(\frac{2R^3}{r^3}+1\big)\sin\theta\cos\phi, \big(1-\frac{R^3}{r^3}\big)\cos\theta\cos\phi, \big(\frac{R^3}{r^3}-1\big)\sin\phi \Big\} \beta B
\end{align}
With the secondary being stationary in $S'$, we can apply the electromagnetic interface conditions in this frame. Using the normal field boundary condition,
\begin{equation}
    \mathbf{E'}.\hat{r}|_{r=R} = -3\beta B\sin\theta\cos\phi = 4\pi\sigma'
\end{equation}
Surface currents responsible for inducing the magnetic dipole field component of Eq. \eqref{eq:apptotB} are given by $\mathbf{K'} = \frac{c}{4\pi}\mathbf{n} \times \mathbf{B'}|_{r=R} = \frac{3c}{8\pi}B\sin\theta \boldsymbol{\hat{\phi}}$. For Lorentz transformations, we also need to convert the surface current density and the surface charge distribution to the form of the current 4-vector.
\begin{align}
    & \mathbf{J'} = \frac{3c}{8\pi}B\sin\theta\delta(r-R) \boldsymbol{\hat{\phi}} \\
    & \rho_{\rm q}' = -\frac{3}{4\pi}\beta B\sin\theta\cos\phi\delta(r-R)
\end{align}
This conversion from surface density to volume density requires the usage of the Dirac-delta function centered at $R$ in the radial direction.

Now, we transform to a frame of reference $S''$ where the secondary neutron star moves with velocity $\beta \mathbf{\hat{y}}$. This means that, $S''$ must be moving with $\boldsymbol{\beta''}=-\beta\mathbf{\hat{y}}$ with respect to the frame $S'$. Note that $S''$ is essentially the frame $S$, but whereas S did not have the secondary neutron star and its associated fields, $S''$ will have all the transformed fields from $S'$. The designation of $S''$ as such is only done for convenience.
The magnetic field exhibits no change for $\beta<<1$, $\mathbf{B''} = \mathbf{B'} - \frac{1}{c}(-\beta\mathbf{\hat{y}})\times \mathbf{E'} = \mathbf{B'} + O(\beta^2/c)$. The electric field $\mathbf{E''} = \mathbf{E'} + (-\beta\mathbf{\hat{y}})\times \mathbf{B'}$.
\begin{align*}
    \mathbf{\hat{y}}\times \mathbf{B'} = &\{\sin\theta\sin\phi, \cos\theta\sin\phi, \cos\phi\}\times \mathbf{B'} \\
    = &-\Big(1+\frac{R^3}{2r^3}\Big)B\sin\theta\cos\phi \mathbf{\hat{r}} + \Big(\frac{R^3}{r^3}-1\Big)B\cos\theta\cos\phi \boldsymbol{\hat{\theta}} \\ &+ \Big(1+\frac{R^3}{2r^3} (\sin^2\theta-2\cos^2\theta) \Big)B\sin\phi \boldsymbol{\hat{\phi}}
\end{align*}
\begin{equation}
    \Rightarrow \; \mathbf{E''} = -\Big\{ \frac{3}{2}\sin\theta\cos\phi, 0, \frac{3}{2}\sin^2\theta\sin\phi \Big\} \frac{\beta BR^3}{r^3}
\end{equation}
The volume current density and the volume charge in $S''$ frame, then, are
\begin{align}
    \mathbf{J''} &= \mathbf{J'} - \rho_{\rm q}'(-\beta\mathbf{\hat{y}}) = \frac{3c}{8\pi}B\sin\theta\delta(r-R)\hat{\phi} + O\big(\beta^2\big) \\
    \rho_{\rm q}'' &= \rho_{\rm q}' - \frac{1}{c}\mathbf{J'}.(-\beta\mathbf{\hat{y}}) \nonumber\\
    &= -\frac{3}{4\pi}\beta B\sin\theta\cos\phi\delta(r-R) - \frac{1}{c} \big(\frac{3c}{8\pi}B\sin\theta\delta(r-R)\big)(-\beta\cos\phi) \nonumber\\
    &= - \frac{3}{8\pi}\beta B\sin\theta\cos\phi\delta(r-R)
\end{align}

As a result, the surface current density and the surface charge density in $S''$ frame are obtained to be $\mathbf{K''}=\frac{3c}{8\pi}B\sin\theta\hat{\phi}$ and $\sigma'' = - \frac{3}{8\pi}\beta B\sin\theta\cos\phi$ respectively.

$\mathbf{E''}$ is the total electric field vector outside the secondary neutron star in $S''$. Using this, we can calculate the component of the electric field in the direction of the magnetic field $\mathbf{B''}$.
\begin{align}
    E''_{\parallel} &= \frac{\mathbf{E''.B''}}{|\mathbf{B''}|} = \frac{1}{|\mathbf{B''}|} \frac{3\beta B^2R^3}{2r^3} \big(1-\frac{R^3}{r^3}\big)\sin\theta\cos\theta\cos\phi \nonumber \\
    &= \frac{3\sin\theta\cos\theta\cos\phi \big(1-\frac{R^3}{r^3}\big)}{\sqrt{4\cos^2\theta \big(1-\frac{R^3}{r^3}\big)^2 + \sin^2\theta\big(2+\frac{R^3}{r^3}\big)^2}}\beta B\frac{R^3}{r^3}
\end{align}
This corresponds to Eq. \eqref{eq:om} herein.

\section{Gap height derivation for lower number density case}
\label{app:goldreichapprox}
We present the lower limit to the analytic gap height, assuming that $E_{\rm gap} = 4 \pi q n h_{\rm gap}$ and $n$ is given by the Goldreich-Julian density due to the motion of the magnetosphere in orbit: $n = n_{GJ} = (2 B)/(q c P_{\rm orb})$, where $B$ is the local magnetic field. 

Starting from Eq. \eqref{eq:gamma_l}, the Lorentz factor of the primaries is:
\begin{equation}
    \gamma(l) = \frac{q E_{\rm gap} l_{\rm acc}}{m_e c^2} \\
\end{equation}
By way of Eqs. \eqref{eq:energy_photons} \& \eqref{eq:pair_threshold} and by substitution we find:
\begin{equation}
\begin{split}
    l_{\gamma, \rm gap} &= 0.4 \rho_{\rm c} m_e c^2 \frac{B_{\rm c}}{B} \frac{\rho_{\rm c}}{3 \hbar c} \bigg( \frac{ m_e c^2}{4 \pi q^2 n_{\rm GJ} l_{\rm acc}^2}\bigg)^3 \\
    &= \frac{0.4 \rho_{\rm c}^2 m_e^4 c^7}{192 \pi^3 \hbar  q^6 n_{\rm GJ}^3 l_{\rm acc}^6} \frac{B_{\rm c}}{B} \\
\end{split}
\end{equation}
Let $k_2 = \frac{0.4 \rho_{\rm c}^2 m_e^4 c^7 B_{\rm c}}{192 \pi^3 \hbar  q^6 n_{\rm GJ}^3 B}$.
Minimizing $h_{\rm gap} = l_{\gamma, \rm gap} + l_{\rm acc}$ we find:
\begin{equation}
\begin{split}
    l_{\rm acc} &= (6 k_2)^{1/7} \\
    l_{\gamma, \rm gap} &= k_2 (6 k_2)^{-6/7} = \frac{k_2^{1/7}}{6^{6/7}}
\end{split}
\end{equation} 
Therefore the $h_{\rm gap}$, in the lower number density limit, is:
\begin{equation}
    \begin{split}
    h_{\rm gap} &= \frac{k_2^{1/7}}{6^{6/7}} + (6 k_2)^{1/7} = \frac{7 k_2^{1/7}}{6^{6/7}} \\
     &= \frac{14}{6^{6/7}} \bigg(\frac{\rho_{\rm c}^2 m_e^4 c^{10} B_{\rm c} P_{\rm orb}^3}{  3840 \pi^3 \hbar q^3 B^4} \bigg)^{1/7} 
    \end{split}
    \label{eq:gap_appendix}
\end{equation}
Where we have included a factor of 2 to account for relative motion of pairs as before. Making use of the fact that $E_{\rm gap} \approx 4 \pi q n_{\rm GJ} h_{\rm gap}$ and $A \approx 4 \pi R_{\rm NS}^2$, we can estimate the radio luminosity as:
\begin{equation}
    \begin{split}
        L_r &= \eta q \Phi_{\rm gap} \dot{N} \\
        &= \eta q E_{\rm gap} h_{\rm gap} n_{\rm GJ} A c \\
        &= 4 \pi \eta q^2  h_{\rm gap}^2 n_{\rm GJ}^2 A c \\
        &= 8 \times 10^{38} \; \eta_{-2} \, h_{\rm gap,3}^2 \, B_{12}^2 \, P_{-3}^{-2} \, R_{\rm NS,6} \; {\rm erg s^{-1}}
    \end{split}
    \label{eq:coherent_lum_appendix}
\end{equation}
We note here that in this derivation of the coherent radio luminosity, there is no explicit dependence on the value of $E_{\parallel}$.

\section{Coherent curvature radiation as an alternative radiation mechanism}
\label{app:cohcurv}
In light of the discovery of FRBs \citep{Lorimer2007,Thornton2013,Petroff2021}, a variety of radiation models have been proposed to explain their origin. A large fraction of the proposed theories rely on highly magnetized neutron stars as progenitors (e.g. \citealt{Keane2012,Popov2013,Katz2016,Belo2017}). Coherent curvature radiation is one such model that has the flexibility to explain many of the observed properties of these enigmatic bursts \citep{Kumar2017,Katz2018,Ghisellini2018,Lu2018,Lu2020,CooperWijers2021}. The model relies on the formation of overdensities (bunches) of pairs,for example due to the two-stream instability \cite{Lu2018,Kumar2020}, such that curvature radiation is emitted coherently. Critics argue that the formation and maintance of particle bunches in spite of repellent electromagnetic forces \citep{Lyubarsky2021}. However, if bunches are formed, one could expect the conditions of the NS-NS merger ($E_{\parallel} \sim 10^{10}$ esu, $B \approx 10^{12}$ G) to be conducive to coherent curvature emission in a similar manner to models of FRBs from magnetospheres. It is however less clear that the coherent curvature mechanism would continuously operate effectively at all times that necessary electromagnetic conditions demanded of $B$ and $E_{\parallel}$ are met. In the following we estimate the coherent radio luminosity of this mechanism during a NS-NS merger inspiral. 
\par
The critical frequency of curvature radiation $\nu_c = \frac{c \gamma^3}{2 \pi \rho_{\rm c}}$ where $\gamma$ is the Lorentz factor of radiating electrons and $\rho_{\rm c}$ is the magnetic field line curvature radius; implying emission observed at $\nu_{\rm obs}$ is emitted by electrons with $\gamma \approx 60 \, \rho_{\rm c, 6}^{1/3} \, \nu_9^{1/3}$. The radiation formation length scale is $\sim \rho_{\rm c}/ \gamma \approx 10^{4}$cm \citep{Lu2018}, much smaller than the spatial extent of $E_{\parallel}$ which is $\sim 10^{6}$cm. As in \cite{Kumar2017}, the isotropic equivalent luminosity can be estimated as:
\begin{equation}
\begin{split}
        L_{\rm iso} &\approx \frac{c^7 q^2 \gamma^{10} n_e^{'2}}{3 \nu_{\rm obs}^6 \rho_{\rm c}^2} \\
        &= 2 \times 10^{37} \: {\rm erg \, s^{-1}} \:  \nu_9^{-8/3} \, n_{\rm e,12}^{'2} \, \rho_{\rm c, 6}^{4/3}
        \label{eq:lum_ccr}
    \end{split}
\end{equation}
Where we have replaced $\gamma$ in terms of $\nu_{\rm obs}$ \& $\rho_{\rm c}$. In the following, we briefly discuss 2 curvature radiation luminosity limits to gauge plausible luminosities in the NS-NS merger scenario. 
\par
In \cite{LuKumar2019}, the authors discuss the maximum luminosity of FRBs in the curvature radiation model due to the rapid production of Schwinger pairs which screen the $E_{\parallel}$ field on very short timescales if $E > E_s \approx \frac{m_e^2 c^3}{q \hbar}$. In the proof, the authors derive generic constraints (i.e. unrelated to extreme Schwinger electric field) on the isotropic equivalent luminosity of coherent emission:
\begin{equation}
\begin{split}
        L_{\rm iso} &< E_{\parallel}^2 \rho_{\rm c}^2 c 
        = 3 \times 10^{42} \: {\rm erg \, s^{-1}} \: E_{\parallel,10}^2 \, \rho_{\rm c, 6}^2
\end{split}
\end{equation}
Where $E_{\parallel}$ is the magnitude of the parallel electric field component and $\rho_{\rm c}$ is the local curvature radius. 
\par
In \cite{Kumar2017}, the authors note that particles emitting coherent curvature radiation induce a perpendicular component to the magnetic field $B_{\perp} = 8 \pi l_{\perp} q n_{\rm e}$, where $l_{\perp}$ is the spatial extend of the coherently radiating particles perpendicular to their motion, and $n_e$ is the number density. For the particles to remain coherent, their momenta should remain aligned to a factor $\frac{\delta \textit{\textbf{p}}}{|p|} < \frac{1}{\gamma}$, where $\gamma$ is the Lorentz factor of the accelerated particles. Particles are in the lowest landau state and move only along magnetic field lines, therefore we should also require that $\frac{B_{\perp}}{B} < \frac{1}{\gamma}$. The results in a maximum luminosity of coherent curvature radiation within a local magnetic field strength $B$ of:
\begin{equation}
\begin{split}
        L_{\rm iso} &< 6 \times 10^{40} \: {\rm erg \, s^{-1}} \: B_{12}^2 \, \nu_9^{-4/3} \, \rho_{\rm c, 6}^{2/3}  \\
\end{split}
\end{equation}
We neglect to include maximum luminosity constraints discussed in \cite{CooperWijers2021} to remain agnostic about the unknown particle number density in the NS-NS merger case. Combining these constraints, we can estimate the maximum isotropic luminosity as:

\begin{equation}
\begin{split}
    L_{\rm iso, max} = \min \big( & 3 \times 10^{42}  \: E_{\parallel,10}^2 \, \rho_{\rm c, 6}^2, \\
    & 6 \times 10^{40} \: B_{12}^2 \, \nu_9^{-4/3} \, \rho_{\rm c, 6}^{2/3} \big) \: {\rm erg \, s^{-1}}
\end{split}
\label{eq:cohcurvlum}
\end{equation}
The number of coherent bunches that may radiate towards the same observer is uncertain, as is the beaming factor of each curvature powered radio burst. In our numerical implementation for the luminosity estimate of the coherent curvature radiation mechanism, we simply show the maximum allowed luminosity directed within a $0.1$ radian solid angle of the observer. The primary result of this subsection is shown in Fig. \ref{fig:GW}, where the black line denotes viewing angle dependent maximum CCR luminosity of a merger involving a $10^{12}$G neutron star. We note that the luminosity profile approximately matches the pulsar-like case with an efficiency of approximately $\eta = 10^{-4}$ and therefore we do not show the CCR luminosity profile in the remainder of our results and discussion. The viewing-angle dependent luminosity profile of CCR is different to the pulsar-like case. This is because it is calculated by finding the maximum possible CCR luminosity (using Eq. \ref{eq:cohcurvlum}) for the set of field lines directed towards each observer, rather than the sum of the luminosity along the all observer-aligned field lines. 

\section{Additional co-detectability plots}
\label{sect:additional_plots}
In this Section we include additional plots as referenced in the main text. In all cases, overplotted are flux sensitivity curves for various efficiency parameters and $B_{\rm s} = 10^{12}$G, assuming a fluence limit of the Square Kilometre Array of 1 mJy ms. 

\begin{figure}
  \centering
{\includegraphics[width=.5\textwidth]{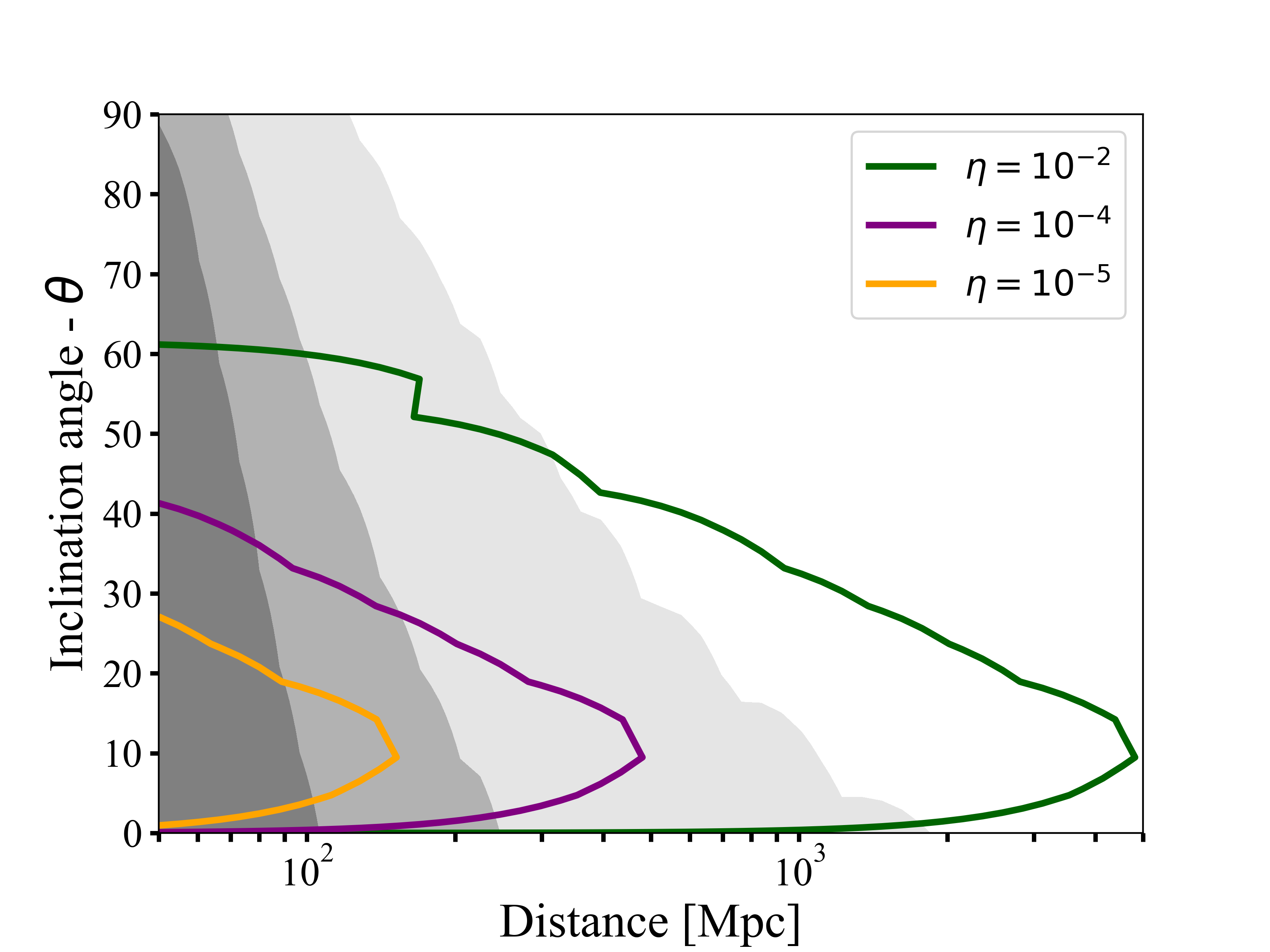}}
\caption{Background colours show MeerKAT detection horizons for peak flux of tophat jets at $\nu = 1.43 \; {\rm Hz}$ assuming detection threshold of $700 \; {\rm \mu Jy}$. Shades of gray correspond to varying circumburst densities (Darkest $n=10^{-5}$; medium $n=10^{-3}$; lightest $n=10^{-1}$).}
\label{fig:afterglow_tophat_MeerKat}
\end{figure}

\begin{figure}
  \centering
{\includegraphics[width=.5\textwidth]{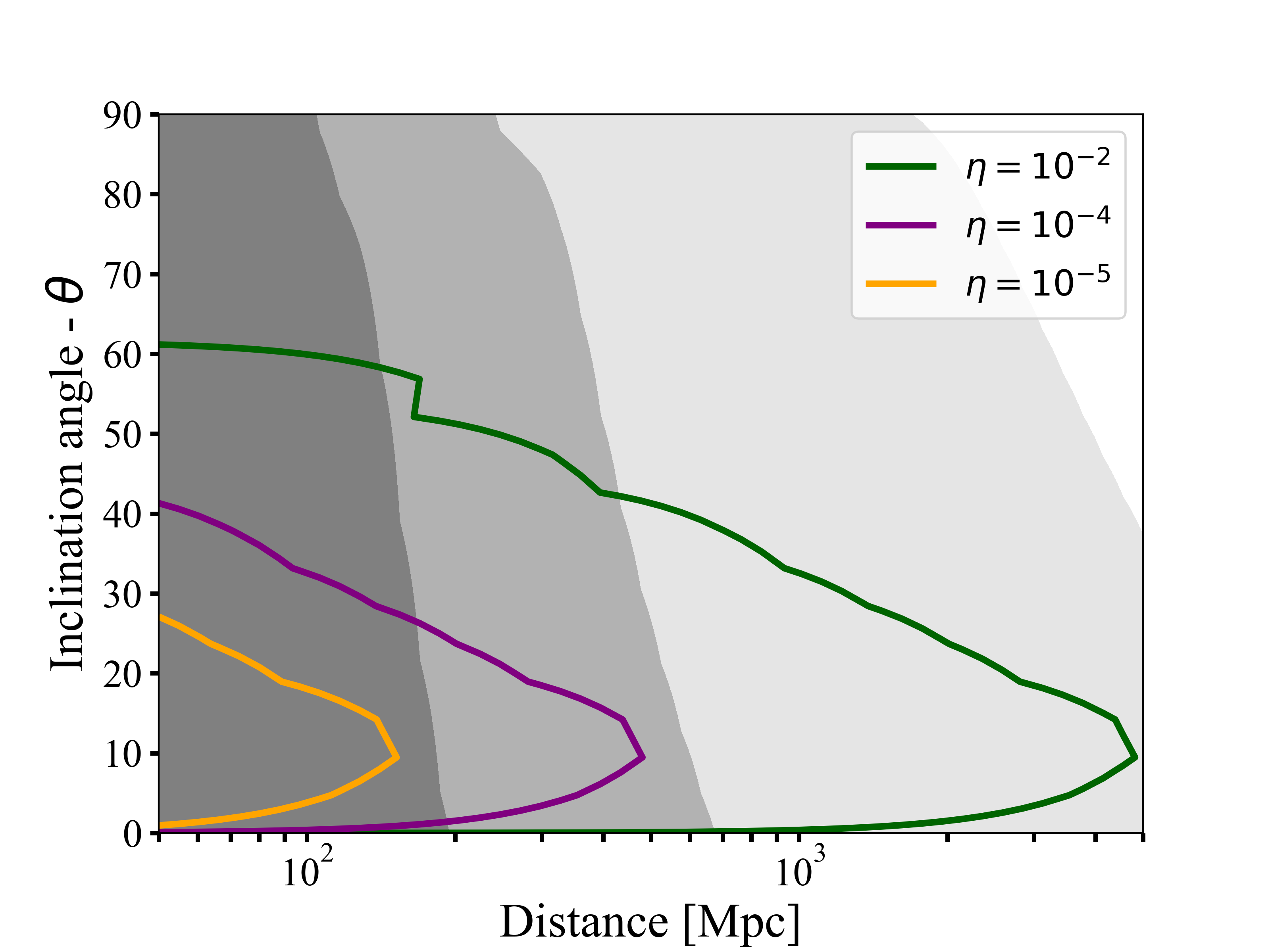}}
\caption{Background colours show SKA detection horizons for peak flux of tophat jets at $\nu = 1.43 \; {\rm Hz}$ assuming detection threshold of $2 \; {\rm \mu Jy}$. Shades of gray correspond to varying circumburst densities (Darkest $n=10^{-5}$; medium $n=10^{-3}$; lightest $n=10^{-1}$). }
\label{fig:afterglow_tophat_SKA}
\end{figure}

\begin{figure}
  \centering
{\includegraphics[width=.5\textwidth]{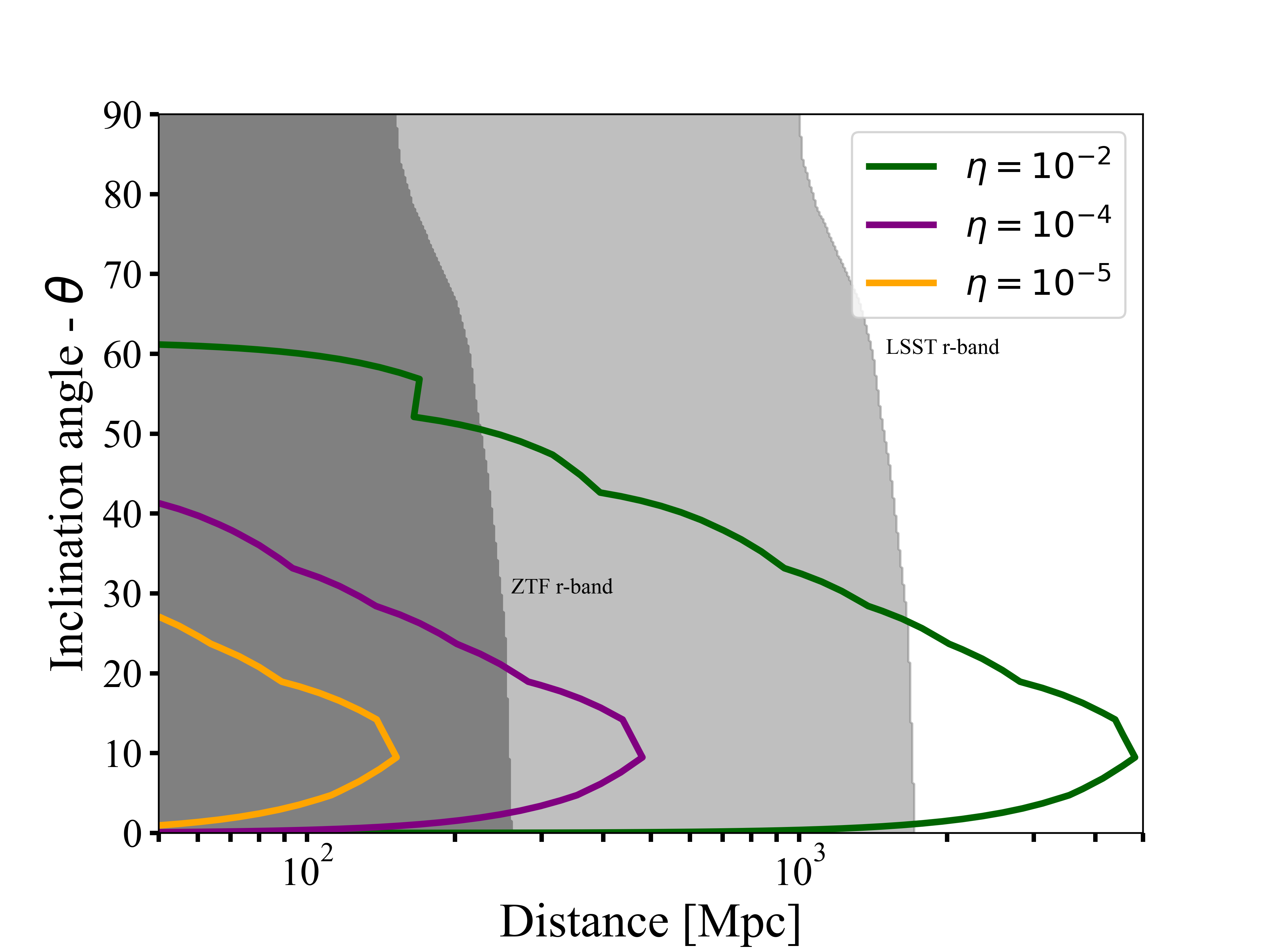}}
\caption[1]{Background colours show ZTF and LSST r-band (darker and lighter shades respectively) detection horizons for peak flux of GW170817-like kilonovae models using models from \cite{Bulla2019}. Coloured contours show viewing angle-dependent SKA detection horizon of pulsar-like coherent pre-merger emission assuming $\eta = 10^{-2}$, $B_{\rm s} = 10^{12}$G.}
\label{fig:kilonovae_r}
\end{figure}

\begin{figure}
  \centering
\includegraphics[width=.5\textwidth]{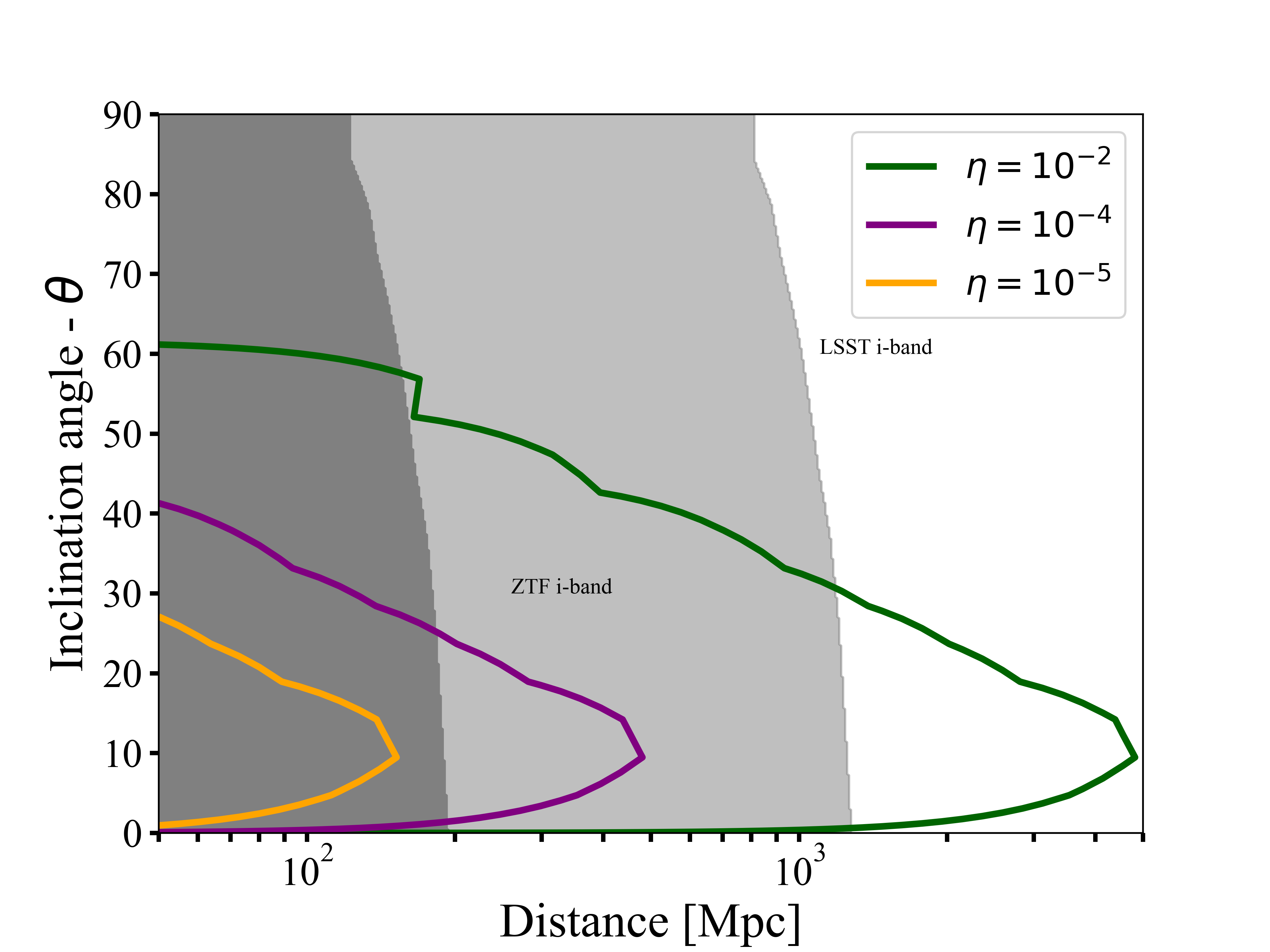}
\caption[1]{Background colours show ZTF and LSST i-band (darker and lighter shades respectively) detection horizons for peak flux of GW170817-like kilonovae models using models from \cite{Bulla2019}. Coloured contours show viewing angle-dependent SKA detection horizon of pulsar-like coherent pre-merger emission assuming $\eta = 10^{-2}$, $B_{\rm s} = 10^{12}$G.}
\label{fig:kilonovae_i}
\end{figure}


\bsp	
\label{lastpage}
\end{document}